\documentclass{aa} 
\usepackage{graphicx,natbib,psfig,amssymb,multirow,sidecap} 
\newcommand{\ltsima} {$\; \buildrel < \over \sim \;$} 
\newcommand{\gtsima} {$\; \buildrel > \over \sim \;$} 
\newcommand{\lta} {\lower.5ex\hbox{\ltsima}} 
\newcommand{\gta} {\lower.5ex\hbox{\gtsima}} 
\newcommand{\Ha} {H$\alpha$}
\newcommand{\Hb} {H$\beta$}

\begin{document} 
\title{On the spectro-photometric properties of the bulk of the
  radio-loud AGN population.}
  
\titlerunning{Spectro-photometric properties of the radio-loud AGN population.}

\authorrunning{R.D. Baldi \& A. Capetti}
  
\author{Ranieri D. Baldi
\inst{1,2}
\and  Alessandro Capetti \inst{2}} 
\offprints{R.D. Baldi}  
\institute{
Universit\'a di Torino, via P. Giuria 1, 10125 Torino, Italy\\
\email{baldi@oato.inaf.it}
\and 
INAF - Osservatorio Astronomico di Torino, Strada
  Osservatorio 20, I-10025 Pino Torinese, Italy\\
\email{capetti@oato.inaf.it}}

\date{}  
   
\abstract{In a previous paper we showed that the radio sources selected by
  combining large areas radio and optical surveys, present a strong deficit of
  radio emission with respect to 3CR radio-galaxies matched in line emission
  luminosity. We argued that the prevalence of sources with luminous extended
  radio structures in high flux limited samples is due to a selection
  bias. Sources with low radio power form the bulk of the radio-loud AGN
  population but are still virtually unexplored.

  We here analyze their photometric and spectroscopic properties. From the
  point of view of their emission lines, the majority of the sample are Low
  Excitation Galaxies (LEG), similar to the 3CR objects at the same level of
  line luminosity. The hosts of LEG are red, massive ($10.5 \lesssim {\rm log}
  \, M_*/M_{\odot} \lesssim 12$) Early-Type Galaxies (ETG) with large black
  holes masses ($7.7 \lesssim {\rm log} \, M_{\rm{BH}}/M_{\odot} \lesssim 9$),
  statistically indistinguishable from the hosts of low redshift 3CR/LEG
  sources. No genuine radio-loud LEG could be found associated with black
  holes with a mass substantially lower than $10^8 M_{\odot}$ or with a late
  type host. The fraction of galaxies with signs of star formation ($\sim
  5\%$) is similar to what is found in both the quiescent ETG and 3CR/LEG
  hosts. We conclude that the deficit in radio emission cannot be ascribed to
  differences in the properties of their hosts. We argue that instead this
  could be due to a temporal evolution of the radio luminosity.

  A minority ($\sim 10\%$) of the sample show rather different properties,
  being associated with low black hole masses, with spiral galaxies, or
  showing a high excitation spectrum. In general these outliers are the result
  of the contamination from Seyfert and from galaxies where the radio emission
  is powered by star formation. For the objects with high excitation
  spectra there is no a clear discontinuity in either the host or nuclear
  properties as they span from radio-quiet and radio-loud AGN.

  }

\maketitle

\section{Introduction}

The advent of large area surveys opened the possibility for the scientific
community to explore large samples of extragalactic sources and to set the
results on several key issues on strong statistical foundations. In
particular, the cross-match of astronomical data from radio and optical surveys
provides a unique tool in the analysis of the radio emitting galaxies, since
it allows us to identify optically large numbers of radio sources, to obtain
spectroscopic redshifts, to determine the properties of their hosts, and to
build up their spectral energy distributions.

In recent years several studies have been indeed carried out on large samples
of radio galaxies in order to investigate the links between the radio
structures, the central engine associated with an Active Galactic Nucleus
(AGN), and the host galaxies. In particular, \citet{best05b} selected a sample
of 2215 low-luminosity radio galaxies by cross-correlating SDSS (Data Release
2), NVSS, and FIRST\footnote{Sloan Digital Sky Survey, \citep{york00},
  National Radio Astronomy Observatory (NRAO) Very Large Array (VLA) Sky
  Survey \citep{condon98}, and the Faint Images of the Radio Sky at Twenty
  centimeters survey \citep{becker95} respectively.}. The resulting catalogue
is by far larger than any previously studied sample of fully optically
identified radio-sources.  This sample (hereafter we refer to it as SDSS/NVSS
sample) is highly (95 \%) complete down to the flux threshold of 5 mJy and
provides a very good representation of radio-galaxies in the local Universe,
up to a redshift of $\lesssim 0.3$, covering the range $10^{38} - 10^{42}$ erg
s$^{-1}$ in radio power.

In a subsequent paper \citep{best05a} studied the properties of the host
galaxies of these radio-loud AGN (hereafter RLAGN) finding that these are
massive galaxies, usually of early Hubble type, with stellar masses in the
range log $(M_*/M_{\odot}) \sim 10^{10}-10^{12}$, and located in richer
environments than normal galaxies.

The SDSS/NVSS sample was then considered by \citet{baldi09} in the context of
the properties of low luminosity radio-galaxies. We show that they display a
strong deficit of radio emission with respect to their nuclear emission-line
luminosity when compared to sources, part of other samples of radio-galaxies,
matched in line luminosity.

This result is particularly intriguing considering that the presence of a
strong correlation between line and radio luminosity is instead well
established (see, e.g.,
\citealt{baum89b,rawlings89,morganti97,tadhunter98,willott99}) from the study
of different samples of radio-galaxies. Both Low and High Excitation Galaxies
(LEG and HEG respectively) obey separately to such a correlation, although the
normalizations for the two classes differ \citep{buttiglione10}. This suggests
that the energy source of the narrow lines is closely linked to the source of
the radio emission. It can be envisaged that a constant fraction of the
available accretion power is channeled into radiative emission (that powers
the emission line regions) and into jet kinetic energy (of which the radio
emission is the electromagnetic manifestation). The study of the SDSS/NVSS
sample suggests instead that total radio luminosity and line emission are
independent from each other.

Nonetheless, considering the properties of miniature radio-galaxies
\citep{balmaverde06b} we found that a link exists between radio {\sl core} and
line emission, extending to low luminosities the analogous relation present in
high power sources. By assuming that the [O~III] luminosity gives an
appropriate bolometric estimate of AGN power, and that radio cores are a good
proxy for the jet power, radio galaxies with similar nuclear properties are
able to produce an extremely wide range of total radio power, of a factor
$\geq$100. Indeed, these radio-galaxies of very low power are all highly
core-dominated, with only feeble extended emission. A similar sample, studied
by \citet{prandoni09} at higher frequency (1.4-15 GHz) and selected at lower
flux cutoff ($>1$ mJy), also shows a substantial fraction of very compact
radio morphologies (d$<$10 kpc) in the same range of radio power covered by
3CR and B2 sample. We argued that the prevalence of sources with luminous
extended radio structures in high flux limited samples is due to a selection
bias, since the inclusion of such objects is highly favored. Core dominated
sources with a low ratio between radio and emission line power form the bulk
of the local radio-loud AGN population but they are still virtually
unexplored. Unfortunately, the images available for the SDSS/NVSS sample are
not sufficient to properly isolate their radio cores.

The aim of this paper is a better understanding of the physical properties of
these RLAGN. In particular we will explore the spectro-photometric properties
of their hosts looking for possible differences that might explain their low
level of radio emission with respect to the classical powerful radio galaxies
matched in line luminosity. Operatively, with the information provided by SDSS
we classify the optical spectra of the AGN on the basis of the excitation
level. Then, we analyze the key properties of the hosts and the AGN, like the
black hole mass, the mass and colors of the hosts, drawing a quantitative
comparison with the 3CR sample. The main result is that the sample shows
indistinguishable properties from those of 3CR sample. This implies that
the reasons for the different radio properties between the bulk of the RLAGN
population and the most studied sample of radio-galaxies must be ascribed to
other mechanisms.

The paper is organized as follows. In Sect.~\ref{sample} we briefly present
the sample selection carried out by \citet{best05a}. In Sect.~\ref{results}
and ~\ref{highz} we analyze the spectro-photometric properties of the
SDSS/NVSS sample considering separately the sub-samples in the redshift ranges
$0.03<z<0.1$ and $0.1<z<0.3$. What emerges from the study is that the sample
is mostly composed of red, massive early-type galaxies with large black hole
mass, with spectra typical of LEG. In
Sect.~\ref{comp3c} we compare their broadband properties with those of the 3CR
sample, finding clear similarity with LEG. The few apparent exceptions to this
general rule are discussed in Sect.~\ref{exceptions}. The summary and
conclusions to our findings are expounded in Sect.~\ref{summary}. In the
Appendix we provide the spectro-photometric diagrams of the sub-sample with
$0.1<z<0.3$.

\section{Sample selection}
\label{sample}

For the analysis we use the sample of RLAGN selected by \citet{best05a}. This
sample is obtained cross-matching the $\sim$212000 galaxies drawn from the
SDSS-DR2 with the NVSS and FIRST radio surveys. Leaving aside 497 sources
identified as star forming galaxies, this radio-selected sample consists of
2215 radio luminous AGN having a radio flux larger than 5 mJy at 1.4GHz.  

The Sloan Digital Survey (\citealt{york00,stoughton02}) with its optical
imaging and spectroscopic survey of about a quarter of the sky provides a
variety of physical parameters for the galaxies of this sample. These include
emission line fluxes, stellar velocity dispersions, optical magnitudes in 5
bands, stellar masses, concentration indices, and strength of stellar
absorption features such as the 4000\AA\ break strength ($D_n(4000)$). These
photometric and spectroscopic parameters have been used for the analysis of
this paper.

\begin{figure*}
\centerline{
\includegraphics[scale=0.75,angle=90]{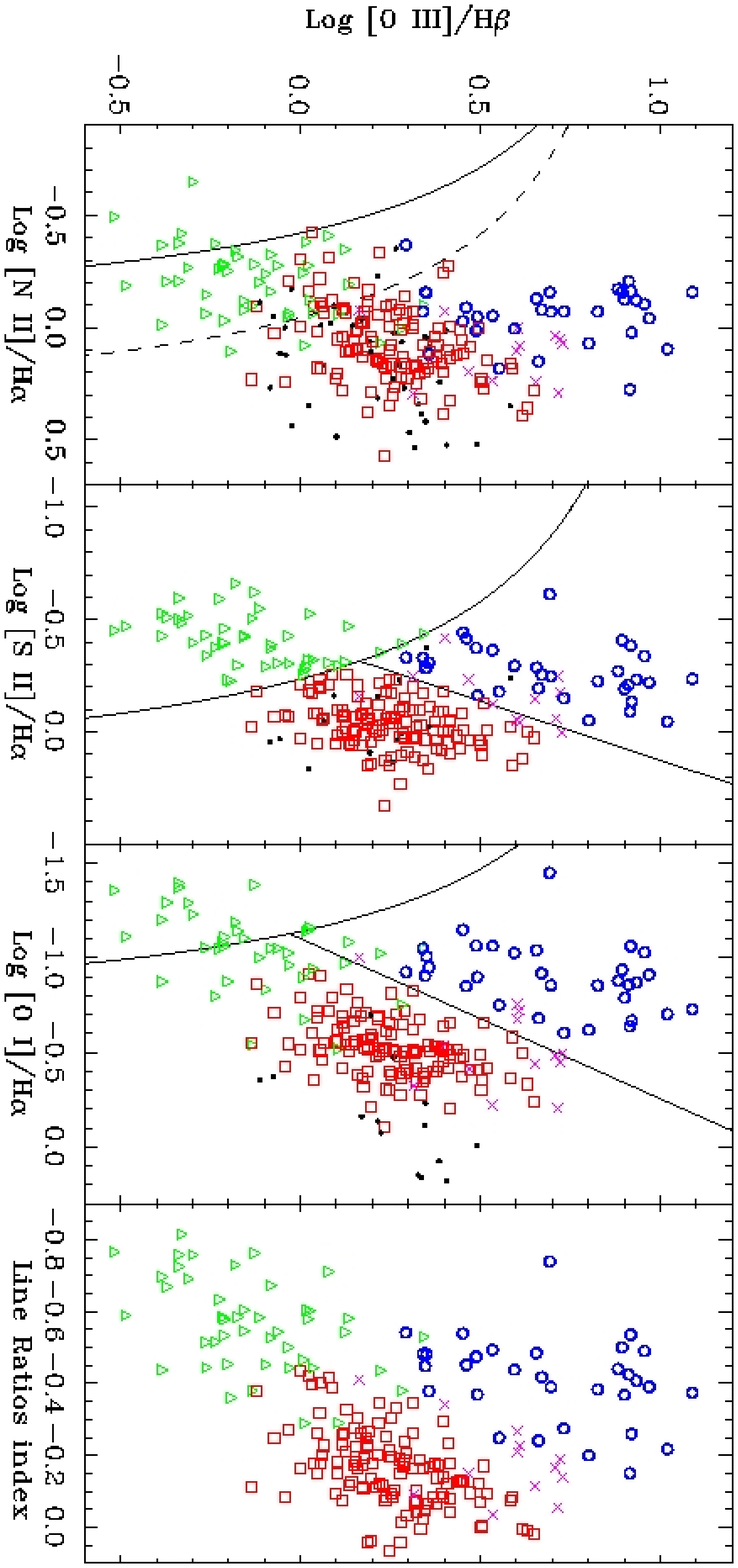}
}
\caption{Spectroscopic diagnostic diagrams for the objects with
  $0.03<z<0.1$: in the first panel, log([O~III]/H$\beta$)
  vs. log([N~II]/H$\alpha$), in the second log([O~III]/H$\beta$)
  vs. log([S~II]/H$\alpha$), and in the third, log([O~III]/H$\beta$)
  vs. log([O~I]/H$\alpha$).  The solid lines separating Star Forming Galaxies,
  LINER, and Seyfert are from \citet{kewley06}; in the first panel the
  region between the two curves is populated by the Composite Galaxies. In the
  fourth panel we give the location of each source derived from the previous
  diagrams in the plane log([O~III]/H$\beta$) vs. Line Ratios Index (see
  text). We mark LEG as red squares, HEG as blue circles, Star Forming
  Galaxies as green triangles, Ambiguous Galaxies as pink crosses, and
  Unclassified Galaxies as black dots.  }
\label{diagn}
\end{figure*}
\section{Results: the sub-sample with $0.03<z<0.1$}
\label{results}

In this section we describe the results obtained considering the sub-sample
of 425 SDSS/NVSS galaxies with $0.03<z<0.1$. The more distant
objects will be instead discussed in Sect. \ref{highz}.
 
\subsection{Optical spectroscopic diagnostic diagrams}

\citet{kewley06} considered a sample of $\sim$85000 emission line galaxies
from the SDSS, using the location of a galaxy in the spectroscopic diagnostic
diagrams, planes formed by pair of emission line ratios, to recognize the
nature of their nuclear emission, e.g. separating active nuclei from star
forming galaxies (e.g. \citealt{baldwin81}). Furthermore, \citet{kewley06}
noted that AGN form branches of different excitation level, i.e. Seyfert and
LINER \citep{heckman80}.

We start our analysis considering the 425 objects with $0.03 <z< 0.1$.  In the
first three panels of the Figure~\ref{diagn} we show the location in the
diagnostic diagrams for the objects that have emission lines detected at
SNR$>$3 separately for each diagram. In the plane log([O~III]/H$\beta$)
vs. log([O~I]/H$\alpha$) there is a well defined ``branch'' of low excitation
galaxies. We also find high excitation AGN but in lower
number. 

\citet{best05a} excluded galaxies in which the radio emission can be accounted
by the presence of star formation\footnote{\citet{best05a} divided the sample
  into two classes: RLAGN and galaxies in which the radio emission is
  dominated by star formation. The division is based on the location of a
  galaxy in the plane of 4000-\AA\ break strength versus radio luminosity per
  unit stellar mass, see Figure \ref{SFGselection}.}. Nonetheless, several
objects with a starburst-like optical spectrum are present in the sample. The
radio emission in these galaxies is possibly powered by the AGN, but their
emission lines likely have a dominant contribution from regions of star
formation.

\begin{table*}
  \caption{Spectral and morphological classification breakdown}
  \label{crit}
  \begin{tabular}{c| c| c c c c c| c}
    \hline 
\hline
                         &                  &Unclass. &LEG    & AG    &   HEG  & SFG    & Total\\
    \hline                                                            
$0.03<z<0.1$             &                  &{\bf213}&{\bf122}& {\bf13}& {\bf32}&{\bf45} & {\bf425} \\
    \hline                                                            
\multicolumn{8}{c}{} \\                                               
\hline                                                                
$C_{r}>2.86$             &$D_n(4000)>1.7$ &181     &    88  &  9    &   4   &  8       & 290\\
and $\sigma(C_{r})<0.2$  &$D_n(4000)<1.7$ &  1     &     7  &   1   &  11   &  9       &  29\\
\hline                                                                
                         & subtotal         &{\bf182}&{\bf95}&{\bf10}&{\bf15} &{\bf17}   &{\bf319}  \\
\hline                                                                
$2.6<C_{r}<2.86$         &$D_n(4000)>1.7$&  14    &    9   &  0    &   1   & 3        & 27    \\
and $\sigma(C_{r})<0.2$  &$D_n(4000)<1.7$&  0     &    4   &  0    &   8   &  8       &   20  \\
\hline                                                                
                         &  subtotal       &{\bf14}  &{\bf13}&{\bf0} & {\bf9} &{\bf11}    &  {\bf47} \\
\hline                                                                
$C_{r}<2.6$              &$D_n(4000)>1.7$&  2     &    3   &  2    &   0   & 2        &  9   \\
and $\sigma(C_{r})<0.2$  &$D_n(4000)<1.7$&  2     &    4   & 1     &   6   &  13      &   26  \\
\hline                                                                
                         & subtotal        & {\bf4} &{\bf7}  &{\bf3} &  {\bf6}&{\bf15}   &  {\bf35}   \\
\hline                                                                
 Total                   &                 &{\bf200}&{\bf115}&{\bf13}&{\bf30} &{\bf43}   &{\bf401}    \\
\hline                                                                
\multicolumn{8}{c}{} \\                                               
\hline                                                                
$C_{r}>2.86$             & ``red''         &176     &  87    &  9    &  6     & 11     &{\bf289}\\
and $\sigma(C_{r})<0.2$  & ``blue''        & 6      &   8    &  1    &  9     &  6     &{\bf30}\\
\hline
  \end{tabular}

\end{table*}

Our optical classification is based on the simultaneous identification of an
object in all three diagnostic diagrams. 212 objects, that have SNR$>$3 for
each of the 6 key emission lines, could be classified (see Table
\ref{crit}). A source is classified as star forming galaxy (SFG), when it is
located below the curved solid lines in even just one diagram.  
This approach is rather conservative and can lead to an overestimate of
the number of SFG. LEG (HEG) are the sources located in the LINER (Seyfert)
region in both [S~II]/H$\alpha$ and [O~I]/H$\alpha$ planes; an ``ambiguous''
galaxy (AG) changes its identification from LEG to HEG (or vice-versa) in the
diagrams involving the [S~II]/H$\alpha$ and [O~I]/H$\alpha$ ratios. The
sub-sample with spectroscopic classification is mostly composed of LEG (122),
while there are 32 HEG, 45 SFG, and 13 AG. The remainder half of the sample
(213 objects) has no optical classification. Their properties will be
discussed in more detail below.

We also estimated the Line Ratios Index (LRI) for our sources, the new
spectroscopic indicator introduced by \citet{buttiglione10}. LRI is defined
as the average of the low ionization lines ratios, i.e. 1/3
(log[N~II]/H$\alpha$ + log[S~II]/H$\alpha$ + log[O~I]/H$\alpha$) and provides
a more stable classification than the individual line
ratios. In the fourth panel of Figure~\ref{diagn} we show the location of the
sources in the [O~III]/H$\beta$ vs. LRI plane, distinguishing on the basis of
the optical classification derived above. The presence of two populations of
low/high excitation level AGN is clearly seen, similar to that observed in the
previous diagnostic diagrams and for the RLAGN sample of 3CR
radio-galaxies \citep{buttiglione10}. As expected, the ambiguous galaxies are
located in an intermediate region between LEG and HEG, but mostly at high
[O~III]/H$\beta$ values.

While the sample analyzed here is radio selected, the active nuclei considered
\citet{kewley06} are mostly radio-quiet. Therefore it is appropriate to
consider also the results of \citet{buttiglione10} on the spectroscopic
properties of radio galaxies from the 3CR sample. Following previous studies,
(e.g. \citealt{hine79,laing94,jackson97}), they separated Low and High
excitation galaxies (LEG and HEG respectively) on the basis of the narrow
emission line ratios.  The spectroscopic separation of RLAGN into HEG/LEG
slightly differs from the radio-quiet Seyfert/LINER classification. In fact,
the branch of LEG reaches [O~III]/H$\beta$ ratios $\sim$0.2 dex higher than
the LINER from \citet{kewley06}. It is unclear whether this is due to a
luminosity mismatch (the 3CR line luminosity is on average $\sim$ 30 times
higher than the SDSS sources) or to a genuine difference between 
radio-quiet and RLAGN in the diagnostic diagrams. Since the objects
considered here have line luminosities similar to the \citet{kewley06} sources
(most of our objects have $10^{39} < L_{\rm{[O~III]}} < 3 \times 10^{40}$), we
decided to adopt their classification of High and Low excitation galaxies. We
will however bear in mind that, with this choice, objects located just above
the Seyfert/LINER separation lines could be instead genuine low excitation
RLAGN.

Let us now focus on the objects that have no optical classification (hereafter
Unclassified Galaxies, UG) because at least one of the key emission lines
(usually \Hb\ and/or [O~III]) is undetected. This subsample is clearly very
important as it is composed of 213 objects, about half of the low redshift
galaxies\footnote{Also in the 3CR sample 29 (28\%) galaxies with $z<0.3$
  cannot be properly identified spectroscopically \citep{buttiglione09,buttiglione10}, and
  they are almost as numerous as LEG (32).}.  First of all, $\sim$ 50 such
objects can still be positioned in at least one of the diagnostic diagrams of
Fig. \ref{diagn}. With only two possible exceptions, they fall into the same
region populated by LEG.

\begin{figure*}
\centerline{
\includegraphics[scale=0.45,angle=90]{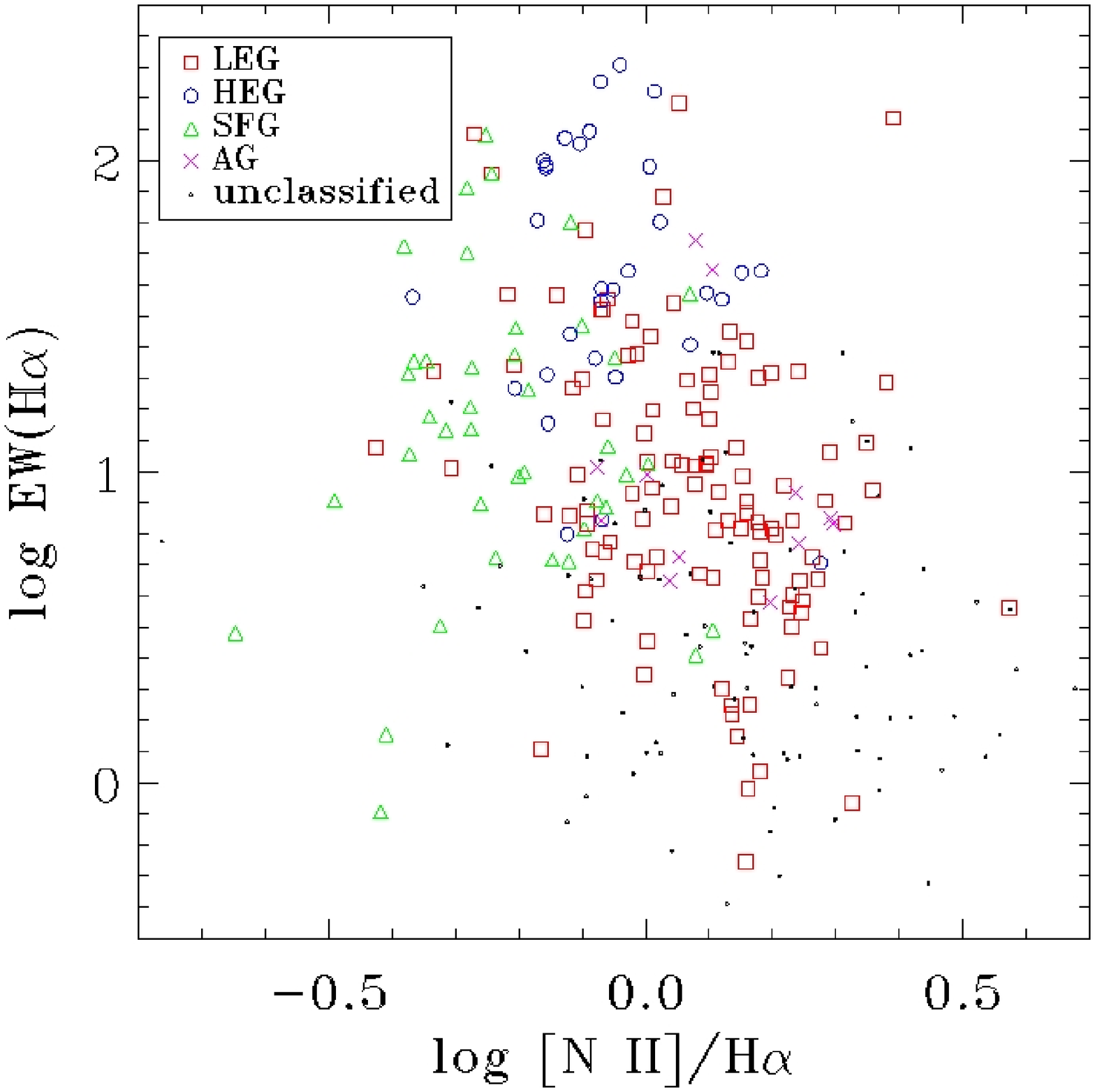}
\includegraphics[scale=0.45,angle=90]{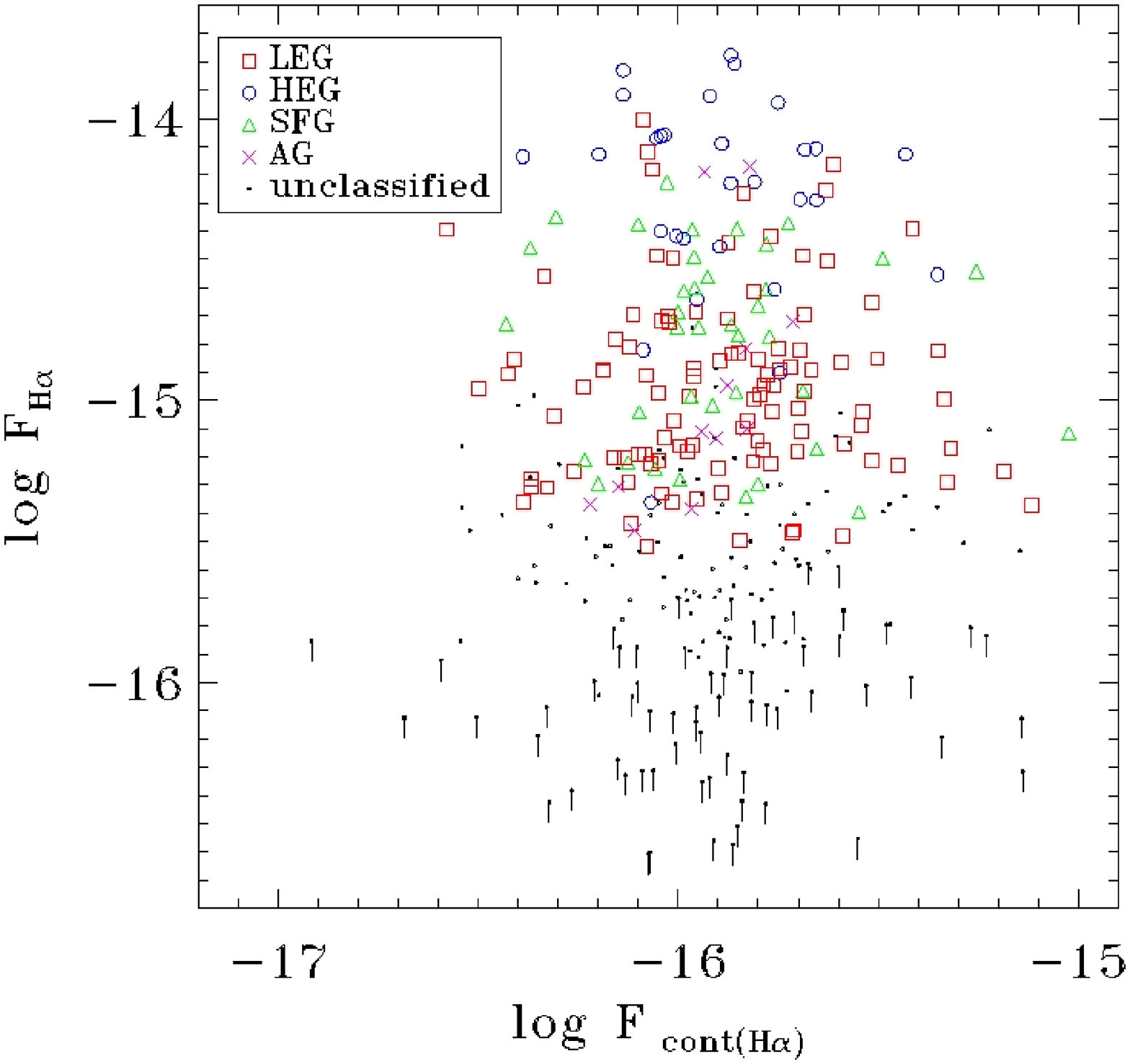}
}
\caption{Left: H$\alpha$ equivalent width against the [N~II]/\Ha\ ratio. Only
  objects with both [N~II] and \Ha\ detected are shown. Right: logarithm of
  the \Ha\ flux versus logarithm of the continuum at the \Ha\ line (in erg
  s$^{-1}$ and erg s$^{-1}$\AA$^{-1}$ units respectively).}
\label{fcfha}
\end{figure*}

\citet{cid09} shown that a good separation between the various spectroscopic
classes can also be obtained by comparing the H$\alpha$ equivalent width (EW)
and the [N~II]/\Ha\ ratio (see Figure \ref{fcfha}). In the SDSS/NVSS sample,
HEG have in general the largest values of EW(\Ha). SFG and LEG span a similar
large range of EW(\Ha), but these two classes are still rather well separated,
particularly at low equivalent widths, with SFG having lower values of
[N~II]/\Ha\ than LEG. The 116 UG present in this diagnostic diagram are, with
only a few possible exceptions, located in the same region of LEG, populating
the lower right corner. This applies also to the UG with only a [N~II]
detection. This strengthens the idea that the unclassified objects belong to
the class of LEG, but with an even lower contrast of the AGN against the host
galaxy emission. In the right panel of Figure \ref{fcfha} we instead compare
the H$\alpha$ flux and the continuum level around this line. This shows 
that UG are located at low \Ha\ fluxes, indicating that
their lower EW(\Ha) is due to a lower line flux and not to a higher
continuum level.

\subsection{Spectro-photometric diagnostic diagrams}
\label{spdd}

We now investigate the spectro-photometric properties of their host galaxies
taking advantage of the information about their optical spectroscopic
classification. We will then consider the host morphology, the stellar age and
the content of young stars, the broad band colors, their mass, and the mass of
their central supermassive black hole (SMBH).

Among the variety of measures provided by SDSS, the concentration index
$C_{r}$ (defined as the ratio of the radii including 90\% and 50\% of the
light in the $r$ band respectively) can be used for a morphological
classification of the hosts.  Early-type galaxies (hereafter ETG) have larger
values of $C_{r}$. Two thresholds have been suggested to define ETG: 
a more conservative value at $C_{r}\geq 2.86$
(e.g. \citealt{nakamura03,shen03}) and a more relaxed selection at
$C_{r}\geq 2.6$ (e.g. \citealt{strateva01,kauffmann03b,bell03}).
\citet{bernardi09} found that the second threshold of concentration index
corresponds to a mix of E+S0+Sa types, while the first mainly selects
elliptical galaxies, removing the majority of Sa's, but also some Es and S0s.
We select from the previous spectroscopically selected sample only the objects
with r-band concentration index measured with an accuracy better than
$\sigma(C_r) < 0.2$. It is composed of 401 objects (see Table \ref{crit}):
most of them (80\%) are secure ETG (with $C_{r}>2.86$) while only a small
fraction (9\%) are most likely late-type galaxies (with $C_{r}<2.6$). The
remaining 11\% of the sub-sample is located in the range $2.6< C_{r}<2.86$.

\begin{figure}
\centerline{
\includegraphics[scale=0.45,angle=90]{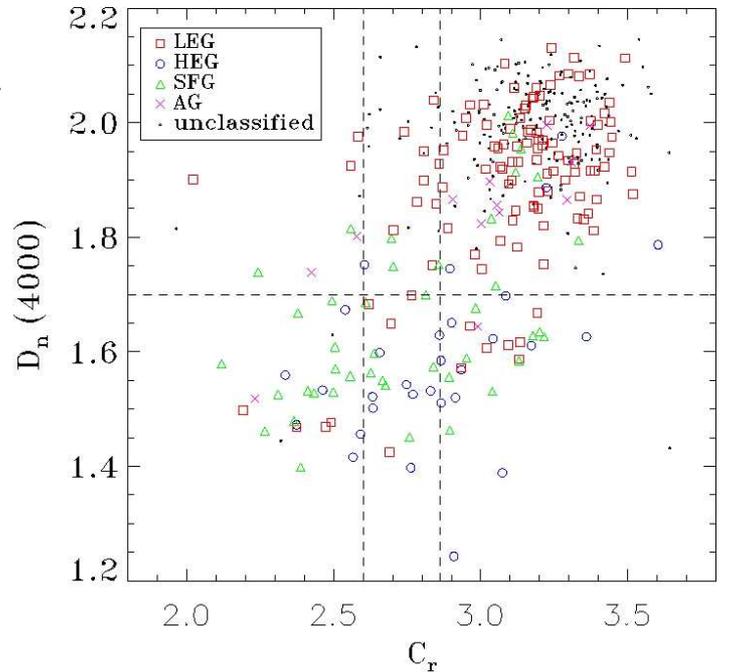}
}
\caption{Concentration index in r band, $C_r$, vs. 4000 \AA\ break,
  $D_n(4000)$, for the 401 objects of the SDSS/NVSS AGN sample with
  $0.03<z<0.1$ and error on the r-band concentration index $<$ 0.2.}
\label{dn4000}
\end{figure}

Another important parameter is the 4000\AA\ break strength, $D_n(4000)$
(defined as the ratio between the fluxes in the wavelength range 3850-3950\AA\
and 4000-4100\AA), that is sensitive to the presence of a young stellar
population \citep{balogh99}. It is instead insensitive to the effects of the
internal reddening, unlike broad-band colors. The distribution of $D_n(4000)$
in the large sample studied by \citet{kauffmann03b} is bimodal, with massive
galaxies (log $M_*/M_{\odot} \gtrsim 10.5$) showing a larger break than less
massive objects. The transition between the two groups is located
approximatively at $D_n(4000) \sim 1.7$.

Figure~\ref{dn4000} shows the distribution of the RLAGN in the
spectro-photometric diagram composed of $C_{r}$ vs. $D_n(4000)$. Most of the
points are located in the region of the plane populated by quiescent
ETG. These are predominantly LEG and unclassified objects. The objects with
$D_n(4000)<1.7$ are spread over a large range of $C_{r}$, encompassing both
early and late-type galaxies. All spectroscopic classes are well represented
in this region of the plane, but with a dominant contribution of SFG and HEG.

Let us now consider the location of the different spectroscopic classes.  The
majority of LEG are hosted in quiescent ($D_n(4000)>1.7$) ETG
(76\% for $C_{r}>2.86$ and 84\% for $C_{r}>2.6$) and this behavior is even
more pronounced in the spectroscopically unclassified sources (with
percentages of 90\% and 98\% respectively). Only a small fraction (13\% among
LEG and 2\% among UG) of these two classes show a possible evidence for star
formation, with $D_n(4000)<1.7$. Similarly, only 7\% of the LEG and 2\% UG
have likely late-type hosts. The ambiguous galaxies are also mostly red and
quiescent.

The situation is radically different considering SFG and HEG.
They cover the whole range of value for $C_{r}$ and show preferentially
low values of $D_n(4000)$: 83\% of the HEG and 70\% of the SFG have
$D_n(4000)<1.7$. 

\begin{figure}
\centerline{
\includegraphics[scale=0.45,angle=90]{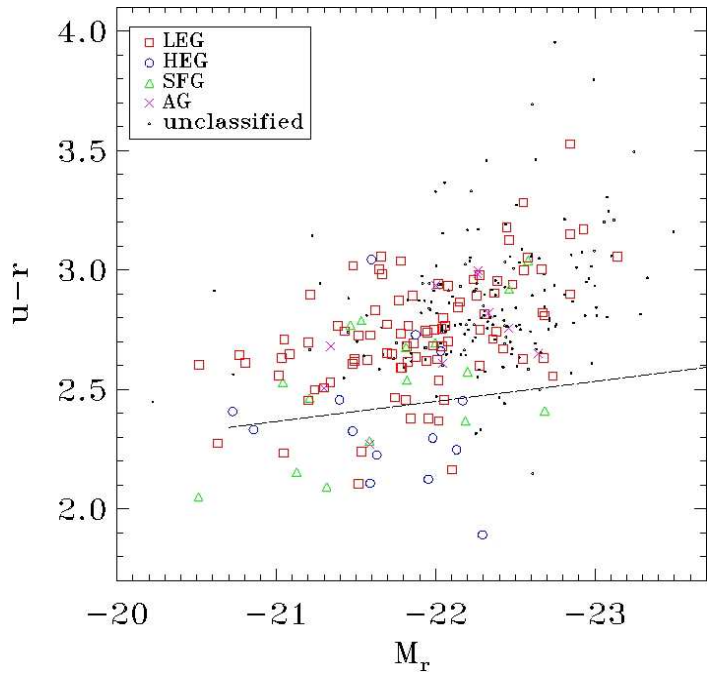}
}
\caption{Absolute $r$ band magnitude, $M_r$, vs. $u-r$ color for the sample of
  secure ETG (i.e. with $C_{r}>2.86$) with $0.03<z<0.1$. The
  dashed line separates the ``blue'' ETG from the red
  sequence, following the definition of \citet{schawinski09}.}
\label{mrur}
\end{figure}

The color index $u-r$ is another parameter able to qualify the star formation
rate present in the host as a whole. In fact, while the $D_n(4000)$ index is
measured over the 3\arcsec\ SDSS fiber centered on the nucleus, the $u-r$
color is estimated over the entire galaxy. In Figure~\ref{mrur} we show the
$u-r$ color vs. the absolute r-band magnitude $M_r$ of the hosts of the sample
limiting to the secure ETG, i.e. $C_{r}>2.86$. Most of the sample is located
at color $u-r>2.5$ and $M_{r}<-21$. However, a tail of bluer galaxies is
present formed mostly by HEG and SFG, but including also a few LEG and UG (see
Table \ref{crit} for a summary). On average, UG are redder and brighter than
LEG.

In Figure~\ref{mrur} the dashed line represents the separation between ``red''
and ``blue'' galaxies introduced by \citet{schawinski09} for a volume-limited
SDSS sample ($0.02< z<0.05$) of ETG, classified morphologically by visual
inspection by the Galaxy Zoo project \citep{lintott08}. This sample shares with
our radio-selected sample the luminosity range, $M_r<-20.7$, straddling the
$M_*$ for low-redshift ETG ($M_{r,*}$ = -21.15, \citet{bernardi03}).  With the
aim of isolating ETG with ongoing star formation, \citet{schawinski09}
determined the 3$\sigma$ offset from the mean of the red sequence in $u-r$
vs. $M_r$ plane and defined as blue ETG those that are below this threshold.

LEG and UG typically do not show sign of an ongoing star formation. In fact,
the fraction of these galaxies having blue u-r color is $\sim$5\% (3\% of the
UG and 8\% of the LEG). In the sample considered by \citep{schawinski09} the
observed fraction of ``blue'' ETG is 5.7\%. This indicates that the fraction
of star forming galaxies\footnote{Note also that in our sample there is the
  possibility that the blue color could be related to a substantial
  contribution of the AGN continuum.} in UG and LEG in our
radio-selected sample is similar to that of the general population of ETG.
Furthermore, the fraction of ``blue'' ETG decreases with luminosity and they
disappear in galaxies with $M_r<-22.5$ in agreement with the results by
\citet{schawinski09}.

Conversely, HEG show a high fraction (60\%) of blue objects and,
not unexpectedly, also SFG are often (35\%) blue.

\begin{figure}
\centerline{
\includegraphics[scale=0.45,angle=90]{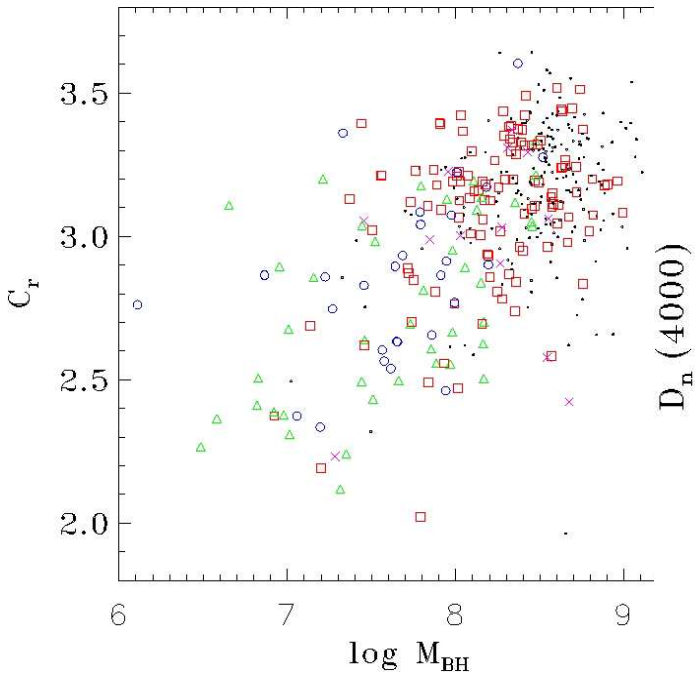}}
\caption{Logarithm of the black hole mass (in solar units) vs. concentration
  index, $C_{r}$, for the sub-sample with $0.03<z<0.1$.}
\label{mbhc}
\end{figure}

SDSS also provides us with the stellar velocity dispersion of the sample, from
which we derive the SMBH masses ($M_{\rm{BH}}$) adopting the relation of
\citet{tremaine02}. To better visualize the $M_{BH}$ distribution for each
class of host galaxies and AGN, in Figure~\ref{mbhc} we show the SMBH mass
against the concentration index. The vast majority of LEG and
UG is associated to ETG (as already shown before) with large
SMBH masses ($8 \lesssim {\rm log} (M_{\rm{BH}}/M_{\odot}) \lesssim 9$). 
A tail of later type galaxies extends toward
lower SMBH masses (reaching ${\rm log} (M_{\rm{BH}}/M_{\odot}) \sim 6.5$) and it
is formed mostly by HEG and SFG.

Let us explore the origin of the radio emission and in particular the
contamination related to star formation.  The criterion applied by
\citet{best05a} to exclude radio sources powered (or strongly influenced) by
star formation is based on the location of a galaxy in the $D_{n}(4000)$
versus $L_{{\rm 1.4GHz}}/M_{*}$ plane, where $M_{*}$ is the galaxy's stellar
mass\footnote{$M_{*}$ is estimated from the $g$ and $z$ SDSS magnitudes
  following following \citet{bell07} resulting in slightly different masses
  from those reported by \citet{best05a}. This explains the fact that some
  sources are located under the threshold curve.}. This allows one to predict
the radio emission per unit mass expected from a stellar population of a given
age.  As seen in Figure~9 of \citet{best05a}, the distribution of galaxies in
this plane is bimodal, but with a large transition region between $1.4\lesssim
D_{n}(4000) \lesssim1.7$. \citet{best05a} associated the radio emission to an
AGN when a galaxy shows an excess $>0.225$ in $D_{n}(4000)$ above the curve
corresponding to the prediction of an aging star formation event lasting 3 Gyr
and exponentially decaying.

\begin{figure}
\centerline{
\includegraphics[scale=0.5,angle=90]{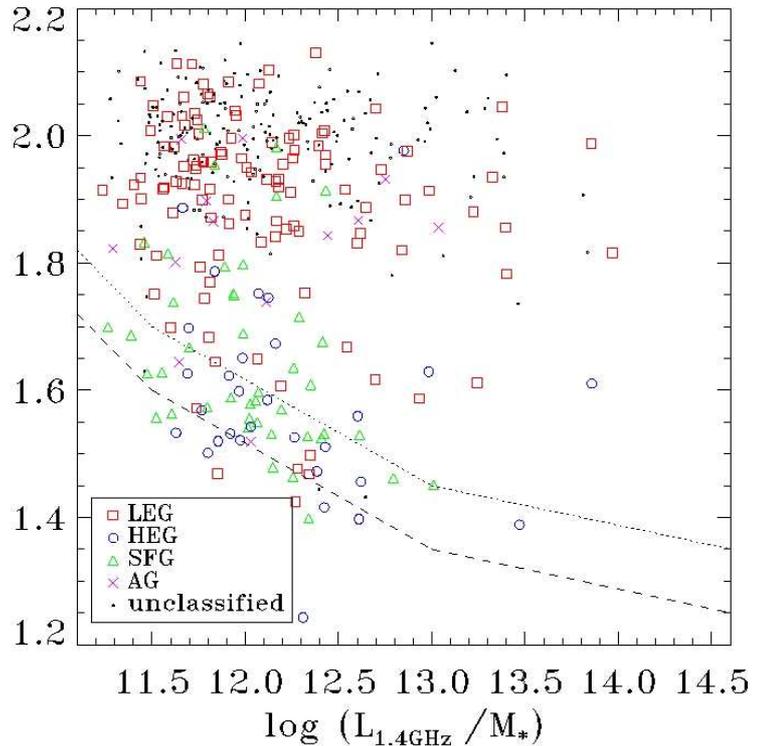}
}
\caption{$D_{n}(4000)$ vs. $L_{\rm 1.4GHz}/M_{*}$ for the galaxies of
  SDSS/NVSS sample with 0.03$<z<$0.1 and $\sigma(C_r) < 0.2$. The dashed curve
  is the empirical separation between the star forming and AGN radio emitting
  sources, performed by \citet{best05a} (read the text for details). The
  dotted curve is shifted 0.1 above in $D_{n}(4000)$ of the dashed curve, in
  order to separate the objects with a possible significant star formation
  contribution to their radio emission.}
\label{SFGselection}
\end{figure}

In Figure~\ref{SFGselection} we show the location of the SDSS/NVSS sample on
$D_{n}(4000)$ versus L$_{1.4GHz}/M_{*}$ plane. LEG and UG are mostly clustered
well above the threshold curve. Conversely, SFG and HEG occupy a region
parallel to the separation curve. The proximity to this curve casts some doubts
on the AGN origin of the radio emission in these two classes. In fact, there
is large mis-match in the scales probed by the NVSS image and SDSS spectrum
that can cause an erroneous identification of the main mechanism of radio
emission. The value of $D_{n}(4000)$ used to derive the age of the stellar
population and to predict the radio emission related to star formation is in
fact appropriate only for the central $3\arcsec$ covered by the SDSS fiber. In
presence of a substantial gradient in stellar age the predicted value of
$L_{1.4GHz}/M_{*}$ for the whole galaxy due to star formation could be
substantially underestimated. This is case, e.g., of spiral galaxies when the
fiber covers only the bulge, while most of the star formation occurs in the
spiral arms. We will return to this issue in Sect. \ref{comp3c} and
\ref{exceptions}.

\begin{figure*}
\centering
\includegraphics[scale=0.70,angle=90]{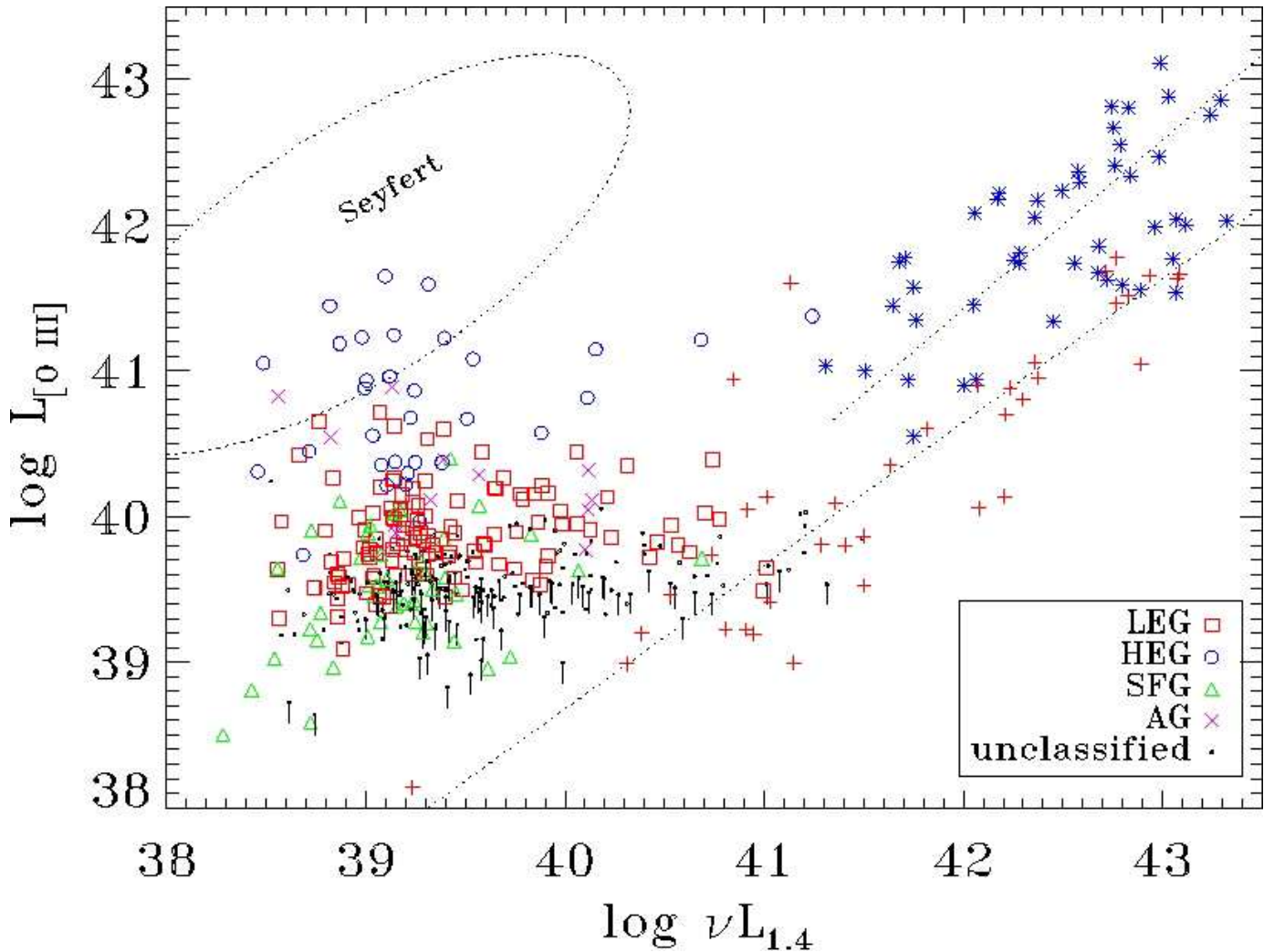}
\caption{Logarithm of the radio vs. [O~III] luminosities (both in erg s$^{-1}$
  units) for the $0.03<z<0.1$ sub-sample.  The two dashed lines reproduce the
  line-radio correlation followed by the LEG and HEG of the 3CR sample (red
  pluses and blue asterisks respectively). The ellipse marks the boundaries of
  the location of Seyfert galaxies (e.g. \citealt{whittle85}).}
\label{lextlo3}
\end{figure*}

Finally, we consider the $L_{1.4~{\rm GHz}}-L_{\rm [O~III]}$ plane
(Figure~\ref{lextlo3}), already used in \citet{baldi09}, now taking advantage
of the additional information on the optical classification of the sources.
In this plane we note a horizontal stratification of the classes: in order of
increasing line luminosity we find SFG and UG, followed by LEG and HEG. Only
LEG and UG reach the line-radio correlation defined by the 3CR sample. Most
HEG show a $L_{1.4~{\rm GHz}}/L_{\rm [O~III]}$ ratio much larger than seen in
LEG and UG.

Summarizing, the SDSS/NVSS
sample is predominantly composed of low excitation AGN hosted by red, quiescent
(from the point of view of star formation), ETG and with large
$M_{\rm{BH}}$. We include in this main class also the spectroscopically
unclassified objects that appear to share these features and are most likely
simply lower luminosity AGN.

High excitation galaxies are also present but represent only a minority ($\sim
10$ \%) of the sample and they usually show ongoing star formation, stronger
emission lines and lower black hole masses than those of LEG, and are often
hosted in late-type galaxies. These properties are very similar to those of
the star forming galaxies (15\% of the sample), with the very important
difference that SFG have a much lower level of [O~III] line luminosity (a
factor of $\sim$ 100) than HEG.

\section{Results: the sub-sample with $0.1<z<0.3$}
\label{highz}

We now briefly summarize the results obtained from the same analysis described
in the previous section, but applied to the objects with higher redshift,
i.e. $0.1<z<0.3$.  We anticipate that, overall, the analysis of more distant
(and more powerful from the point of view of their line and radio
luminosities) SDSS/NVSS sources confirms the main results obtained for the
lower redshift objects.

In Fig. \ref{diagnhz} we show the diagnostic spectroscopic diagrams. Not
surprisingly the fraction of unclassified galaxies is even larger than at low
redshift, reaching 85\%.  In fact, focusing on the 1684 objects with an error
on the concentration index smaller than 0.2, there are 1438 UG, 130 LEG, 37
HEG, 53 SFG, and 26 AG. The dominance of UG makes difficult to establish
whether the breakdown in the other classes changes with redshift. However,
considering the upper limits to the \Ha\ fluxes for UG (see Fig. \ref{allhz1},
upper left panel), similarly to what we found for the near sample, they are
substantially offset from HEG, suggesting an identification of UG with a
mixture of LEG and SFG.  Thus the fraction of HEG appears lower than in
the near sample being reduced from $\sim 7\%$ to $\sim 2\%$.

In Fig. \ref{allhz1} (upper right panel) we report the analogous of
Fig. \ref{dn4000} for the higher redshift objects showing the concentration
index vs. the 4000 \AA\ break. The bulk of the sample (formed by UG and LEG)
is shifted at slightly lower $C_r$ and lower $D_n(4000)$ as expected for the
impact of seeing in these more distant objects \citep{hyde09} and for the
smaller age of the stellar population. The population with $D_n(4000) < 1.7$
is still found to be dominated by HEG and SFG, that also form the bulk of the
objects with likely late-type morphology.

The location of many of the HEG and SFG in Fig. \ref{allhz1} (lower left
panel), where we compare $D_{n}(4000)$ and $L_{\rm 1.4GHz}/M_{*}$, is close to
the line marking the division between a radio emission associated with an AGN
and star-formation. However, the fraction of HEG at higher redshift with a
clear AGN-origin for the radio emission is larger than that at lower redshift
(Fig.~\ref{SFGselection}). Conversely, most LEG and UG at higher $z$ still
show an AGN as dominant process for the production of the radio emission.

In Fig. \ref{allhz1} we also present (lower right panel) the distributions of
the $M_{\rm{BH}}$ for each spectroscopic class (the filled portion represents
the contribution of galaxies with $C_r<2.86$). For all classes they are
similar to those of the lower z objects (compare with Fig. \ref{histombh})
with the only exception of SFG that have $M_{\rm{BH}}$ larger by a factor of
$\sim$ 10.  Apparently, the dominant population of SFG moves from low mass
spiral galaxies, to more massive early-type. Unfortunately for this redshift
range we do not have a well defined reference sample of non active ETG to
contrast with our results.

Finally, in Fig. \ref{allhz3} we compare the radio and the [O~III]
luminosities, similarly to what we did in Fig. \ref{lextlo3}.  The selection
of more distant objects causes a general shift toward higher luminosities on
both axis, but the stratified position of the various classes is confirmed.

\section{Comparison with 3CR radio-galaxies}
\label{comp3c}

In this Section we will draw a quantitative comparison between the
spectro-photometric properties of the SDSS/NVSS sources with those of
the powerful radio-galaxies part of the 3CR sample.

\subsection{Low excitation galaxies}
Since from the point of view of the optical spectra, most of the SDSS/NVSS
radio sources are classified as low excitation galaxies, we start with a
comparison with the LEG of 3CR sample\footnote{We consider as `LEG' sources
  all those classified as low excitation galaxies, see \citet{buttiglione10},
  from the optical spectra independently of their radio morphology.}.

\begin{figure*}
\centerline{
\includegraphics[scale=0.15]{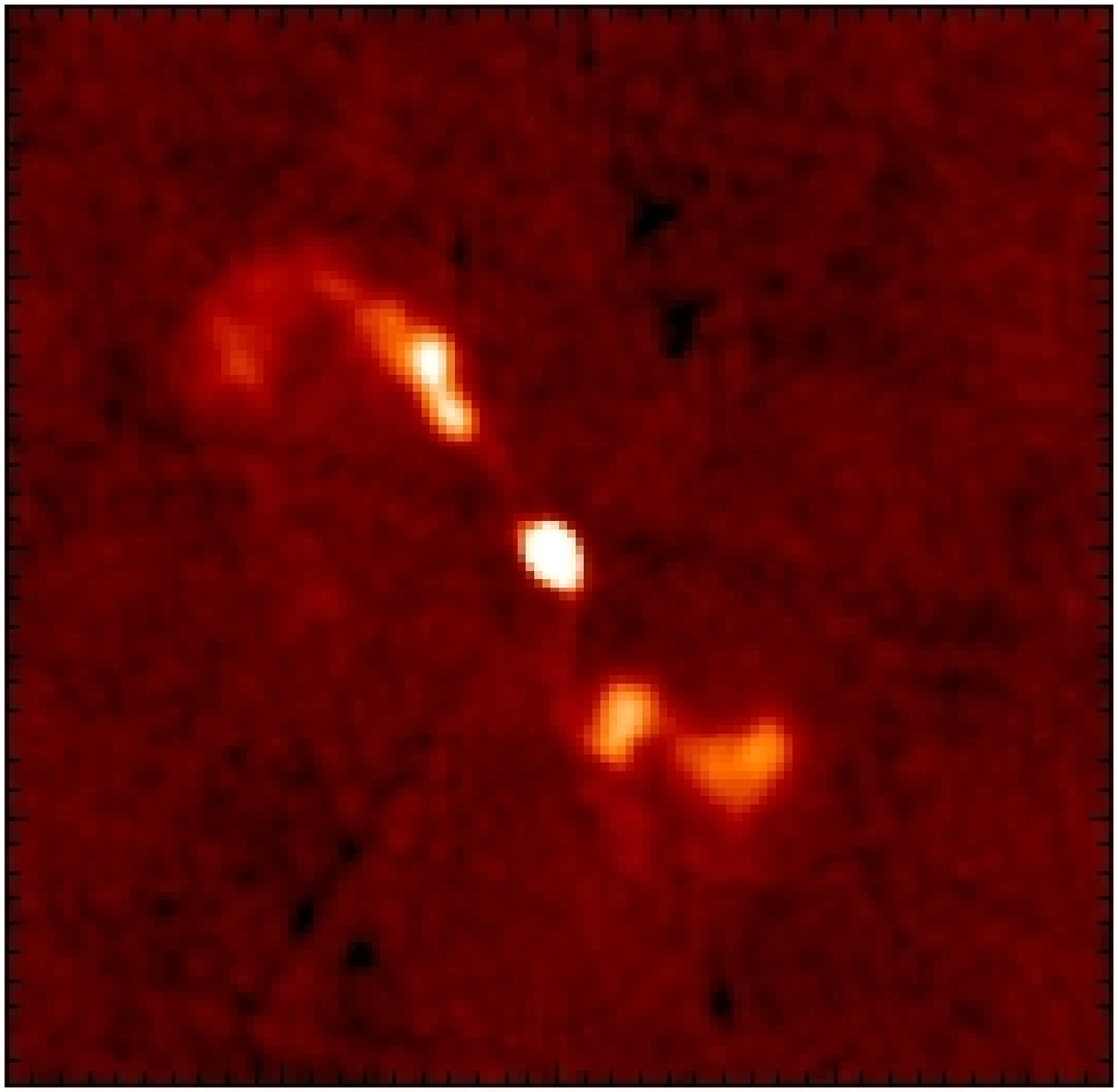}
\includegraphics[scale=0.15]{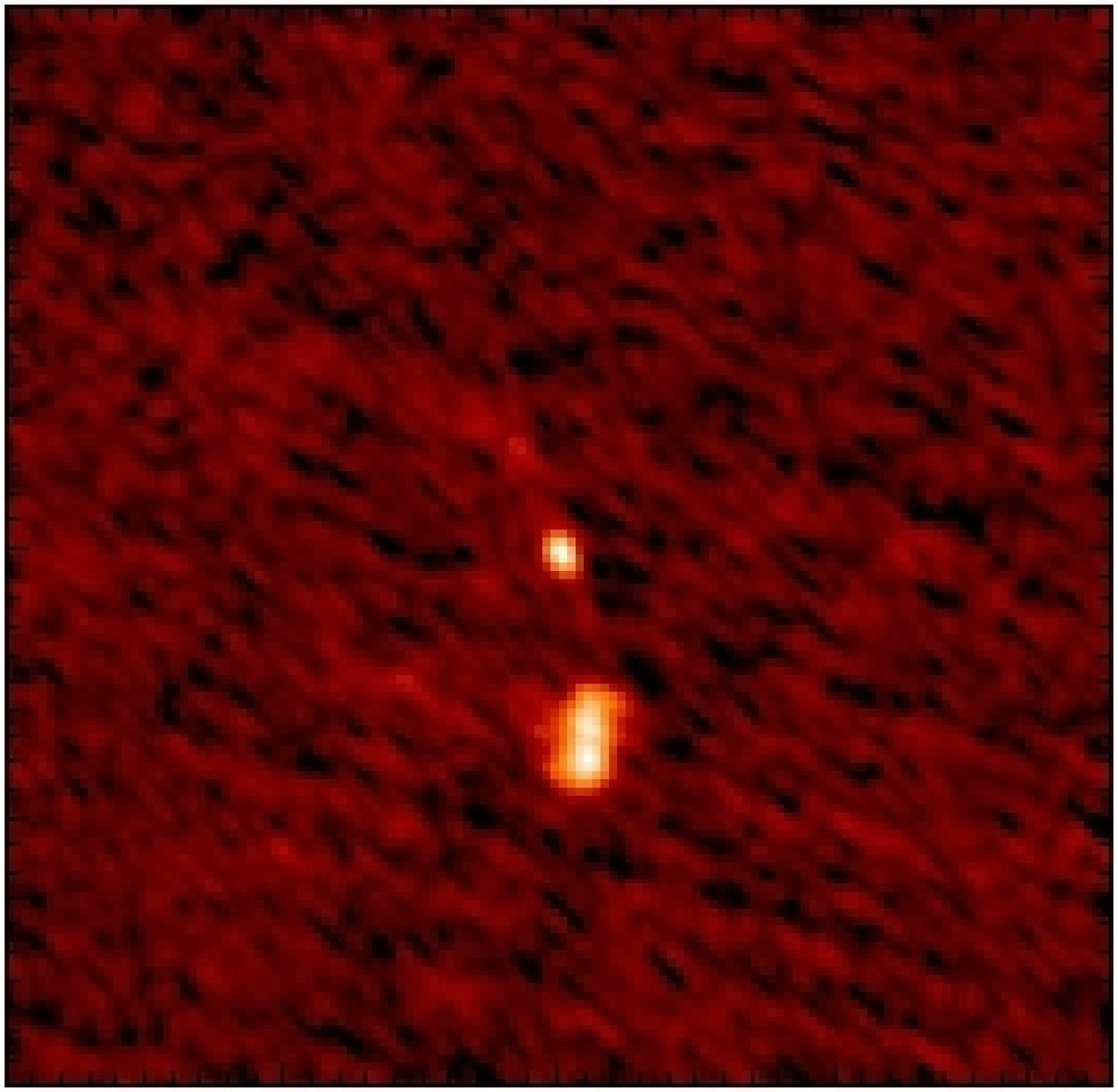}
\includegraphics[scale=0.15]{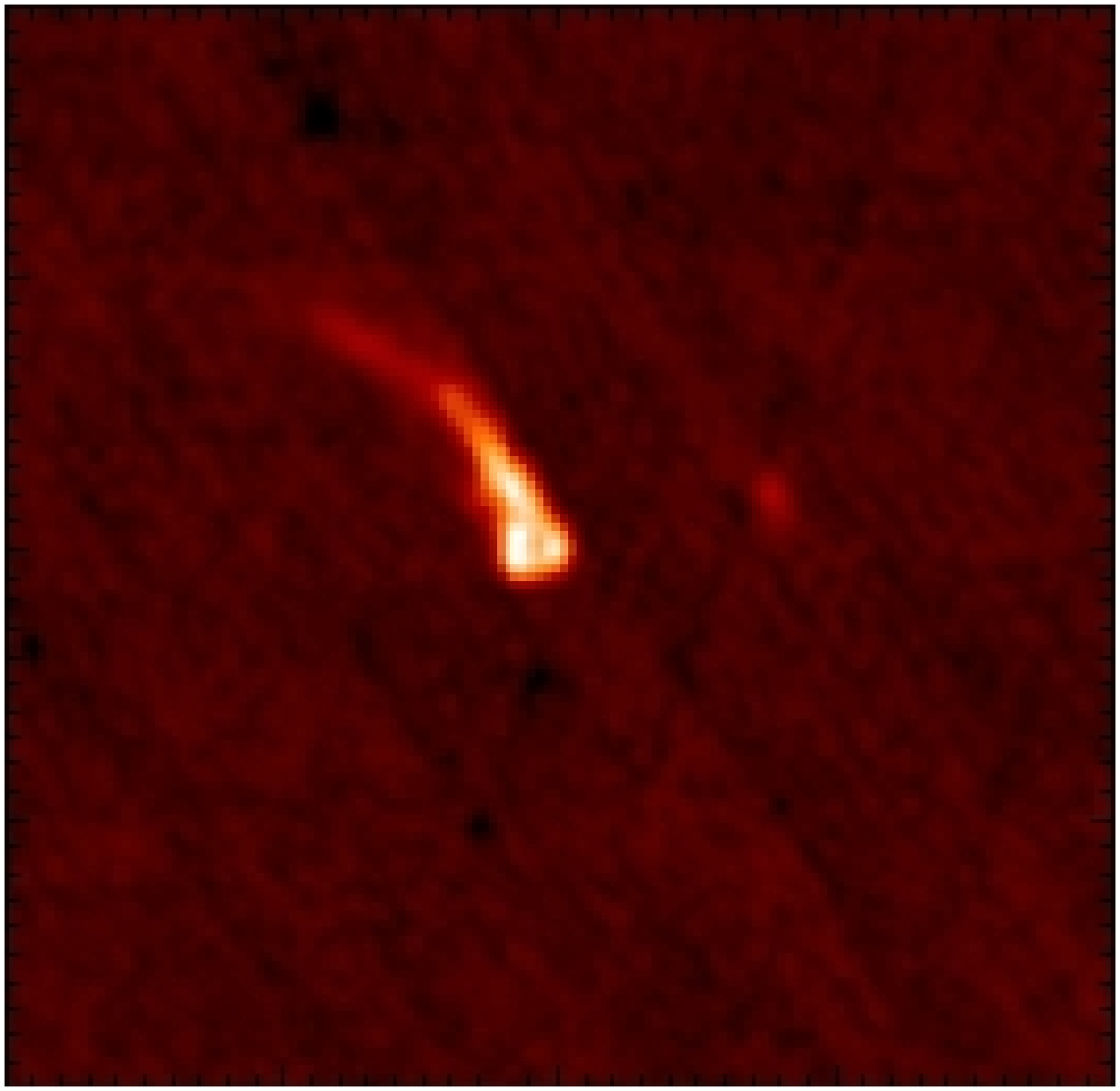}
\includegraphics[scale=0.149]{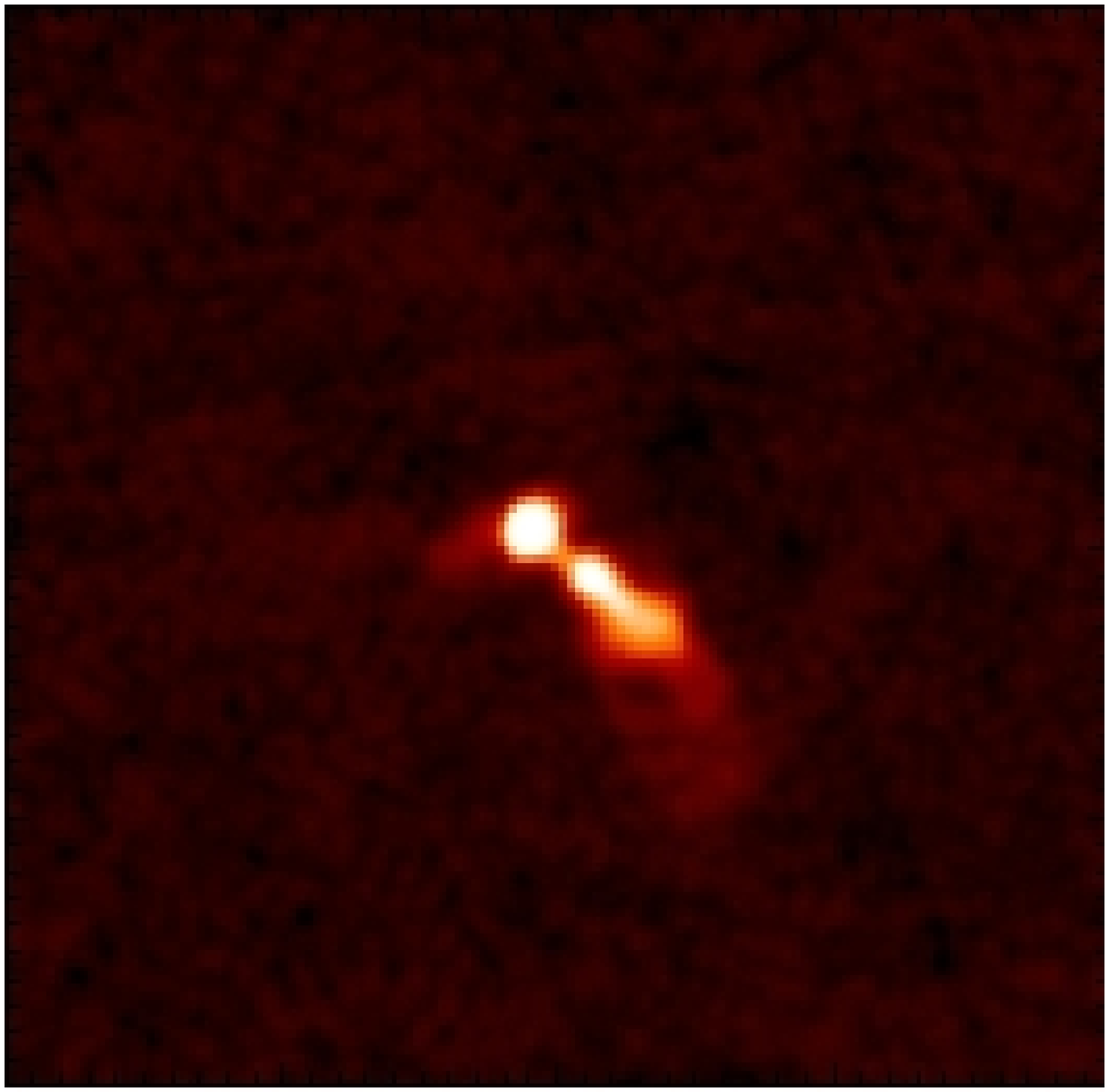}
\includegraphics[scale=0.15]{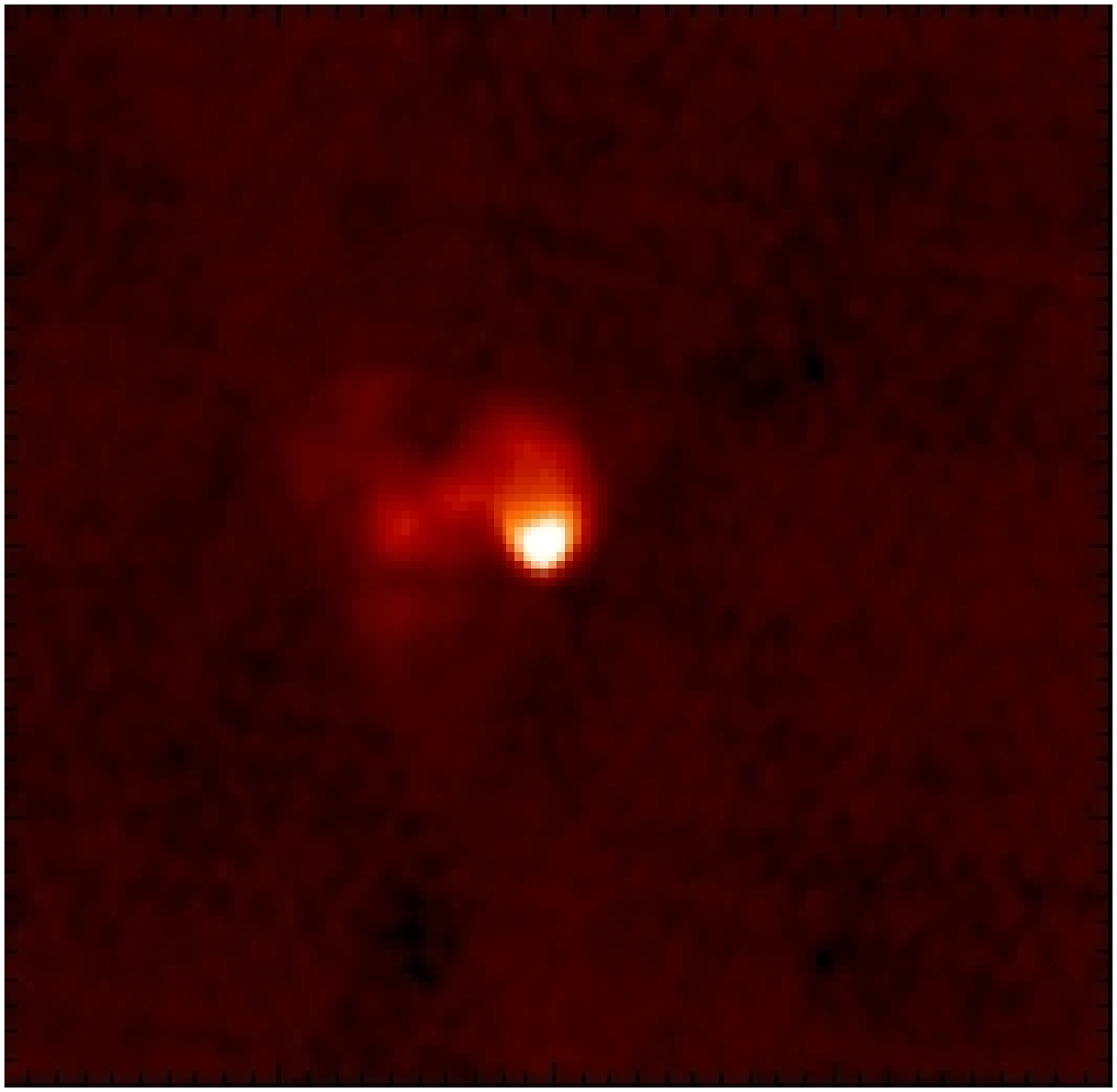}
\includegraphics[scale=0.1495]{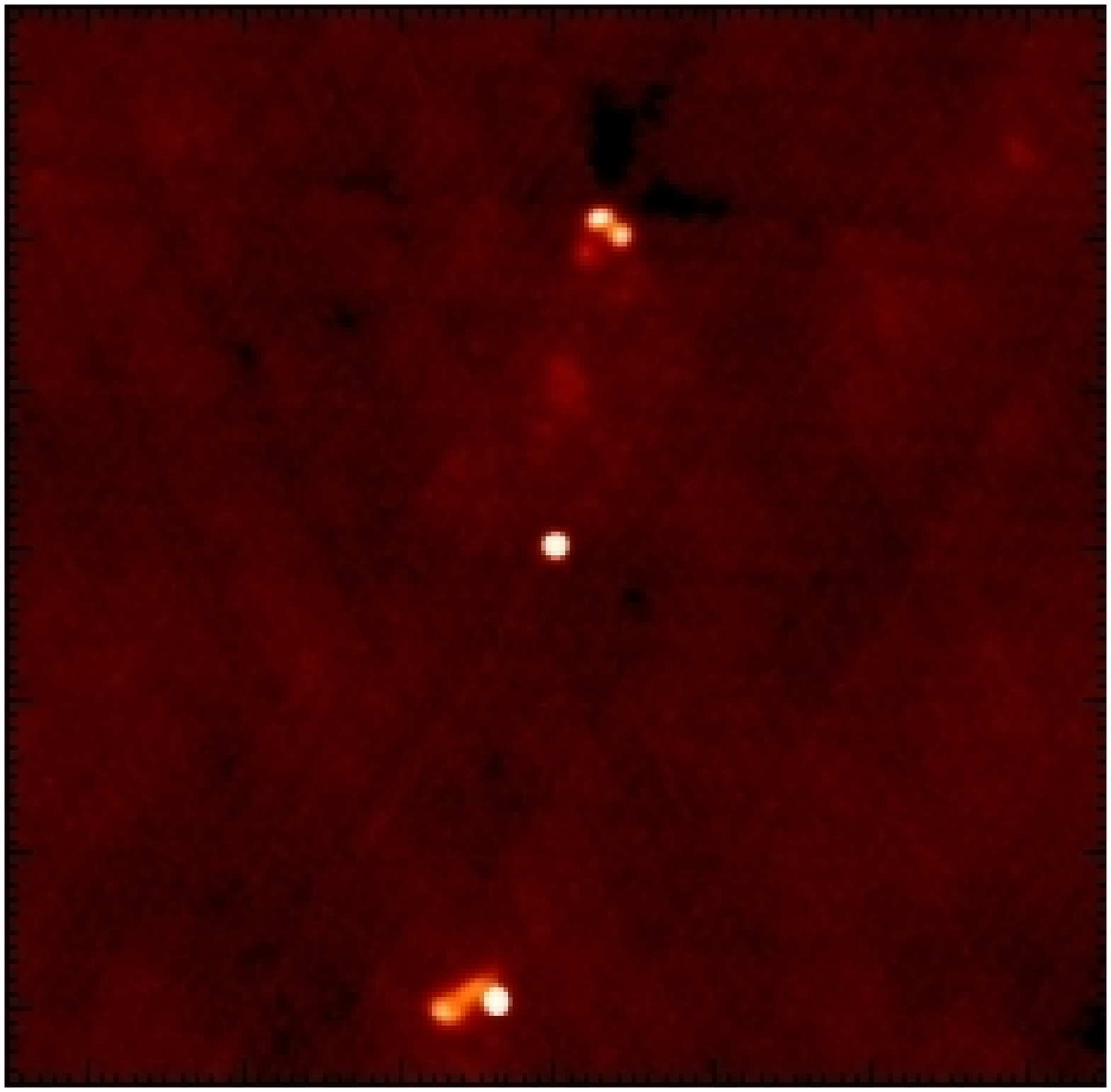}}

\centerline{
\includegraphics[scale=0.15]{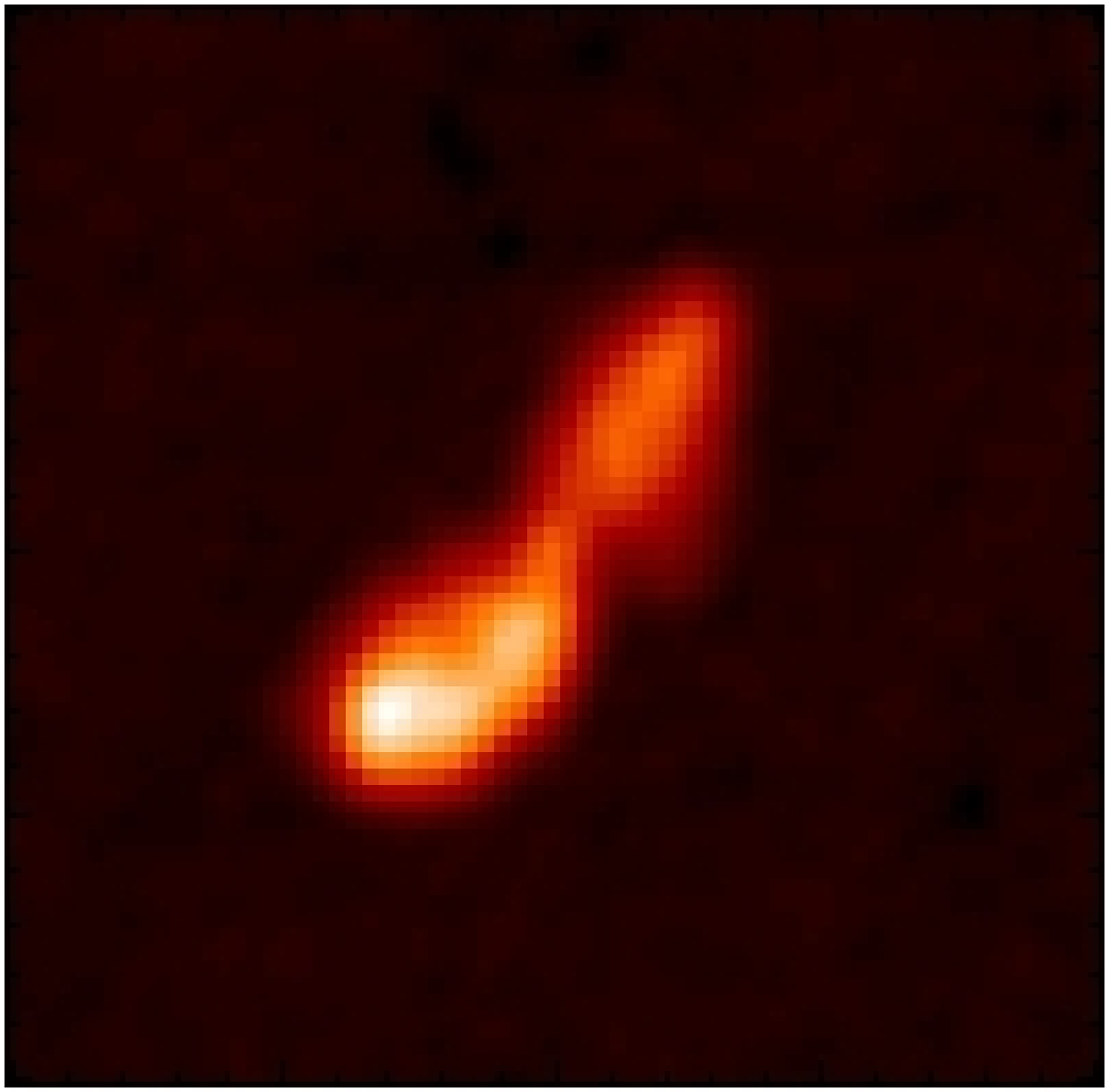}
\includegraphics[scale=0.15]{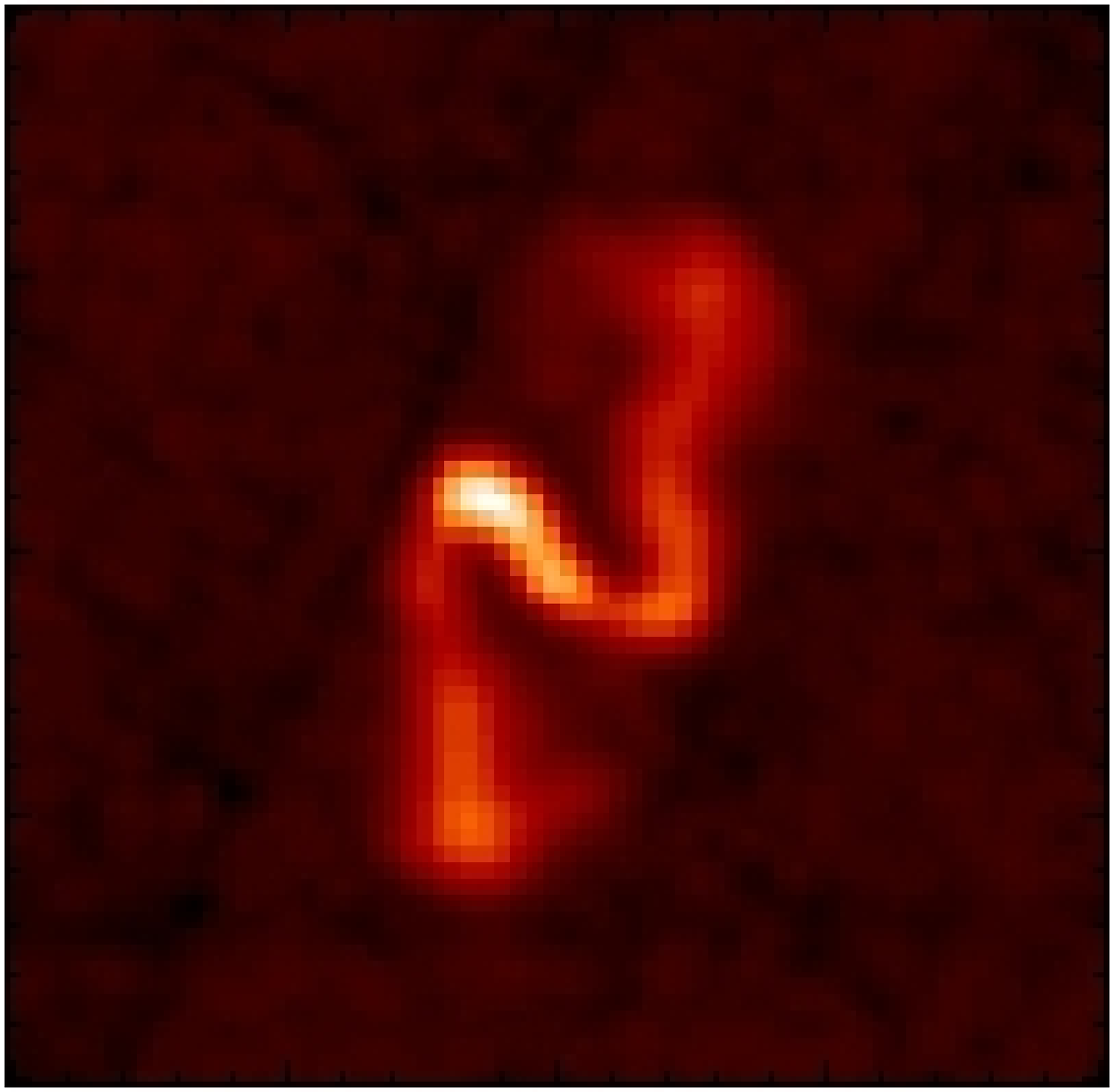}
\includegraphics[scale=0.15]{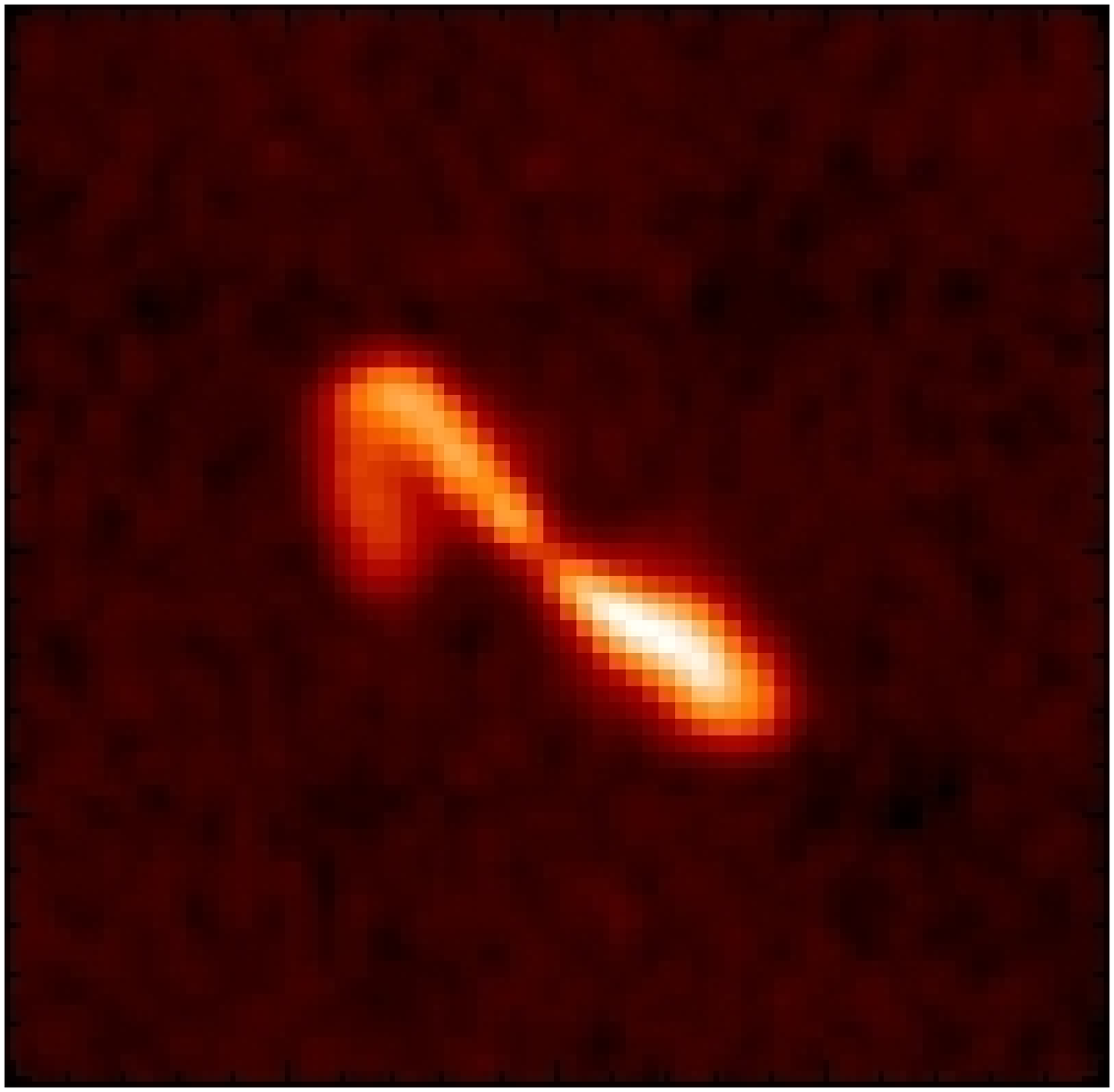}
\includegraphics[scale=0.15]{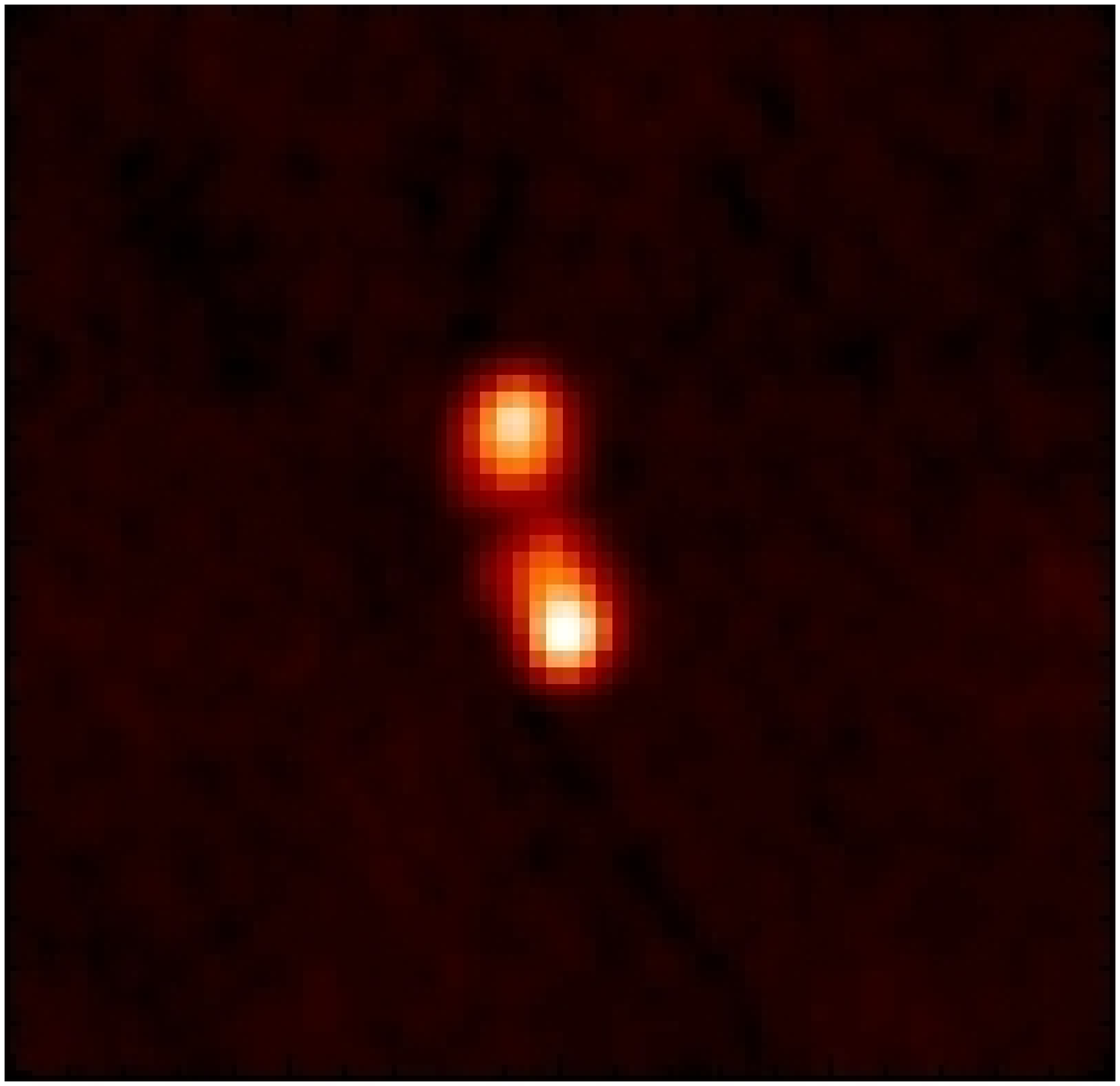}
\includegraphics[scale=0.15]{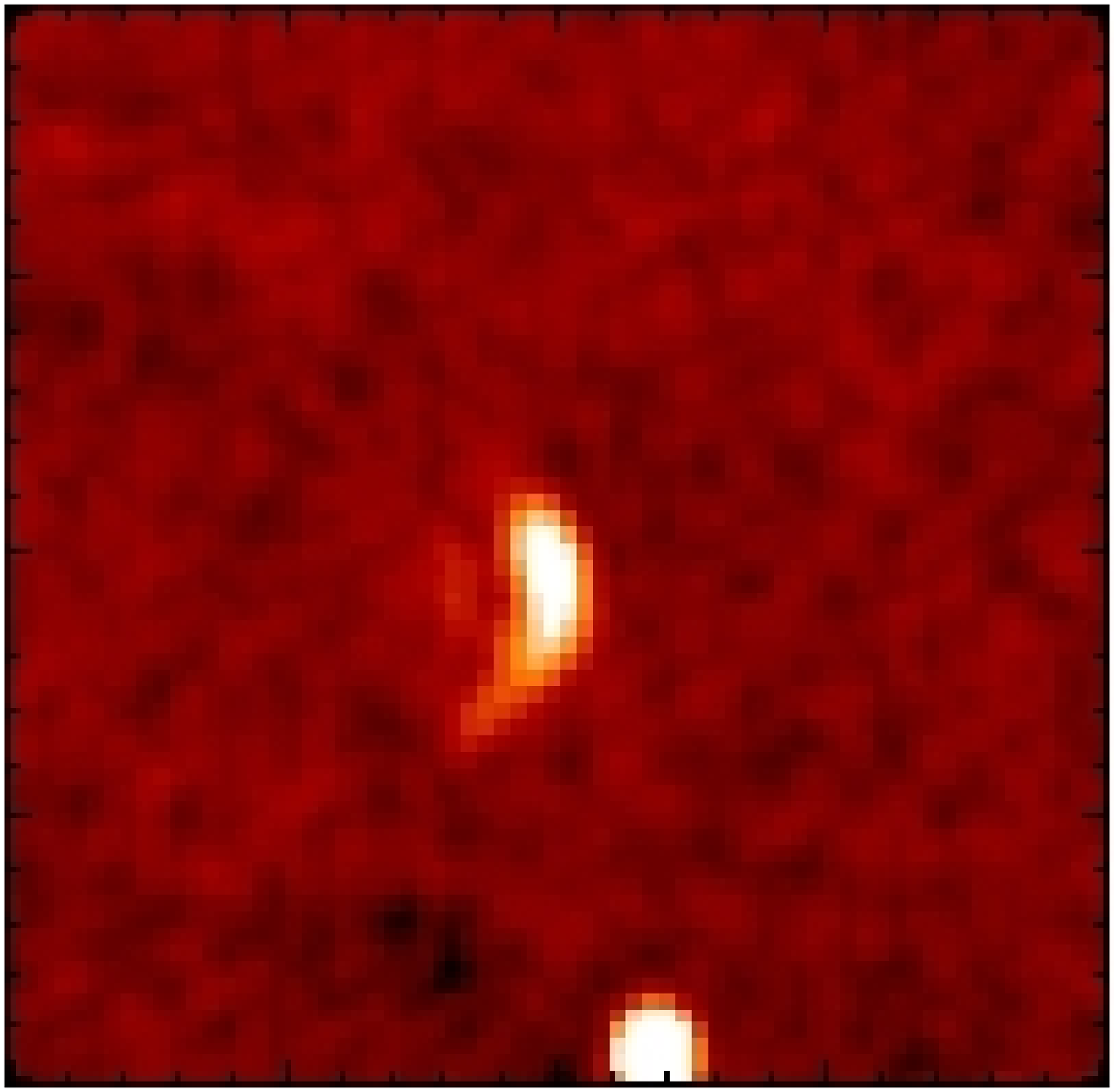}
\includegraphics[scale=0.15]{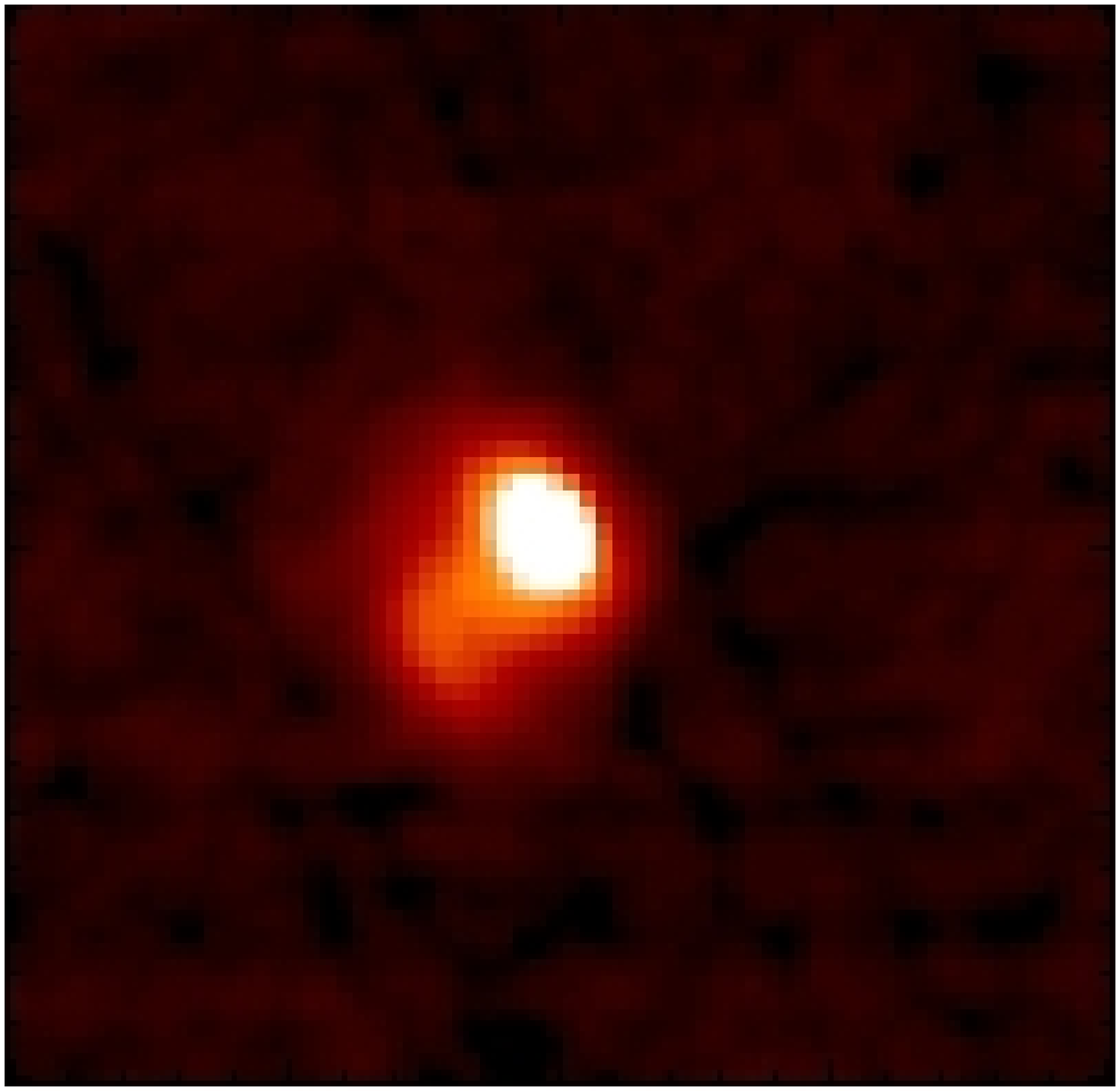}}

\caption{FIRST images of the 12 brightest SDSS/NVSS radio sources with $z<0.1$
  associated with UG and LEG, with a [O~III] detection. The fields of view are
  $4\arcmin \times 4\arcmin$ and $2\arcmin \times 2\arcmin$ for the top and
  bottom row respectively. The physical sizes range from 45 to 400 kpc.}
\label{first}
\end{figure*}

We here compare the radio morphology of the two samples. We visually inspected
all UG and LEG at $z<0.1$, finding that the majority of them is unresolved or
barely resolved at the 5$\arcsec$ FIRST resolution, corresponding to a limit
to their size of $\lesssim$ 10 kpc.  Nonetheless we found 6 (5 \%) LEG and 58
(27\%) UG with well resolved radio structures, all of these having $\nu L_r
\gtrsim 2 \times 10^{39}$ erg s$^{-1}$. The fraction of extended radio sources
increases with radio luminosity and, at the higher radio powers, most of them
are well resolved (see Fig. \ref{first}). All morphological classes are
represented, including twin-jets and core-jet FR~I, Narrow and Wide Angle
Tails, and FR~II with clear hot spots. These classes are all found among the
3CR/LEG that, however, are all well resolved with physical sizes ranging from
15 to 400 kpc, in the same range of the SDSS/NVSS sources shown in
Fig. \ref{first}.

\begin{figure}
\centerline{
\includegraphics[scale=0.6,angle=90]{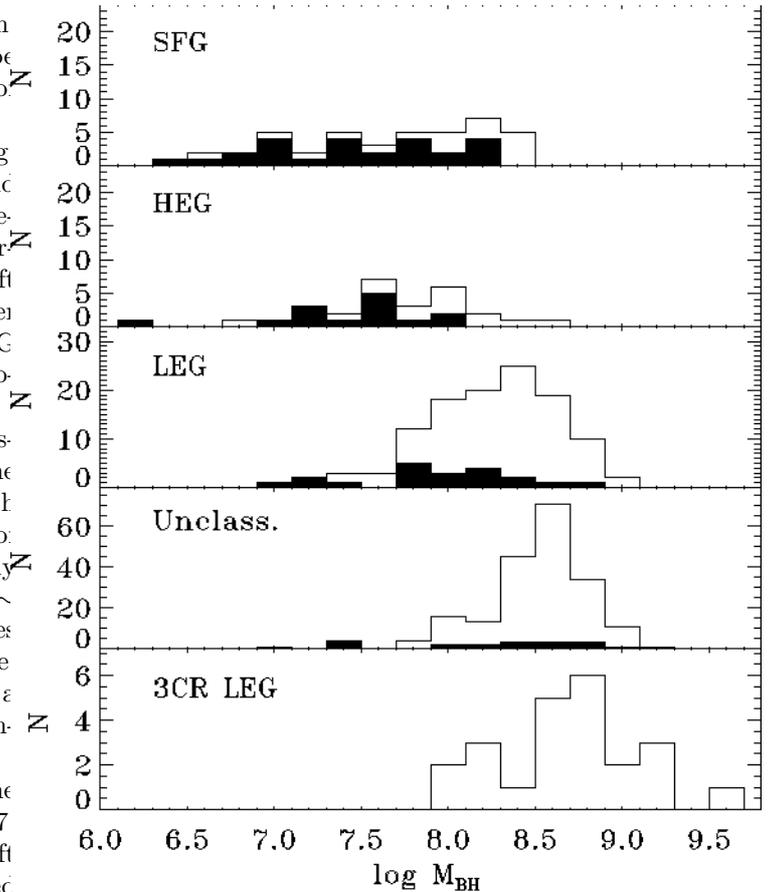}}
\caption{Distributions of the SMBH mass (in solar units) for the various
  spectroscopic SDSS/NVSS sub-samples compared to that of
  LEG part of the 3CR sample. The filled portion of the histograms represent
  the contribution of galaxies with $C_r<2.86$.}
\label{histombh}
\end{figure}

In Fig. \ref{histombh} we analyze the SMBH mass distribution of each
spectroscopic class in comparison with that of 3CR/LEG. These powerful radio
sources have black hole mass in the range ($8.0 < {\rm log}
(M_{\rm{BH}}/M_{\odot}) < 9.5$) similar to those of the LEG and UG of the
SDSS/NVSS sample. In fact, the vast majority of these objects have a black
hole mass in the range $7.7 < {\rm log} (M_{\rm{BH}}/M_{\odot}) < 9.5$. A low
mass tail, with log $(M_{\rm{BH}}/M_{\odot}) < 7.5$, is present among LEG and UG but
amounting to only 3\% of the galaxies and mostly composed of late-type
galaxies (these sources will be discussed in more detail in 
Sect. \ref{exceptions}).
If we consider only the ETG, the median logarithm of 
$M_{\rm{BH}}$ in solar units are 8.30, 8.52, and 8.44 for LEG, UG and the
combined ``LEG+UG'' sample respectively. $M_{\rm{BH}}$ in 3CR/LEG are only
slightly larger than in UG and LEG, with a median of ${\rm log} \,
M_{\rm{BH}}/M_{\odot} = 8.67$. A Kolmogorov-Smirnov test indicates that the
distributions of SMBH for UG+LEG and 3CR/LEG are not statistically different
at a significance limit of 99\%.  As already noted, SFG and HEG have instead
lower $M_{\rm{BH}}$ and a large contribution of late type galaxies.

\begin{figure}
\centerline{
\includegraphics[scale=0.4,angle=90]{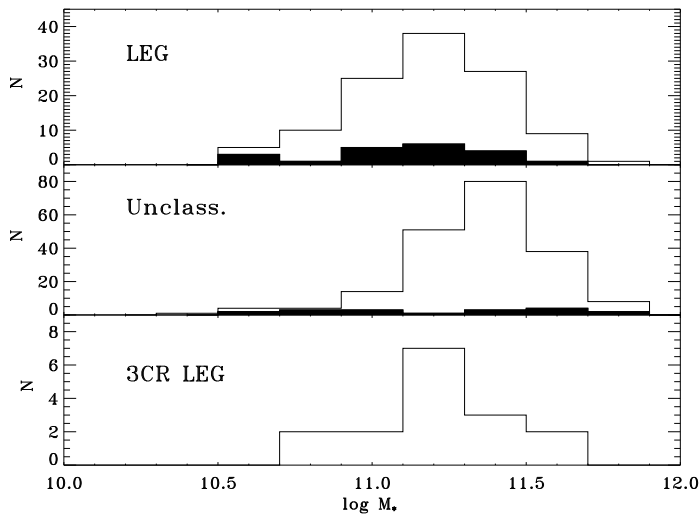}
}
\caption{Distributions of the stellar masses (in solar units) for the samples
  of LEG and unclassified galaxies in the SDSS/NVSS sample, compared to that
  of LEG part of the 3CR sample with z$<$0.1. The filled portion of the
  histograms represent the contribution of galaxies with $C_r<2.86$.}
\label{histoms}
\end{figure} 

In Fig. \ref{histoms} we show the distribution of the host masses,
estimated from the $g$ and $z$ SDSS magnitudes \citep{bell07} 
restricting to LEG and UG sources, and of the 17 3CR/LEG with available SDSS
images. They share the same host mass range $10.5 < {\rm log}
(M_{\rm{*}}/M_{\odot}) < 12$.  The median host masses are
log $(M_{\rm{*}}/M_{\odot}) =$11.18, 11.37 and 11.26 for LEG, UG and 3CR/LEG
respectively. Using a Kolmogorov-Smirnov test, we found that these small
differences are not statistically significant.

Considering now the morphology of the host galaxies, 3CR/LEG are found in
elliptical galaxies (e.g. \citealt{martel99,madrid06}) similarly to the
typical hosts of SDSS/NVSS galaxies. 

From the point of view of their broad band color, we consider the location of
the 3CR/LEG in the $M_r$ vs. $u-r$ plane (see Figure \ref{mrur}) looking for
blue galaxies. All the 3CR/LEG sources are red, with the exceptions of 3C~236
and 3C~084. The color of the latter object is however dominated by the bright,
highly polarized, and variable optical nucleus \citep{martin76}, typical of
BL~Lac objects. Instead 3C~236 shows UV knots related to star formation
\citep{odea:3c236}, indicating the presence of only 1 blue galaxy among the 16
3C/LEG. This is in full agreement with the finding of \citet{baldi08} based on
the analysis of their near UV HST images.

Summarizing, the hosts of 3CR/LEG and those of LEG in the SDSS/NVSS sample are
typically red quiescent early-type galaxies, with distributions of black holes
and stellar masses statistically indistinguishable.  The most striking
difference between the two samples is related to the radio morphology, i.e. to
the large fraction of objects ($\sim80$\%) in the SDSS/NVSS that, despite
having a line luminosity matched to the 3CR/LEG, do not show extended radio
structures.

\subsection{High excitation galaxies}
\label{hegsect}

About half of the 3CR sources with $z<0.3$ are HEG
\citep{buttiglione10}. They are all bright radio-sources (log $\nu L_{178} $
[erg s$^{-1}$] $ \gtrsim 41$) with a FR~II morphology. About 1/3 of them show
broad emission lines.  Their distribution of SMBH masses covers the range
${\rm log} \, M_{\rm{BH}}/M_{\odot} = 8.1 - 9.5$ with a median of 8.75
\footnote{The black hole masses for 3CR/HEG have been estimated from their
  H-band absolute magnitude (after removal of the nuclear contribution
  estimated by \citealt{baldi10}) as tabulated by \citet{buttiglione10},
  converted into K-band luminosity assuming H-K = 0.21 \citep{mannucci01}, and
  using the $L_{K}-M_{BH}$ relation \citep{marconi03}.}.

In the SDSS/NVSS sample the HEG are instead a minority, 7\% in the sub-sample
of nearby objects ($0.03<z<0.1$) and only 2\% in the ``far sample'' with
($0.1<z<0.3$) for a total of 67 objects.\footnote{All these objects only show
  narrow emission lines. In fact \citet{best05a} excluded the objects
  classified as QSO (that instead generally do show broad lines) by the
  automated SDSS classification pipeline, since the presence of a bright
  nucleus would have prevented a detailed study of the host properties.}

Considering their radio morphology, only a minority ($\sim$ 20 \%) have
resolved structures in FIRST, and only 7 of them are clear FR~II. These are
all high power sources $\nu L_r \gtrsim 10^{41}$ erg s$^{-1}$, a low limit
similar to that found for 3CR/HEG.

HEG in the SDSS/NVSS sample are generally blue ($\sim$ 60 \%) and with a
strong level of star formation (25 out of 30 have $D_n(4000) < 1.7$), in line
with 3CR/HEG \citep{baldi08,smolcic09}. But SDSS/NVSS HEG are hosted almost in
the same fraction by early and late-type galaxies, at odds with the elliptical
morphology of 3CR/HEG hosts. Furthermore, their SMBH masses distribution is
shifted by about a order of magnitude (the median is ${\rm log} \,
M_{\rm{BH}}/M_{\odot} = 7.88$) toward lower masses than 3CR/HEG. In addition,
in the $D_{n}(4000)$ vs. $L_{\rm 1.4GHz}/M_{*}$ plane (Figure
\ref{SFGselection}) two thirds of HEG are located close (with an offset $<$
0.1 in $D_{n}(4000)$) to the boundary of the region characteristic of the
objects in which the radio emission is powered by star formation. Although
these are certainly AGN, given their emission line ratios, they might enter
the SDSS/NVSS sample only because of a high level of star formation that
powers their radio emission. This should be taken into account when dealing
with AGN radio luminosity functions derived from this sample.

\begin{figure}
\centerline{
\includegraphics[scale=0.5,angle=90]{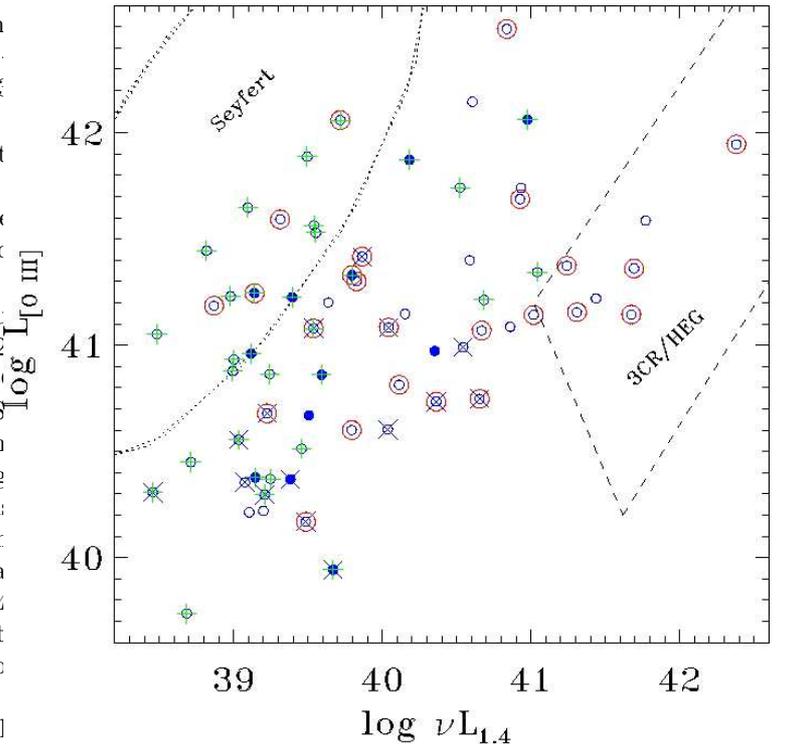}
}
\caption{ Logarithm of the radio vs. [O~III] luminosities for the HEG with
  $0.03<z<0.3$. The symbols, in addition to the empty circles, correspond to:
  1) filled points are late-type galaxies; 2) large red circles have
  black hole masses log $M_{\rm{BH}}/M_{\odot} > 8$; 3) the green pluses are
  the objects close to the empirical separation between the star forming and
  AGN radio emitting sources; 4) the blue crosses are the objects that would
  be classified as LEG adopting to separation HEG/LEG separation derived for
  3CR galaxies \citep{buttiglione10}. We draw as a comparison the location of
  other classes of high excitation AGN: the ellipse marks the boundary of the
  region characteristic of Seyfert \citep{whittle85}, while the dashed region
  comprises the 3CR/HEG \citep{buttiglione10}.}
\label{heg}
\end{figure} 

Let us focus on the location of SDSS/NVSS HEG (including all objects with
$0.03<z<0.3$) in the $L_{\rm 1.4GHz}$ vs $L_{\rm [O~III]}$ plane
(Figure~\ref{heg}). They form the high end of the luminosity distribution in
emission lines ($L_{\rm [O III]} \gtrsim 10^{40}$ erg s$^{-1}$) and their
radio total luminosities span a large range (from 10$^{38.5}$ to 10$^{42.5}$
erg s$^{-1}$). At high emission line luminosities ($L_{\rm [O III]} \gtrsim
10^{41}$ erg s$^{-1}$) a few objects are located in the region characteristic
of 3CR/HEG, but HEG also reach the locus of Seyfert galaxies
\citep{whittle85}, without any apparent discontinuity.  Moving from high to
low radio luminosities, there is a trend in the host properties: high $L_{1.4
\rm{GHz}}$ sources are exclusively ETG (empty circles), while those of lower
radio luminosity, falling in the Seyfert region, are still mostly ETG but with
a contribution from late-types (filled circles) and with many objects with a
likely possibility of a star-formation origin for their radio emission (``+''
symbols). The black hole mass (large red circles mark galaxies with log
$M_{\rm{BH}}/M_{\odot} > 8$) does not seem to have any simple relation with
the location of the objects in the $L_{\rm 1.4GHz}$ - $L_{\rm [O III]}$ plane.

At lower line luminosities, the situation is somewhat different. In the 3CR,
there are no HEG below a line luminosity of 10$^{41}$ erg s$^{-1}$ and
similarly there are no SDSS/NVSS HEG extending the correlation between line
and radio luminosities defined by 3CR HEG toward lower powers. Among the
objects with $L_{\rm [O III]} \lesssim 10^{41}$ erg s$^{-1}$ there is also a
substantial fraction of sources that would be classified as LEG by adopting
the HEG/LEG classification of RLAGN (crossed circles). The presence of low
mass, late-type galaxies is also widespread.

Thus SDSS/NVSS HEG appear to be rather different from those found in the 3CR
sample, with only a few of them showing properties consistent to those of
these powerful FR~II radio-galaxies. SDSS/NVSS HEG are instead more often
associated with lower mass and later type galaxies, similar to the host of
Seyfert galaxies with which they share also the location in the line vs. radio
luminosity plane. The combination of these characteristics suggests that the
SDSS/NVSS HEG sub-sample contains a large fraction of sources that would be
considered classically as radio-quiet AGN and a minority of 
RLAGN. 

Interestingly, several HEG lie in between Seyfert and 3CR/HEG (i.e., between
radio-quiet and radio-loud AGN) without a clear discontinuity in either the
host or the AGN properties.

\subsection{Star forming galaxies}
\label{sfgsect}

In the 3CR sample there is only one galaxy classified as SFG\footnote{This is
  3C~198, a bright FR~II, hosted by a blue elliptical galaxy ($C_r$ = 3.1),
  with $D_n(4000)= 1.1$, and with ${\rm log} \, M_{\rm{BH}}/M_{\odot} =
  8.2$.}. The fraction SFG in the SDSS/NVSS sample is substantially higher,
  $\sim 11\%$ in the low-$z$ sample, although we remind that the method used
  to classify the SFG might lead to an overestimate of the number of these
  objects. This difference might be related to the higher AGN power of the 3CR
  sources that makes more difficult to see the effects of star formation in
  their optical spectra. The exclusion a priori of SFG might also bias our
  estimate of the fraction of ``blue'' ETG derived in Sect.~\ref{spdd}.

However, the properties of SFG are in general not compatible with the
interpretation that they are simply the result of a combination of a radio
loud LEG nucleus with a higher level of star formation. In fact, they usually
have lower $M_{\rm{BH}}$ than LEG/UG and are mostly found in late-type
galaxies. Furthermore most SFG are close to the curve that separates objects
in which the radio emission is powered by an AGN and by star-formation (see
Figure \ref{SFGselection}) and generally well offset from LEG/UG, arguing for
a substantial contamination to their radio luminosity from non nuclear
processes.

Only a minority of SFG ($\sim$ 5 - 10 objects, depending on the diagram
considered) have spectro-photometric properties (leaving aside the emission line
ratios) similar to those of LEG/UG. These could be indeed LEG/UG from the
nuclear point of view, but with a higher star formation rate. In particular 6
SFG are hosted in ETG with a blue color (see Figure \ref{mrur}). Including
them among the blue galaxies their overall fraction would be raised to 7\%,
still consistent with the the fraction of star formation in quiescent ETG
\citep{schawinski09}. 

Conversely, SFG can be in part the result of the combination of a HEG nucleus
  with a high level of star formation. Indeed, the properties of the hosts of
  the two classes are similar. A varying relative contribution of the active
  nucleus and of star formation to the emission lines can account for the
  broad range of diagnostic line ratios covered by SFG and HEG.

\section{Are there exceptions to the rule?}
\label{exceptions}
The SDSS/NVSS sources are thus composed mainly
of objects with low excitation spectra and their host galaxies are mostly
red\footnote{As discussed before, this is the general rule for the hosts of
  low excitation RLAGN, and the fraction of objects with signs of star
  formation are very similar in the 3CR and SDSS/NVSS samples as well as in
  quiescent (non AGN) ETG.}, massive early-type galaxies, harboring SMBH of
very large masses. These characteristics are shared by the LEG drawn from the
3CR sample. 

As shown in the previous sections, there is a minority of radio sources
associated with galaxies showing a HEG or SFG emission line spectrum and these
are also found in smaller hosts, often of late type. However these sub-classes
suffer from a strong contamination from radio-quiet AGN or from objects
whose radio emission is powered by star formation.

At this stage it is important to establish whether there are exceptions to
these general rules answering to these questions: can a RLAGN be associated
with a small black hole? Are there RLAGN hosted in spiral galaxies?  We will
focus on the ``near sample'' ($0.03<z<0.1$) for which we have better
spectro-photometric information and that is still sufficiently large to
contain relatively rare outliers, when compared to, e.g., the 34 3CR in the
same redshift range.

The black hole mass distribution of the 315 LEG (or UG) has a small tail of
low $M_{\rm{BH}}$, while in the 3CR there is sharp cut-off in the black hole
mass distribution at ${\rm log} (M_{\rm{BH}}/M_{\odot}) \sim 7.7$. We will
then focus on the 11 SDSS/NVSS objects with ${\rm log} (M_{\rm{BH}}/M_{\odot})
< 7.5$. There are also 11 LEG/UG with $C_{r}<2.6$, thus likely late-type
galaxies (4 of them also have ${\rm log} (M_{\rm{BH}}/M_{\odot}) < 7.5$ and
are in common with the sample of galaxies of low SMBH mass). Furthermore,
spiral hosts could also be found among the objects with $C_{r}>2.86$ since
$\sim$ 1/3 of Sa and 13 \% of Sb galaxies make this selection (see
\citealt{bernardi09}).

Before we can conclude that any of these are genuine exceptions we examine
their properties in closer detail. We then performed an object by object
inspection of the position in spectro-photometric diagnostic diagrams, and of
their optical and radio images.

\begin{figure*}
\centerline{
\includegraphics[scale=0.175]{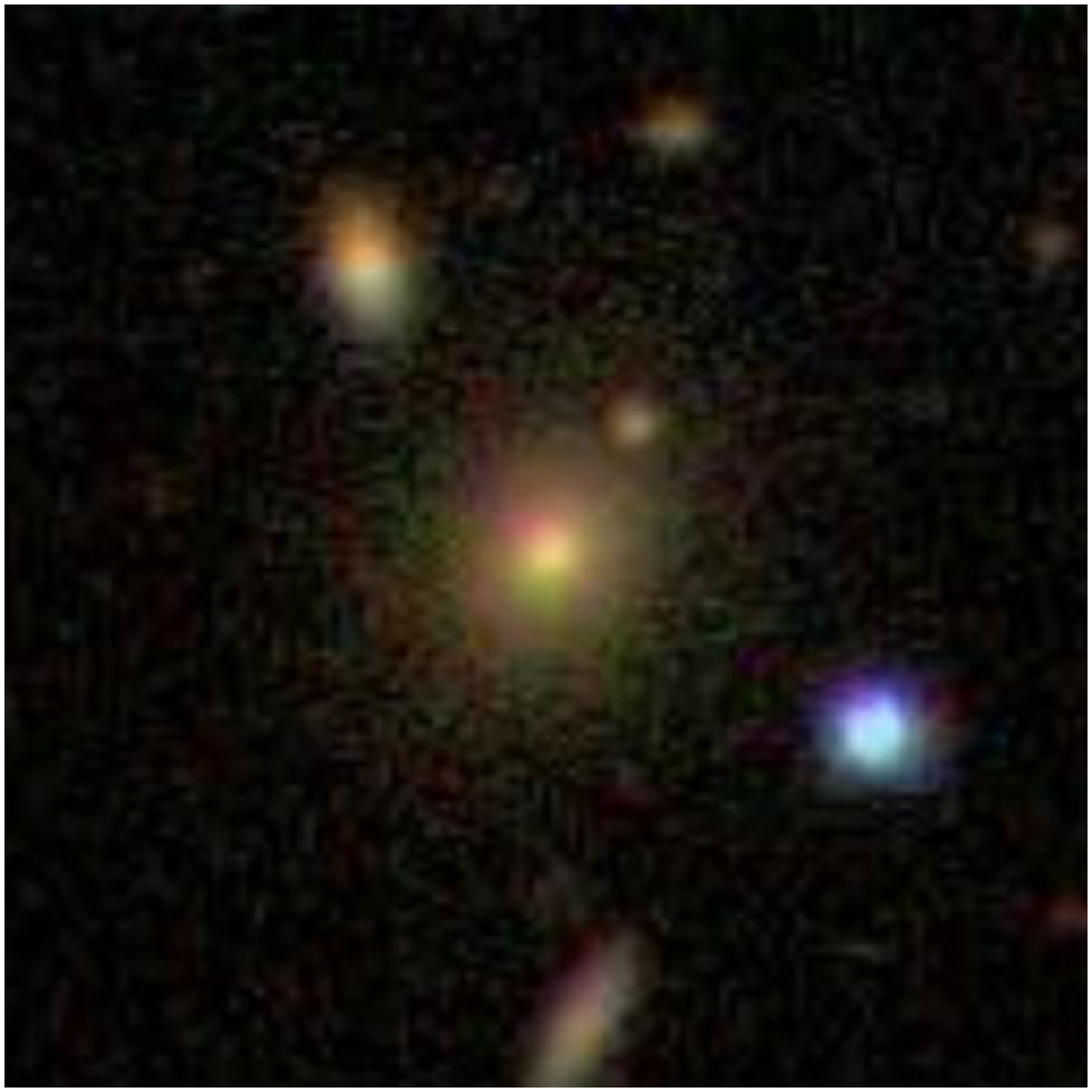}
\includegraphics[scale=0.175]{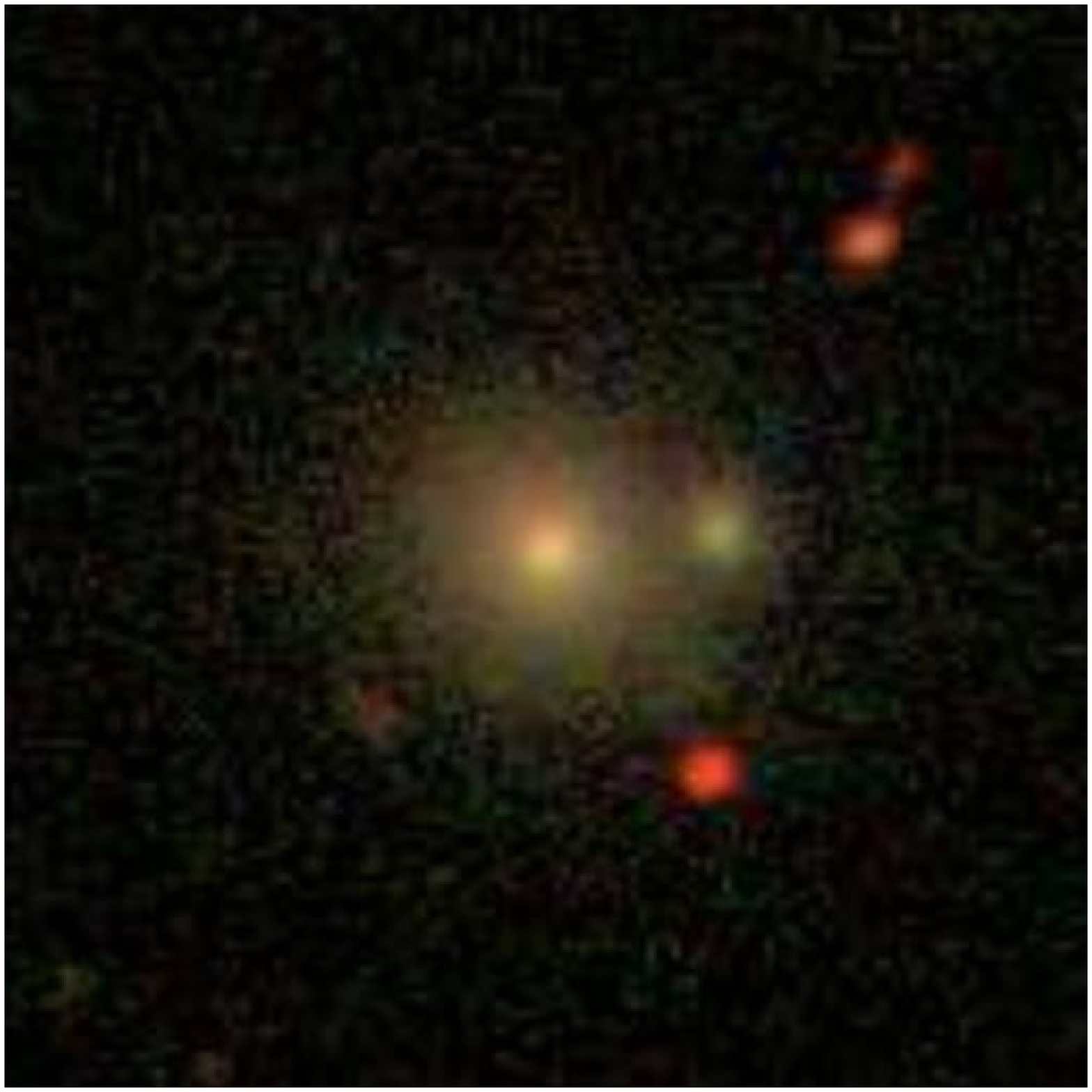}
\includegraphics[scale=0.175]{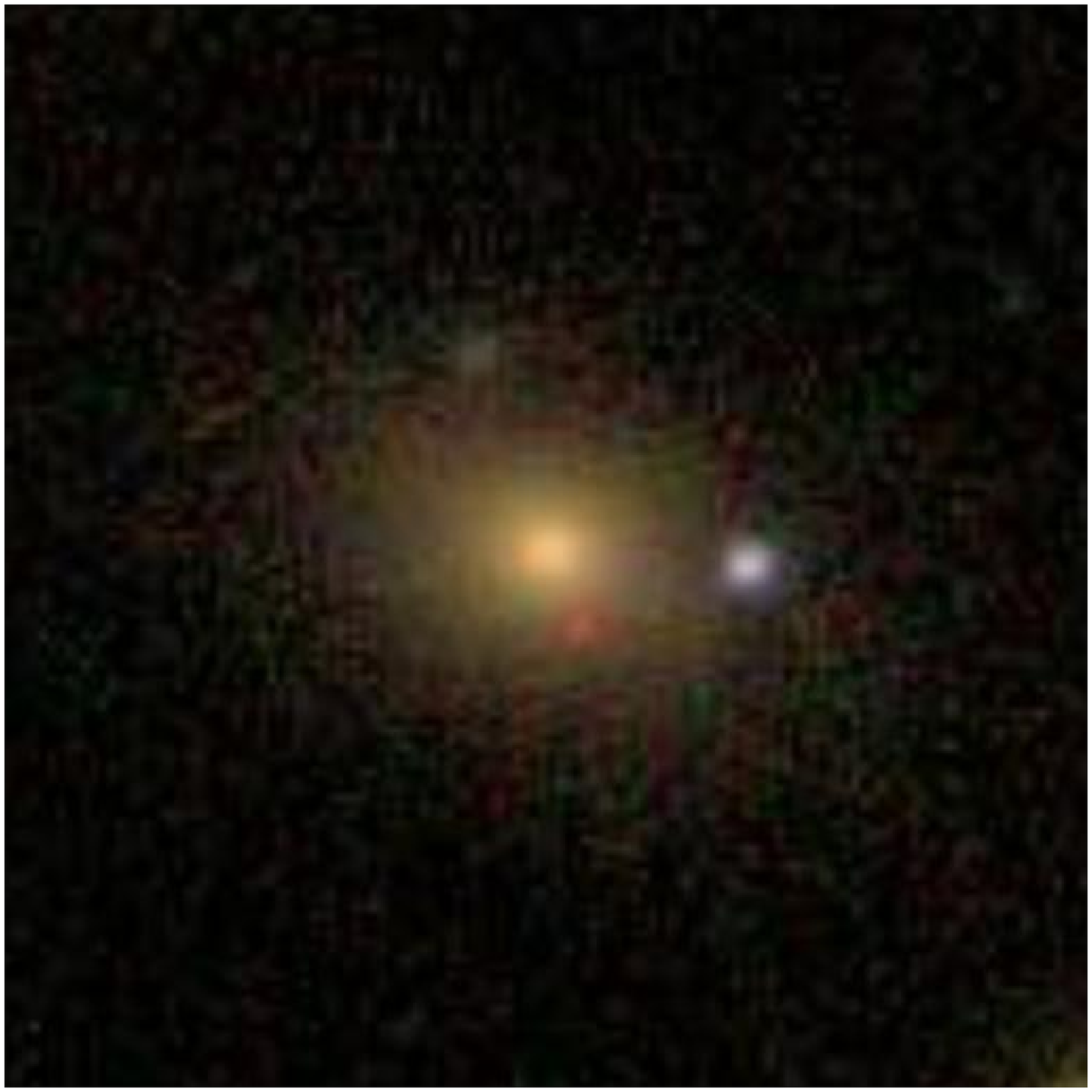}
\includegraphics[scale=0.175]{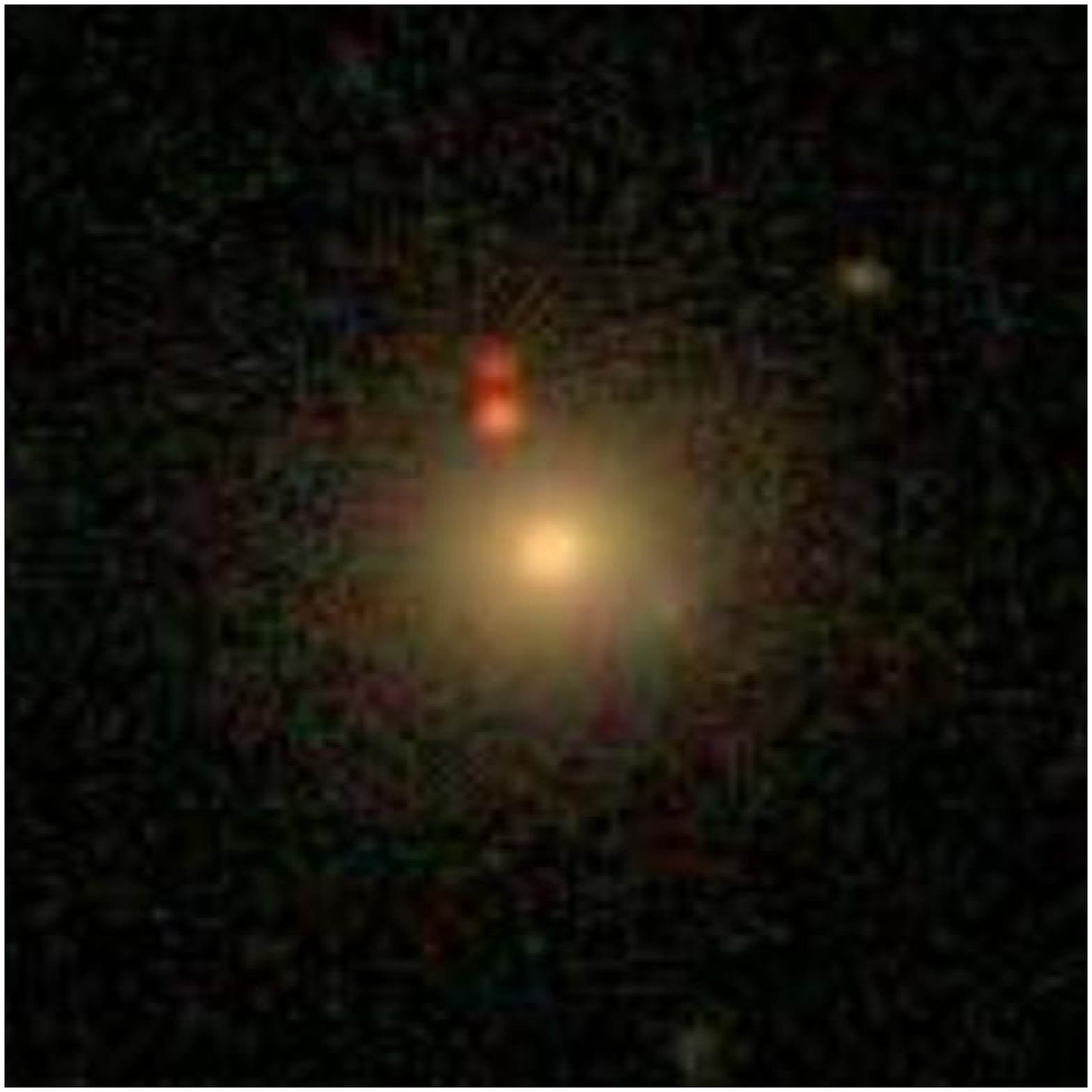}
\includegraphics[scale=0.175]{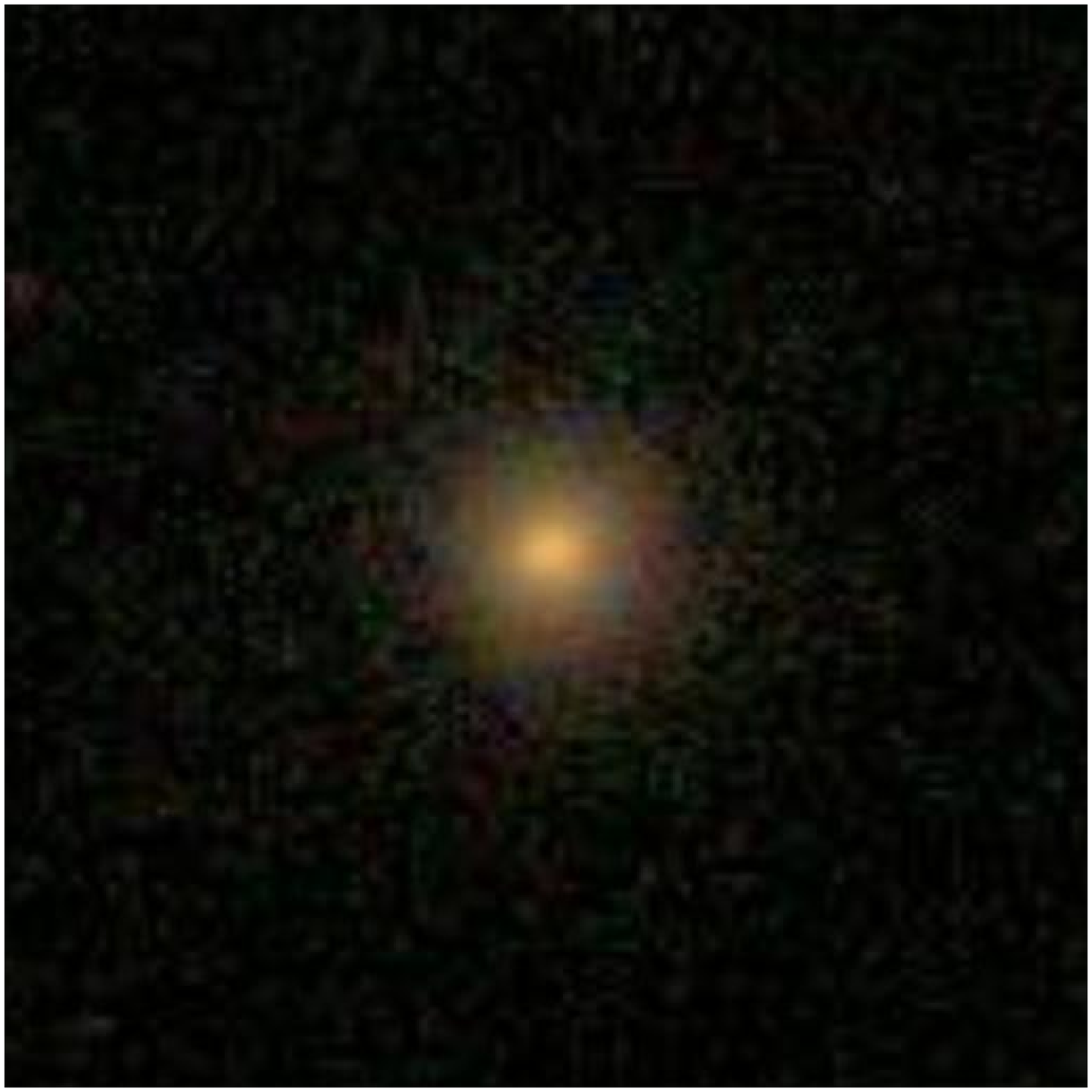}}

\centerline{
\includegraphics[scale=0.175]{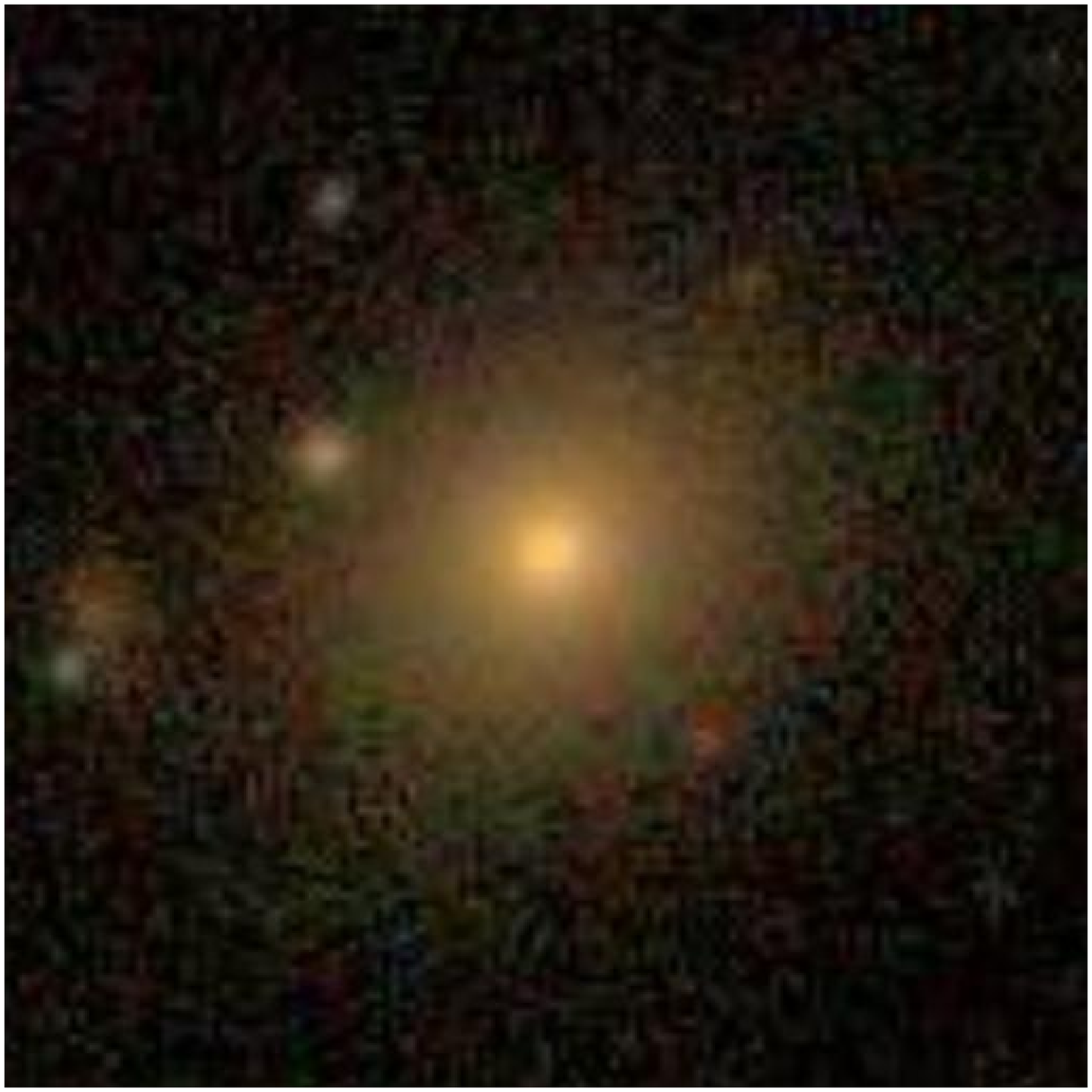}
\includegraphics[scale=0.175]{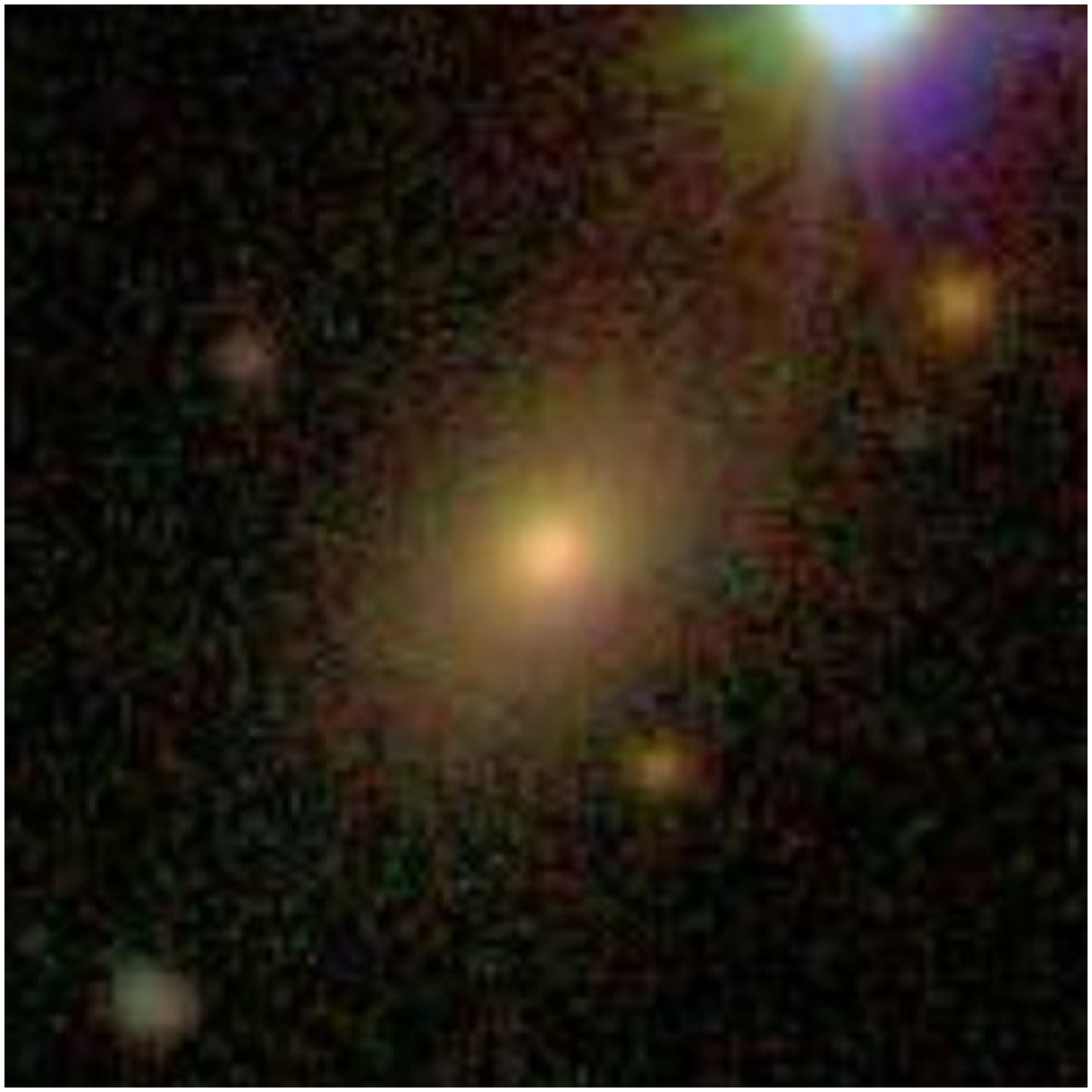}
\includegraphics[scale=0.175]{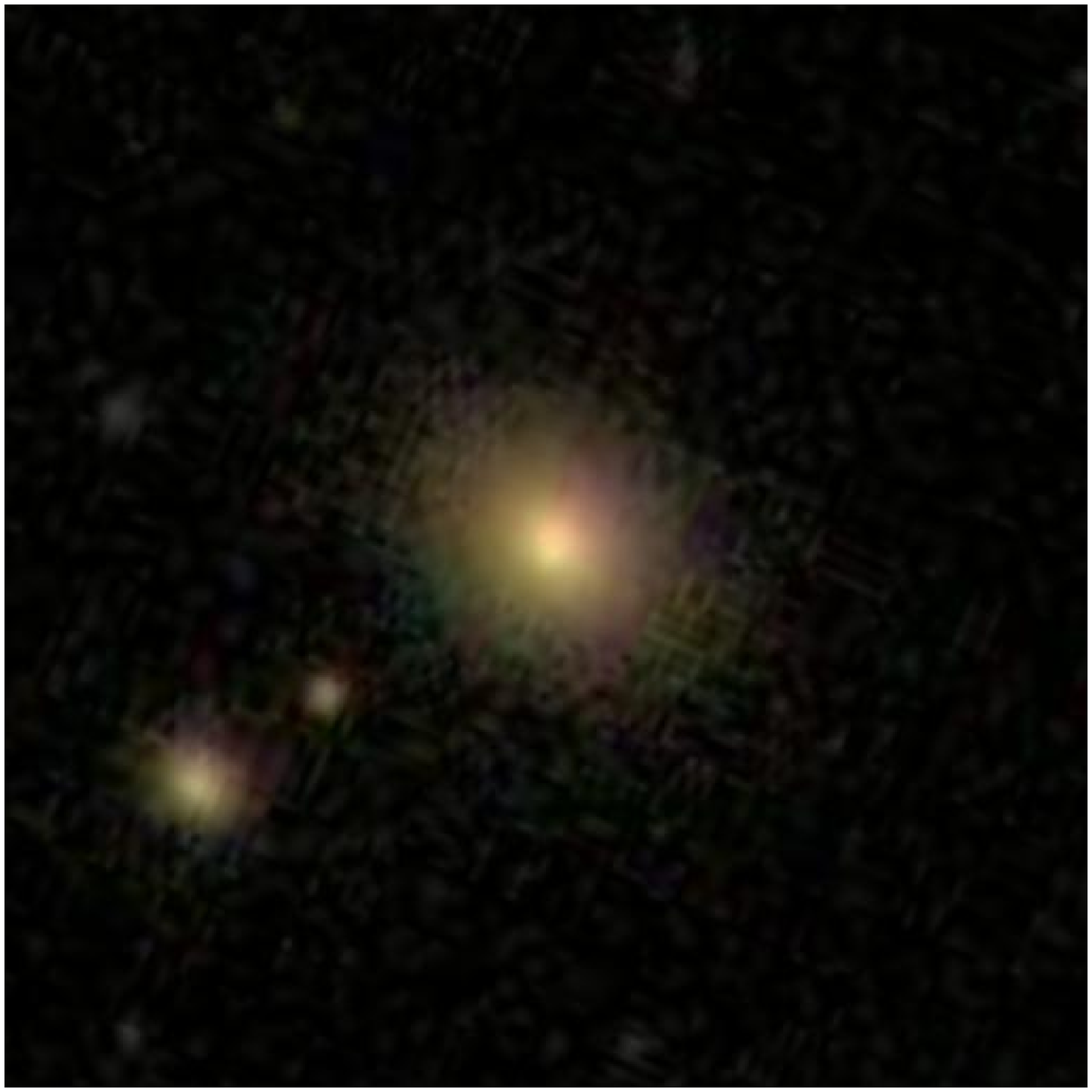}
\includegraphics[scale=0.175]{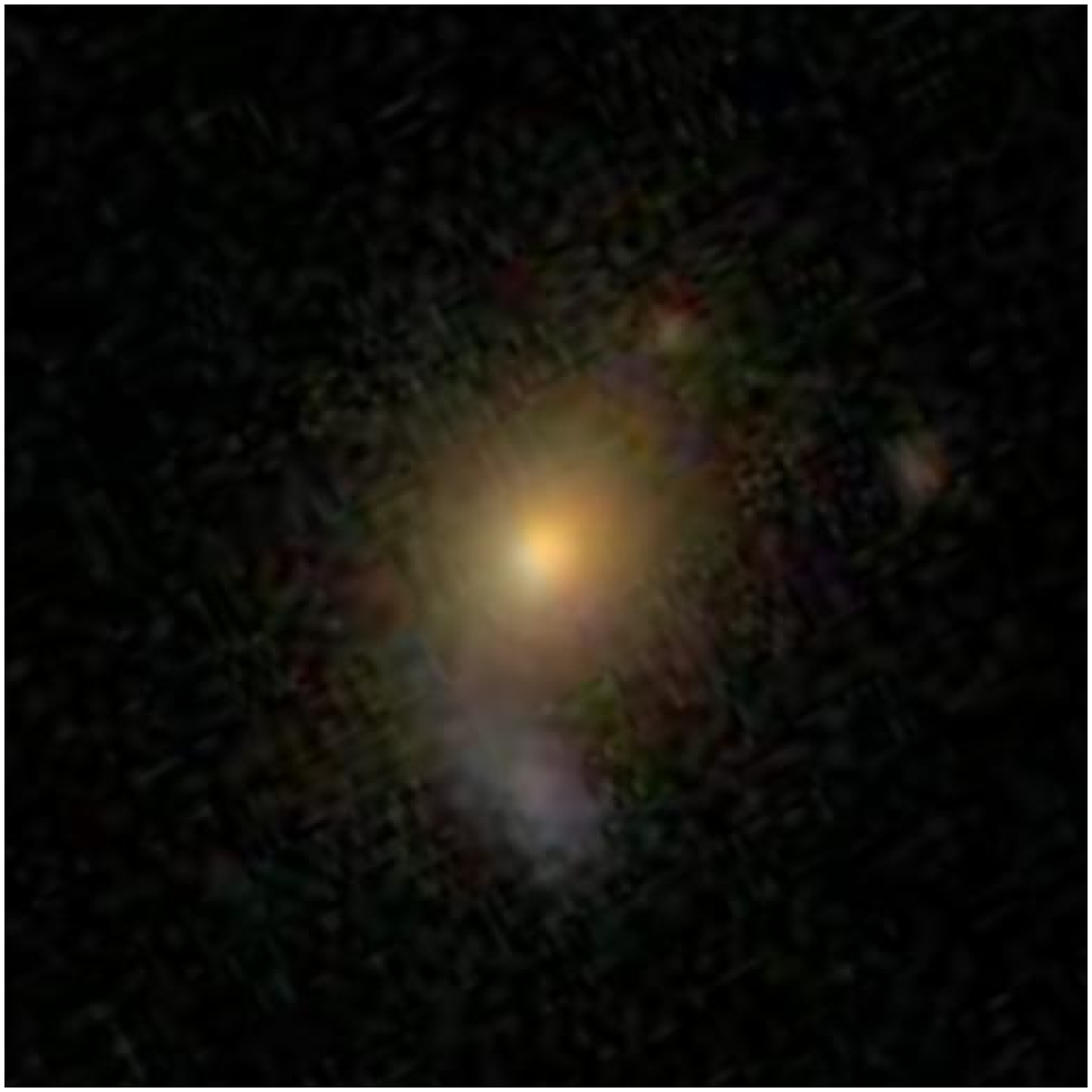}
\includegraphics[scale=0.175]{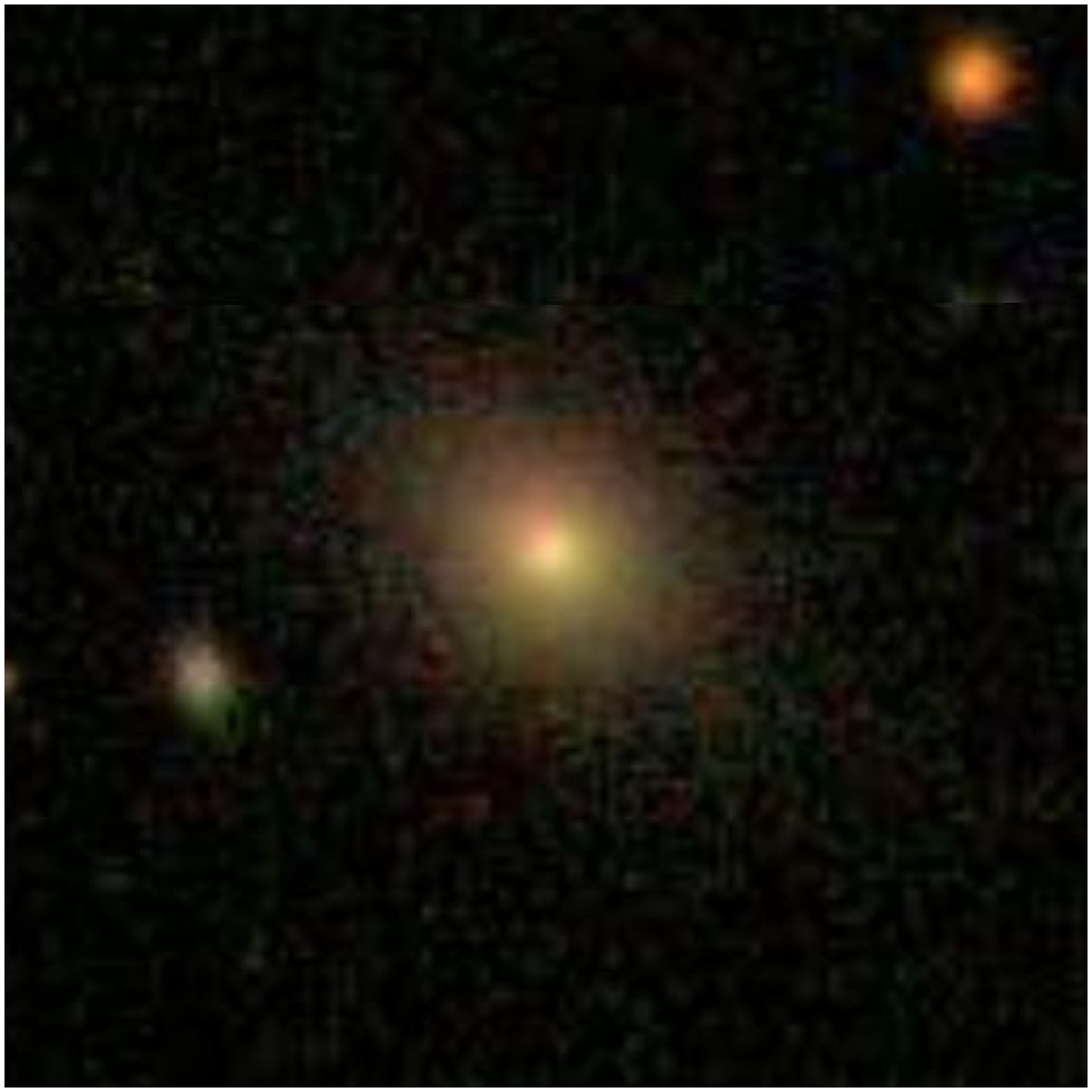}}

\centerline{
\includegraphics[scale=0.175]{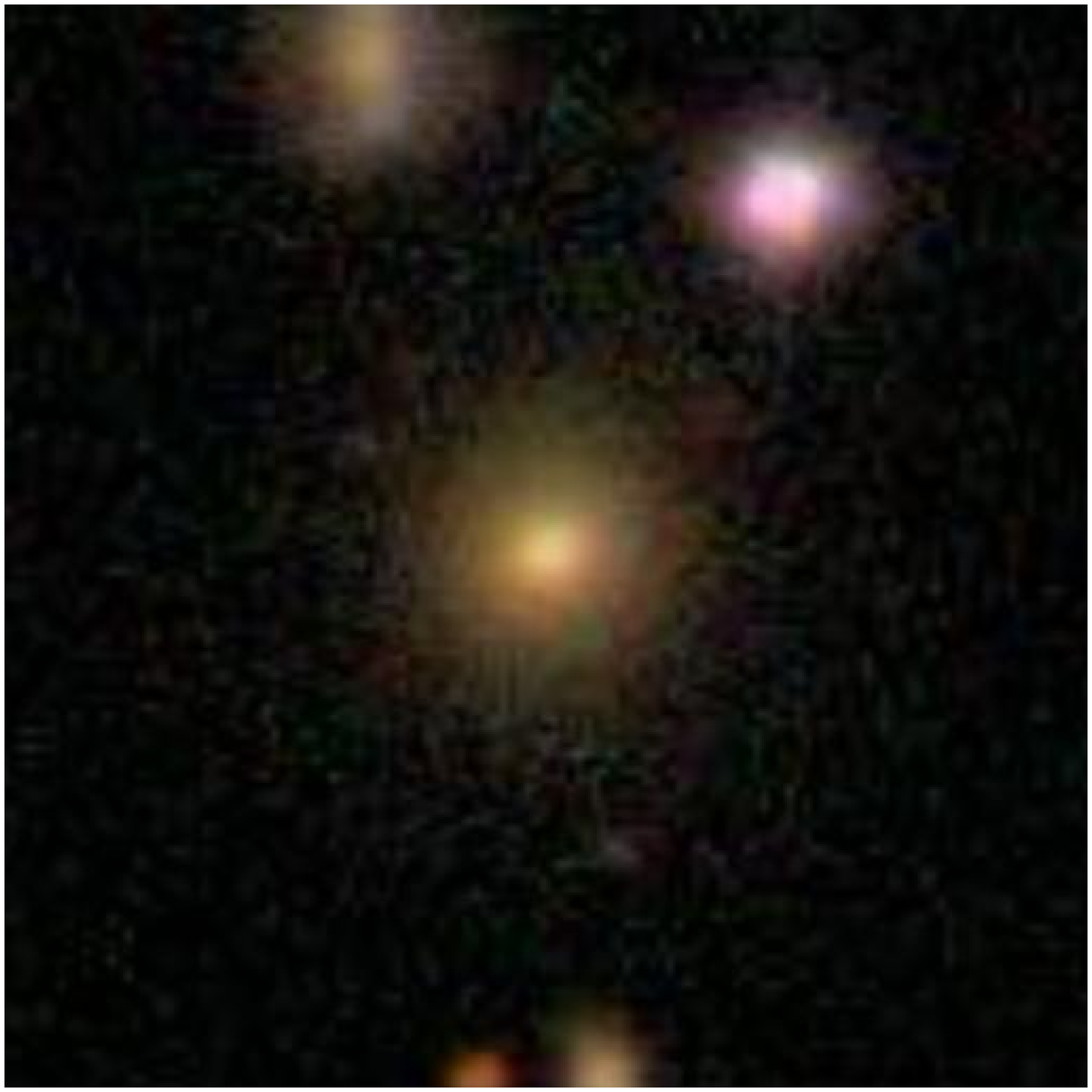}
\includegraphics[scale=0.175]{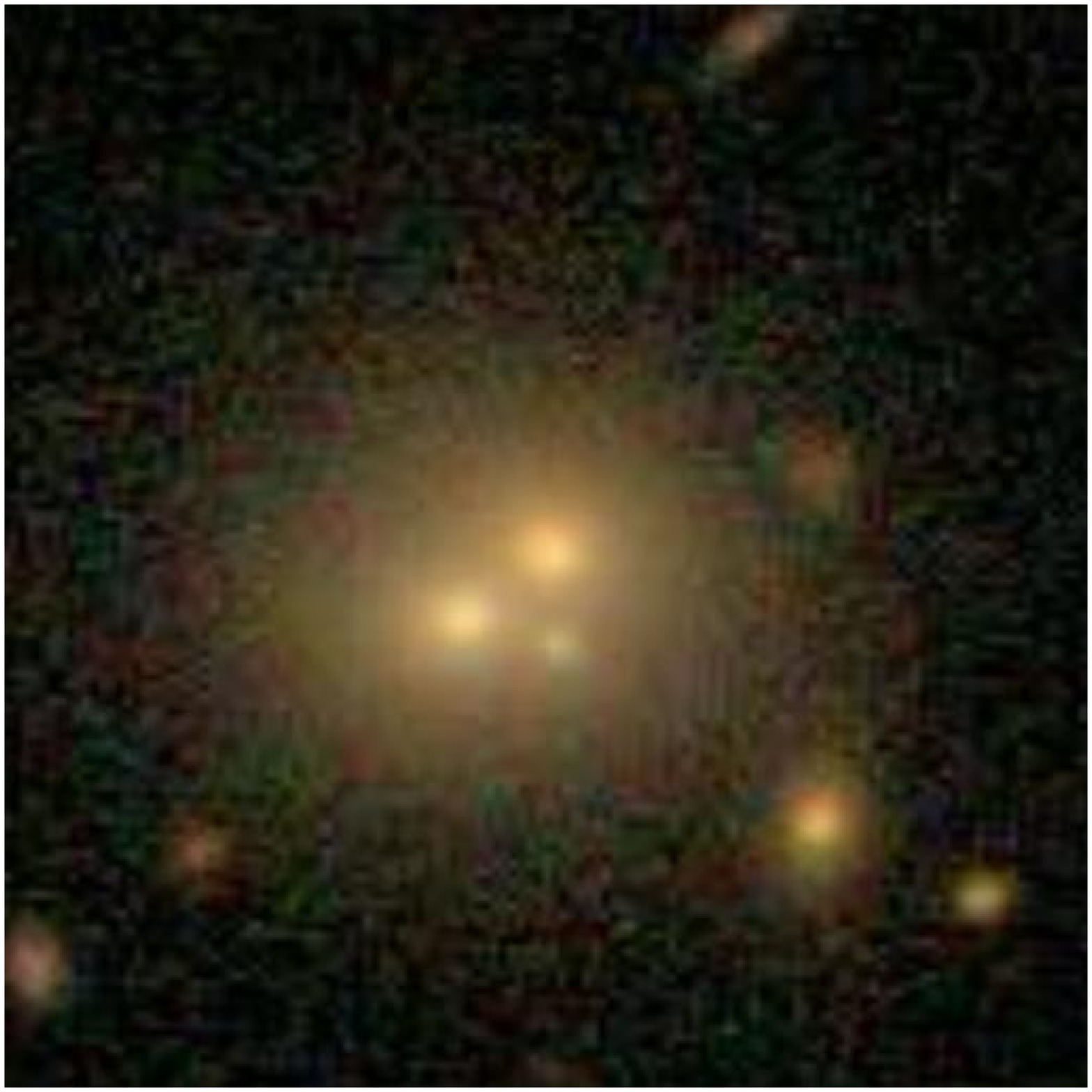}
\includegraphics[scale=0.175]{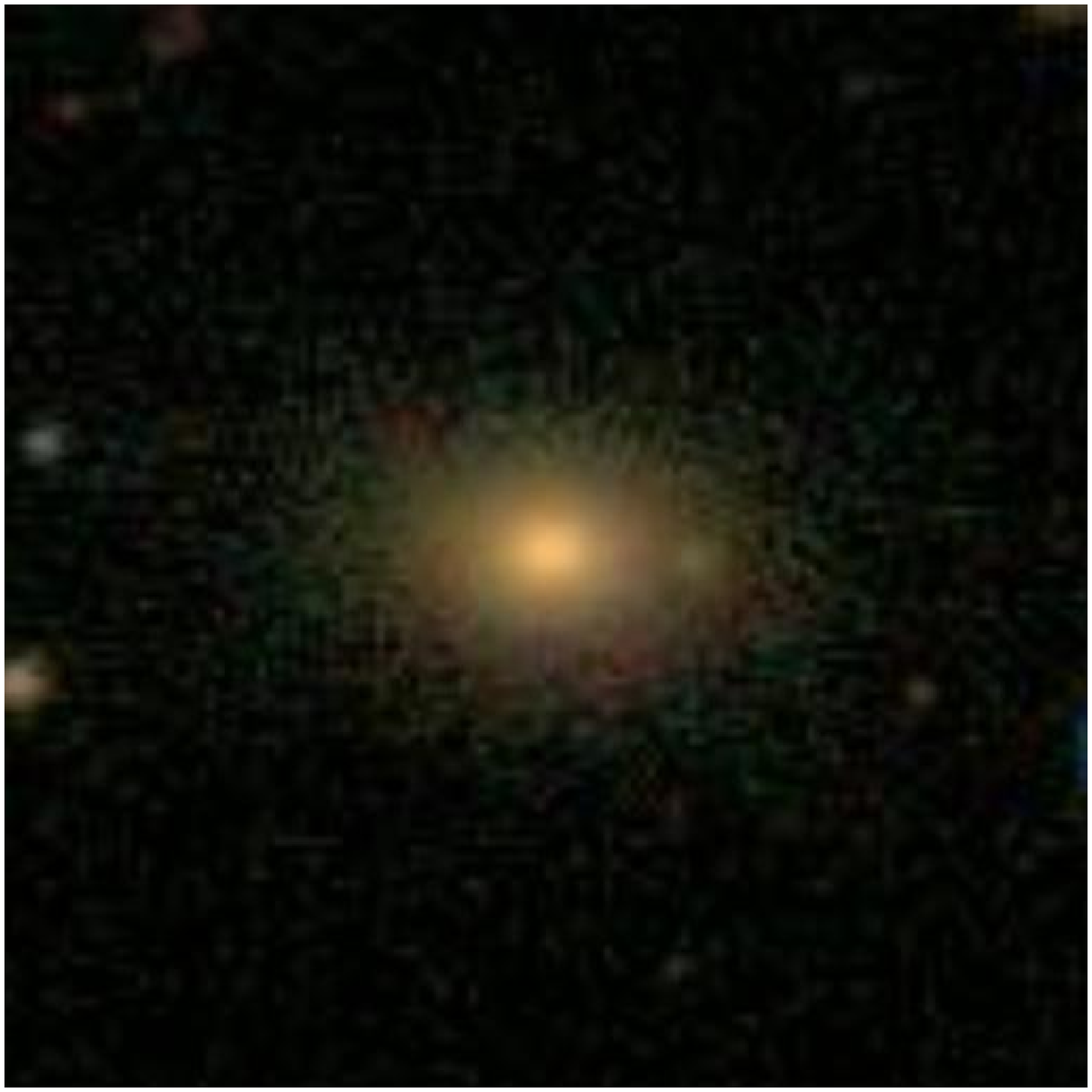}
\includegraphics[scale=0.175]{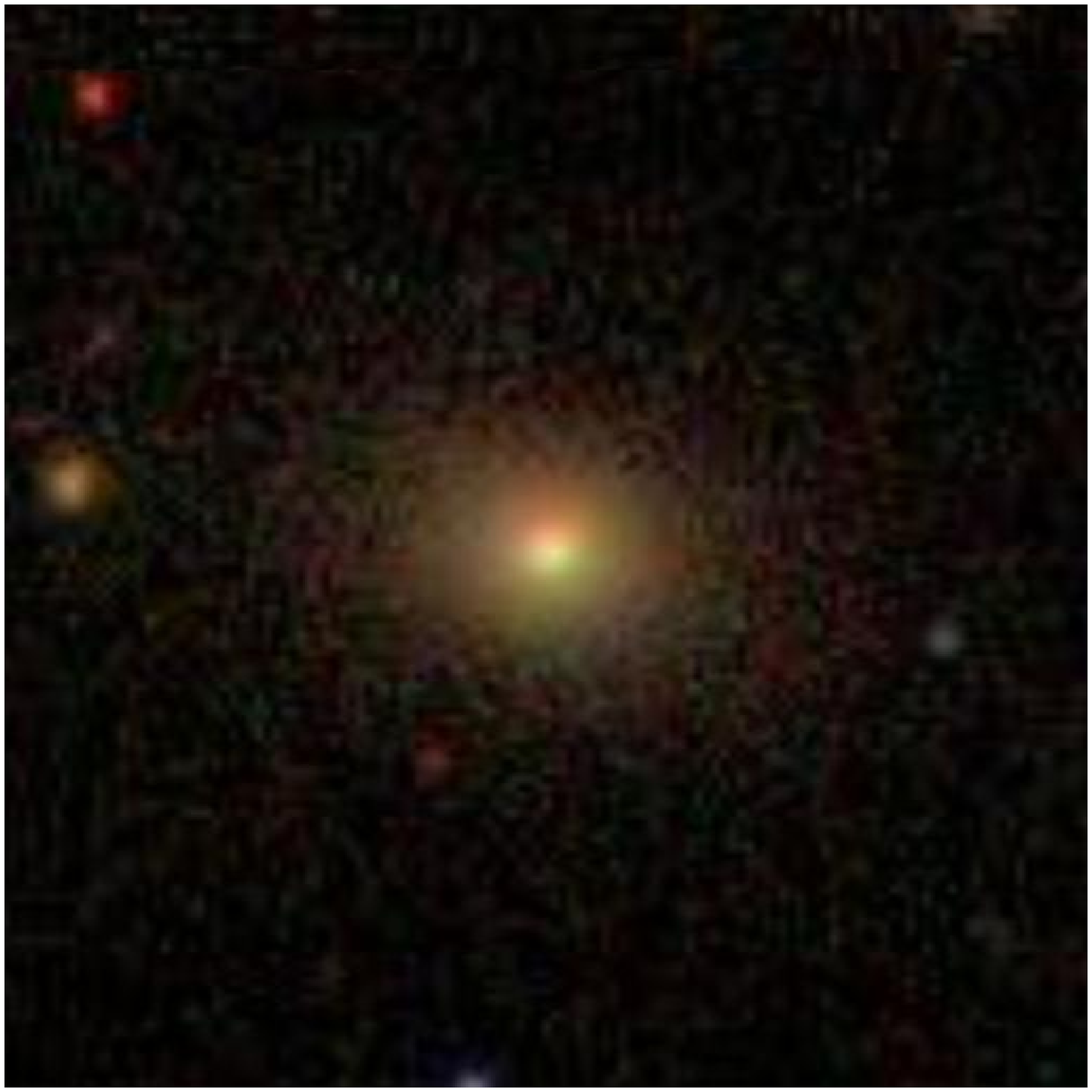}
\includegraphics[scale=0.175]{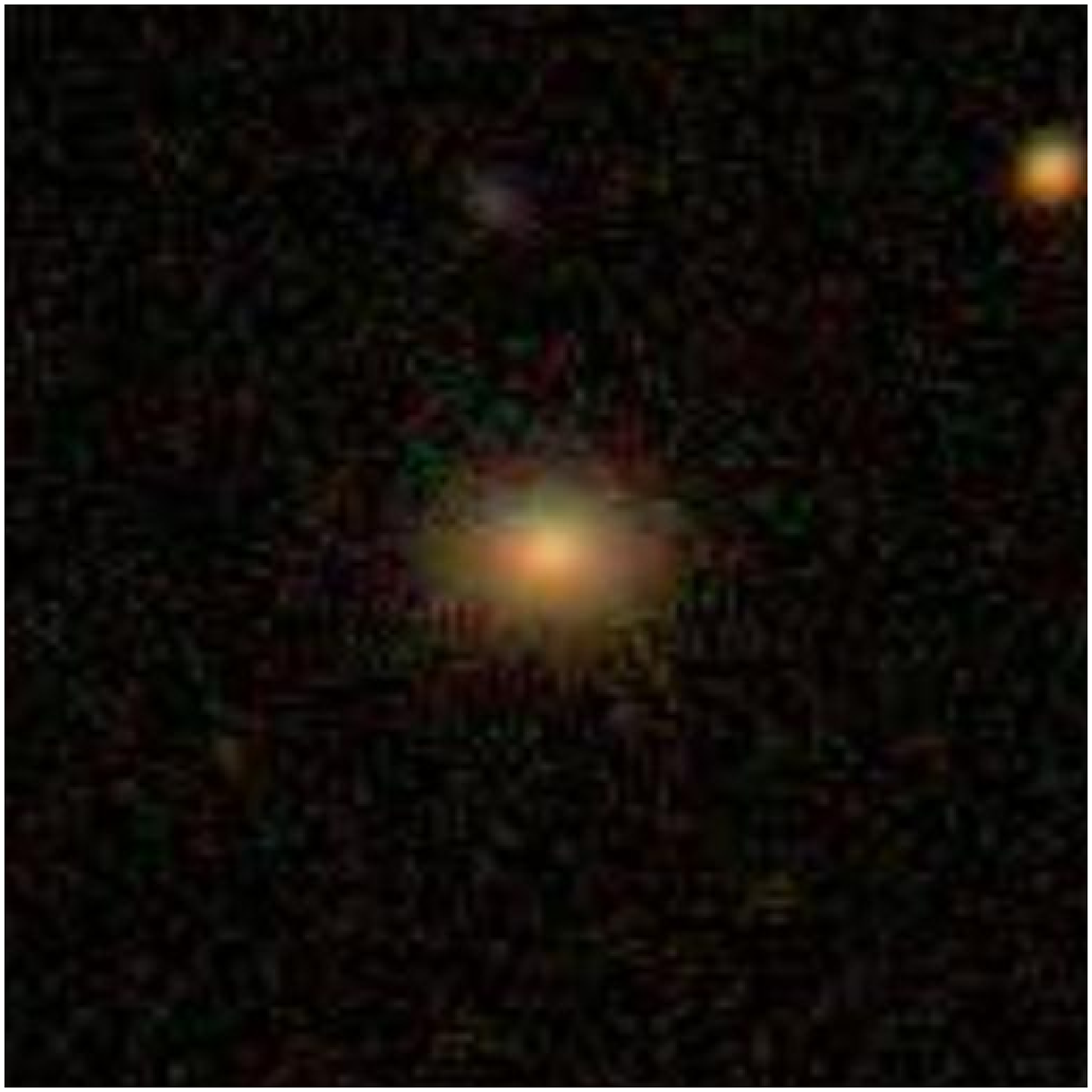}}

\centerline{
\includegraphics[scale=0.175]{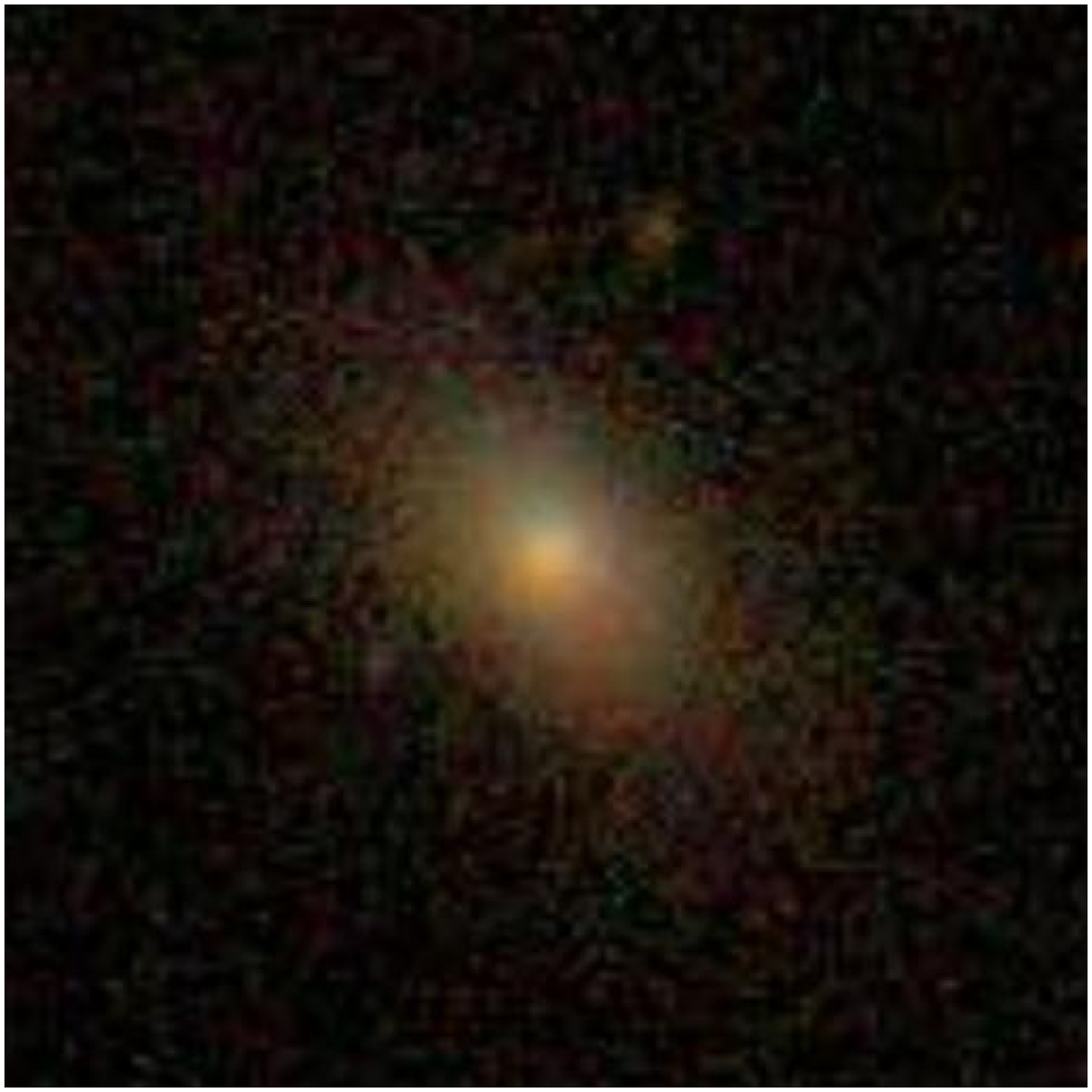}
\includegraphics[scale=0.175]{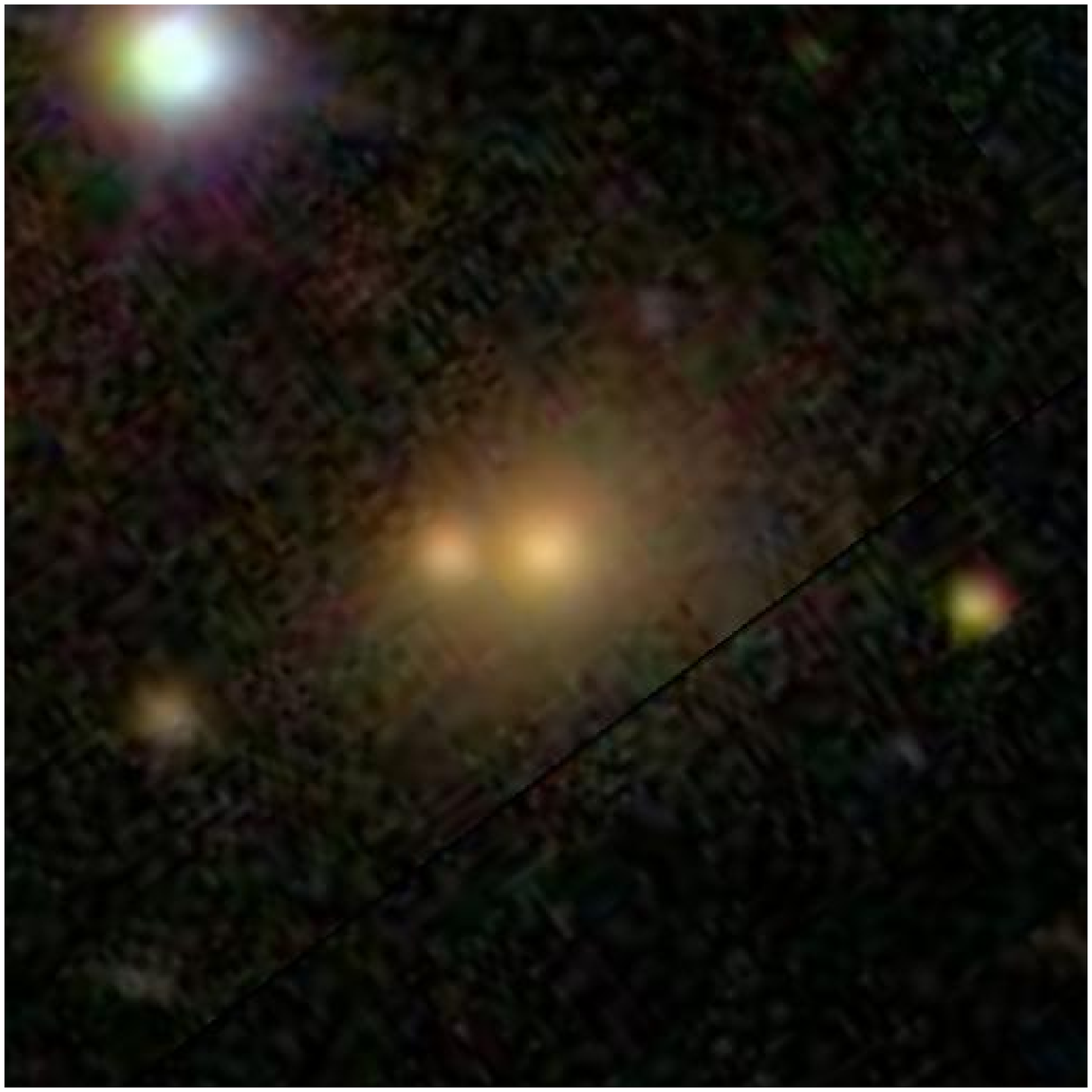}
\includegraphics[scale=0.175]{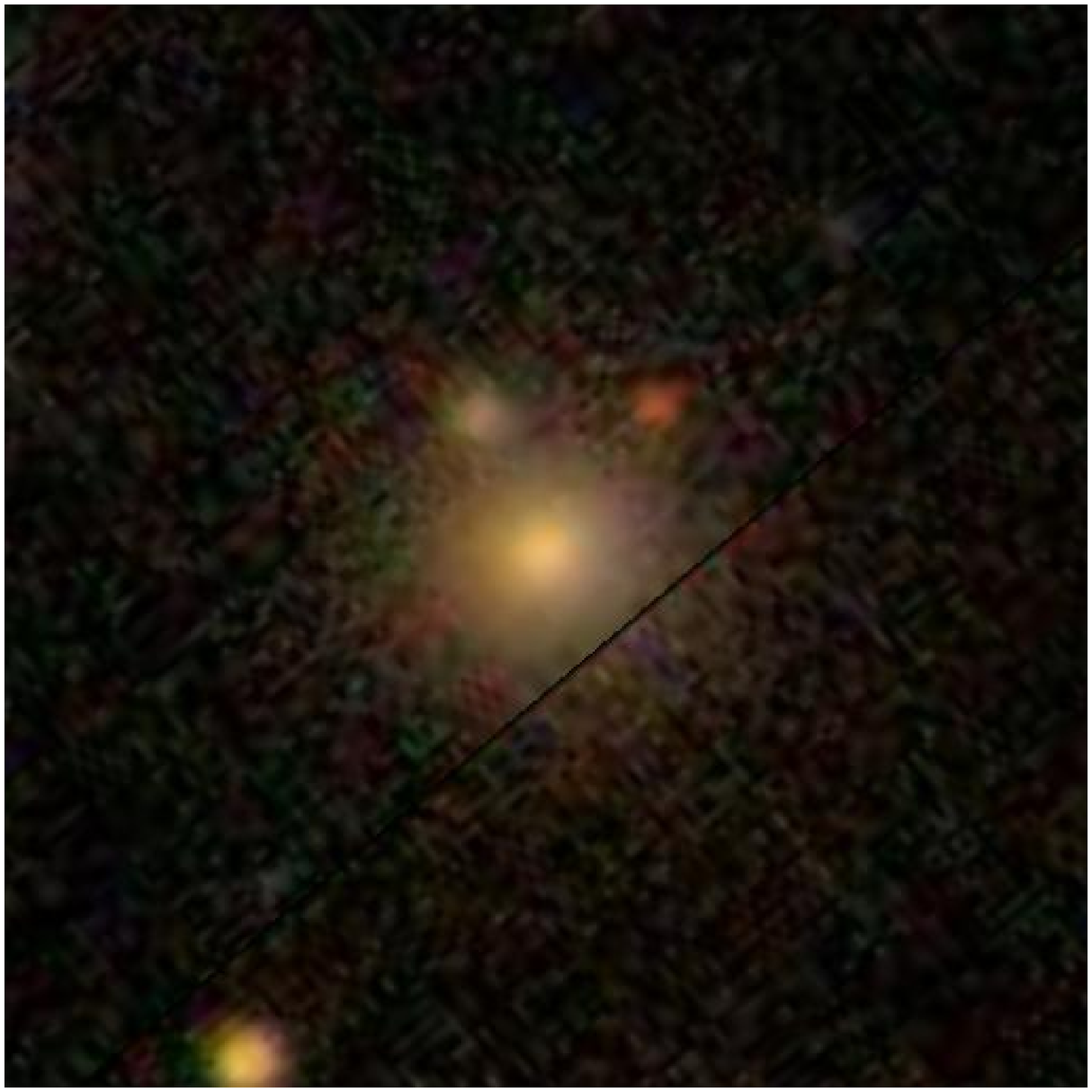}
\includegraphics[scale=0.175]{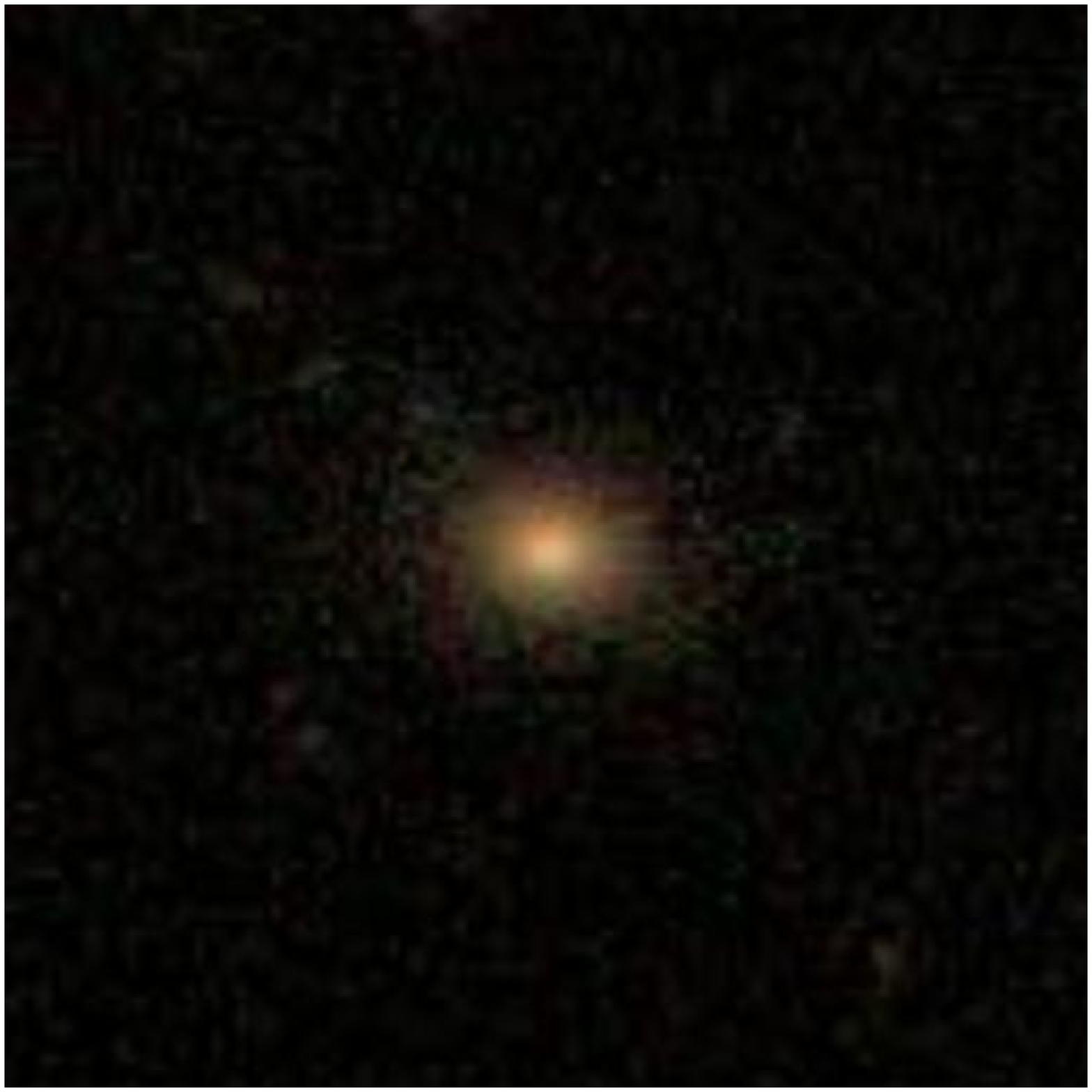}
\includegraphics[scale=0.175]{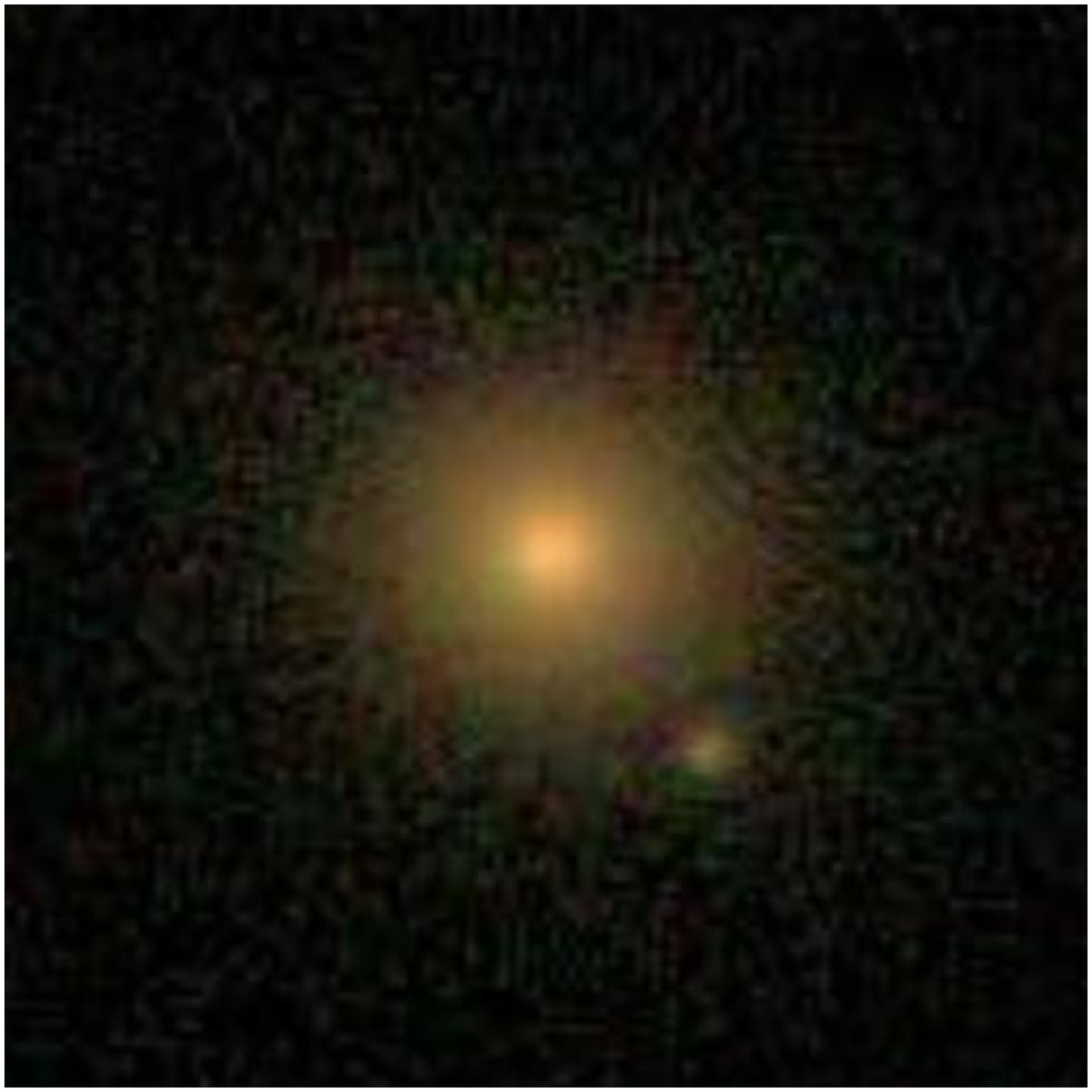}}

\bigskip

\centerline{
\includegraphics[scale=0.175]{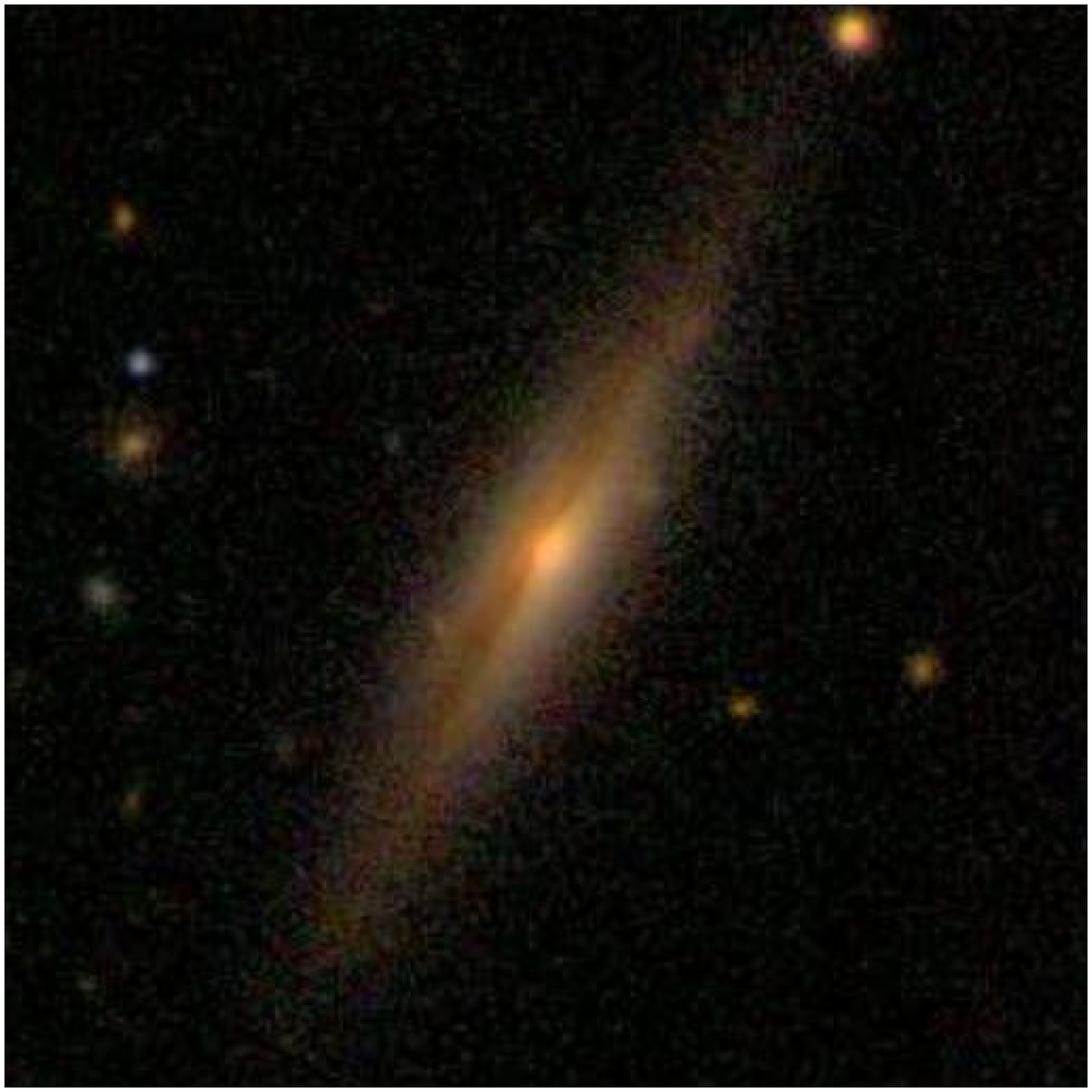}
\includegraphics[scale=0.175]{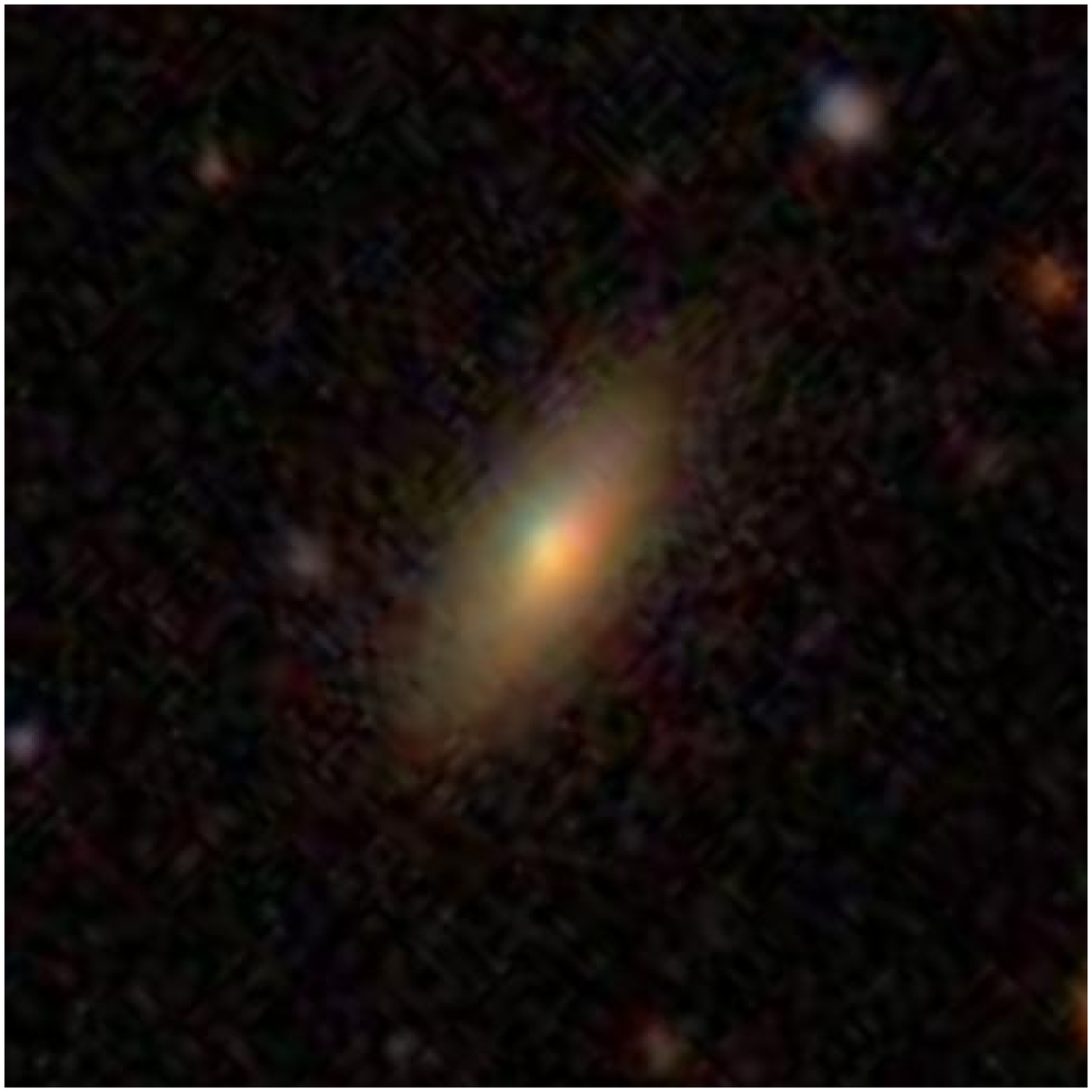}
\includegraphics[scale=0.175]{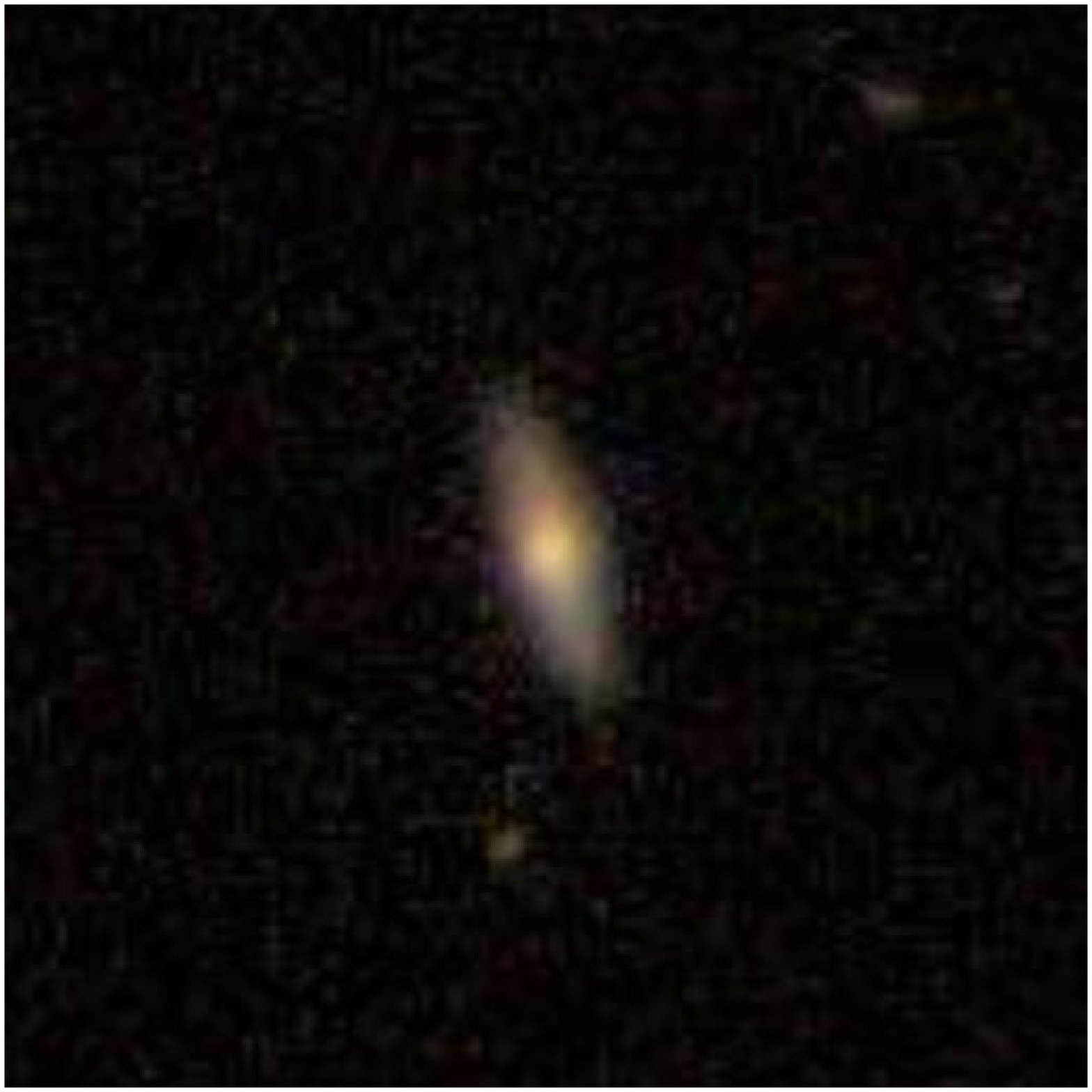}
\includegraphics[scale=0.175]{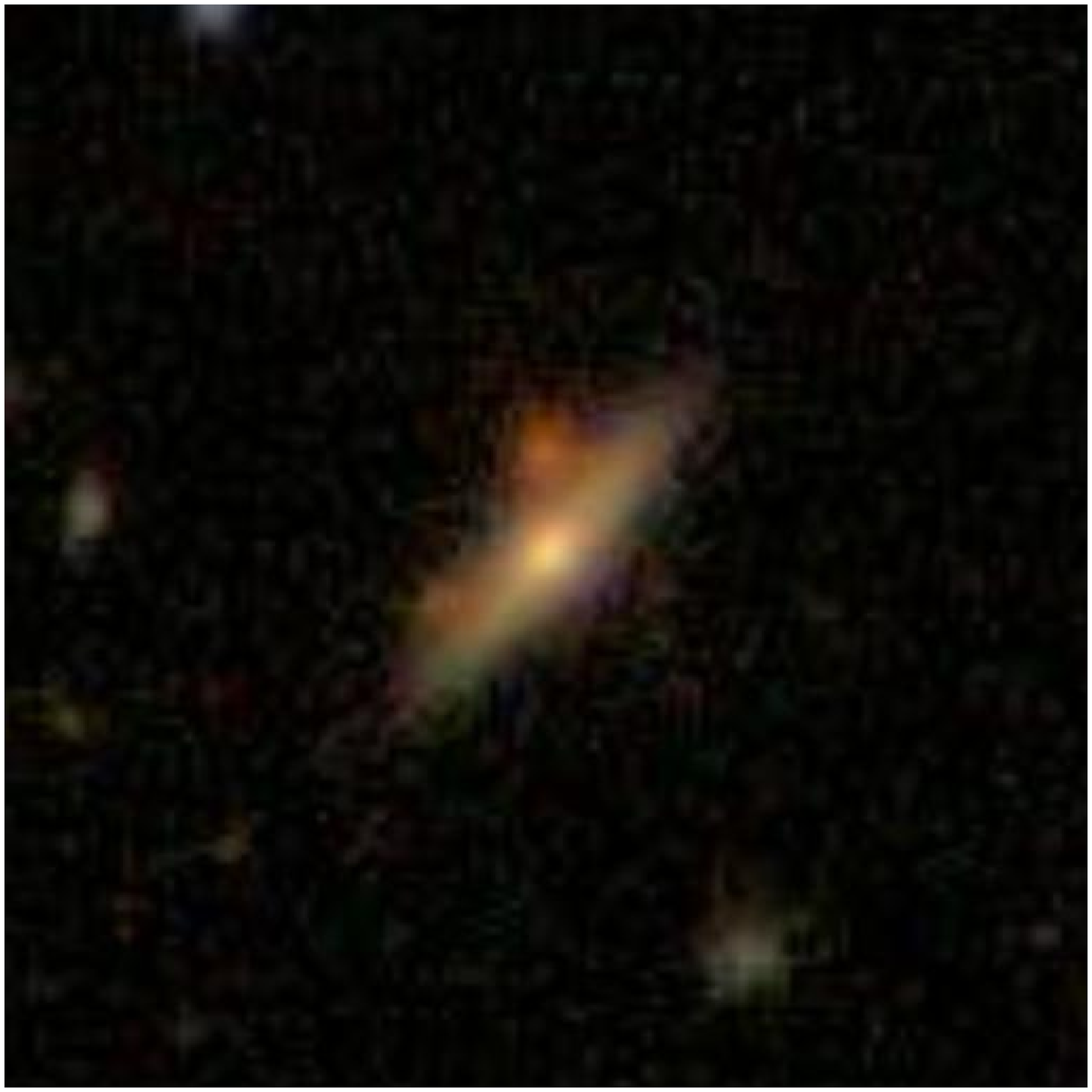}}

\caption{First four rows: $50\arcsec \times 50\arcsec$ SDSS images of 20
  randomly selected LEG or UG in the redshift bin $0.09 < z < 0.1$. All of
  them can be classified by visual inspection as ETG.  Bottom row: 4 galaxies
  whose Hubble type cannot be defined since they are edge-on.}
\label{xfig}
\end{figure*}

First of all we looked for spirals in the sub-sample with $C_{r}>2.86$.
\citet{bernardi09} showed that late and early type galaxies have different
distributions of axial ratios, but with a substantial overlap. Thence
ellipticity cannot be used to assess the possible contribution of spirals to
the RLAGN population. To explore this issue we are left only with direct
visual inspection of the SDSS images. We then considered the 277 galaxies with
$z<0.1$ having $C_{r}>2.86$, $\sigma(C_r) < 0.2$, and a UG or LEG
classification. Although we extended this method to a larger redshift than
used by e.g. the Zoo Project, we are confident that the quality of the SDSS
images is sufficient for a robust identification of their Hubble type. This
conclusion can be drawn looking at the images in Figure \ref{xfig}, showing 20
galaxies selected at random from those in the largest redshift bin considered,
$0.09 < z < 0.1$, all of them clear ETG. Probably the high reliability of
visual inspection at these relative high redshift is due to the fact that we
are dealing, in most cases, with very massive objects whose morphology remains
well defined even at $z \sim 0.1$.

We only found 4 galaxies whose type cannot be defined securely since they are
seen almost edge on (see the bottom row in Figure \ref{xfig}) and that, in
principle, could be spirals. However, if indeed massive late type galaxies
were associated with RLAGN, in our analysis we should have found also their
face-on counterparts. This leads us to conclude that the 4 galaxies under
scrutiny are most likely the small fraction of S0 expected to be seen edge-on.

We then consider the 11 galaxies with $C_{r}<2.6$. The first issue is
to establish whether their radio emission is indeed powered by an AGN and not
by star formation. As shown in Figure~\ref{SFGselection}, 9 LEG/UG are
located close (with an offset of $<$ 0.1 in $D_{n}(4000)$) to the boundary of
the region characteristic of the objects in which the radio emission is
powered by star formation; 6 of them are galaxies with $C_{r}<2.6$.  As
discussed in Sect. \ref{spdd} this casts doubts on the AGN origin of the radio
emission in these objects.

This leaves us with only 5 galaxies with $C_{r}<2.6$ with radio emission of
apparently secure AGN origin. Visual inspection of their SDSS images
clearly indicates that 4 of them ($\sim \% 1$ of the sample) are clearly
ETG. This is not surprising since \citet{bernardi09} estimated that the
fraction of galaxies visually classified as ETG having $C_{r}<2.6$ is
$\sim$2\% for Es and $\sim$7\% for S0s. Limiting to the most massive ones
($M_r \lesssim -21$) these fraction are approximately halved. Only one galaxy
is indeed a spiral (SDSS J170007.17+375022.2) at a redshift of 0.063. However,
the $3\arcsec$ SDSS fiber covers only its bulge, returning a $D_{n}(4000)$ =
1.75, typical of a rather old stellar population. This is the value used to
locate this source in the $D_{n}(4000)$ vs. $L_{\rm 1.4GHz}/M_{*}$ plane. We
argue that a substantial fraction of its radio emission can originate from
younger star within the spiral arms, an idea supported by the lack of a
central source in the FIRST image. 

We finally consider the 7 galaxies with low $M_{\rm{BH}}$ (4 were already
discussed among the likely late type). Although the radio emission of one of
them is probably dominated by star formation, another 3 possibly make the 5
mJy threshold only thanks to the contribution of unrelated nearby radio
sources, there are 3 galaxies behaving like the rest of the LEG/UG population
is all diagnostic diagrams, but with $M_{\rm{BH}} \sim 2 - 3 \times 10^7
M_{\odot}$.  One of them (SDSS J133453.37-013238.5) is associated with a radio
source of clear FR~II morphology extended over $\sim 90$ kpc, an unmistakable
sign of the presence of a RLAGN.

Nonetheless, the quoted SMBH masses are not actual measurements, but estimates
derived from the stellar velocity dispersion, $\sigma_*$. Since this relation
has an intrinsic scatter, in a population with a low-mass cut-off in black
hole masses, a tail of low $\sigma_*$ is indeed expected. Without a prior
knowledge of SMBH masses distribution it is difficult to estimate whether our
fraction of low $\sigma_*$ objects ($\sim$ 1\% of the sample) is consistent
with this interpretation. We then used the relation between the host and SMBH
masses derived by \citet{marconi03} to obtain a independent $M_{\rm{BH}}$
measurement. Their host masses are in the range log $M_*/M_{\odot} =10.54 -
10.99$ corresponding to ${\rm log} (M_{\rm{BH}}/M_{\odot}) \sim 7.93 - 8.37$.

Summarizing, we failed to find any convincing RLAGN associated with a spiral
galaxy or with black holes with a mass substantially lower than to $10^8
M_{\odot}$.

However a link between $M_{\rm{BH}}$ and radio luminosity exists. In fact, the
  normalization of the radio luminosity function of galaxies of different SMBH
  masses shows a dependence as $\rho \sim M_{\rm BH}^{1.6}$
  \citep{best05b}. Thus AGN associated with smaller SMBH are expected to be
  less frequent, at a given threshold in radio luminosity, than in SMBH of
  higher mass. The stellar mass function of ETG is essentially constant (in a
  log-log representation) for $M_*/M_{\odot} \lesssim 10^{11}$
  (e.g. \citealt{bernardi09}), corresponding to $M_{\rm{BH}}/M_{\odot} \sim
  2\times10^{8}$, having adopted the ratio $M_{\rm{BH}}/M_*$ from
  \citet{marconi03}. The number of LEG/UG radio-galaxies with $7.7 \lesssim
  {\rm log} M_{\rm{BH}}/M_* \lesssim 8.7$ is $\sim$ 300 and the number of {\sl
  expected} radio-galaxies in the SDSS/NVSS sample with $6.7 \lesssim {\rm
  log} M_{\rm{BH}}/M_* \lesssim 7.7$ is $\sim 300\times10^{-1.6}\sim7$.  The
  failure to find any of such low mass objects indicates that the link between
  radio luminosity and $M_{\rm{BH}}$ breaks toward low SMBH masses.

\begin{figure*}
\centerline{
\includegraphics[scale=0.175]{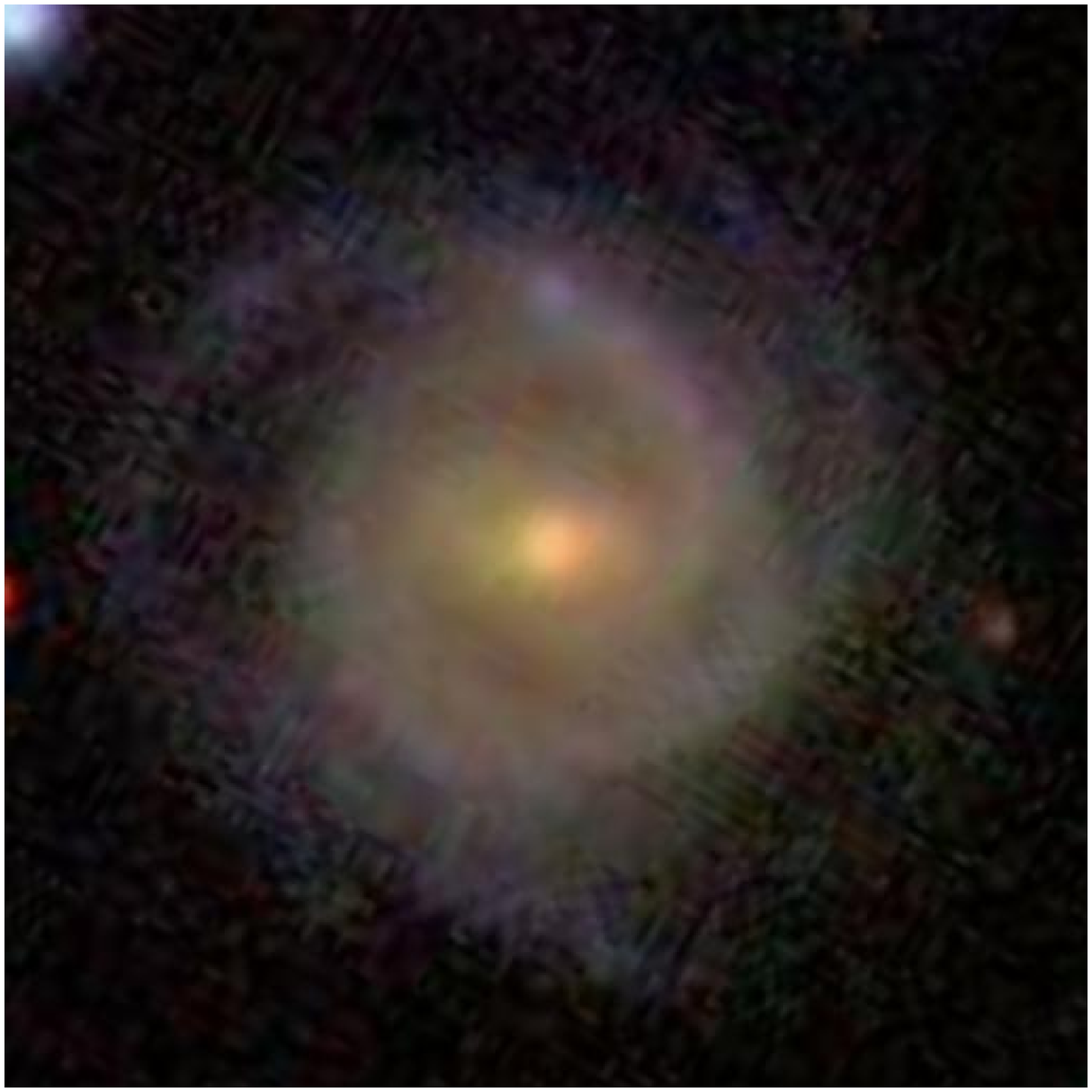}
\includegraphics[scale=0.175]{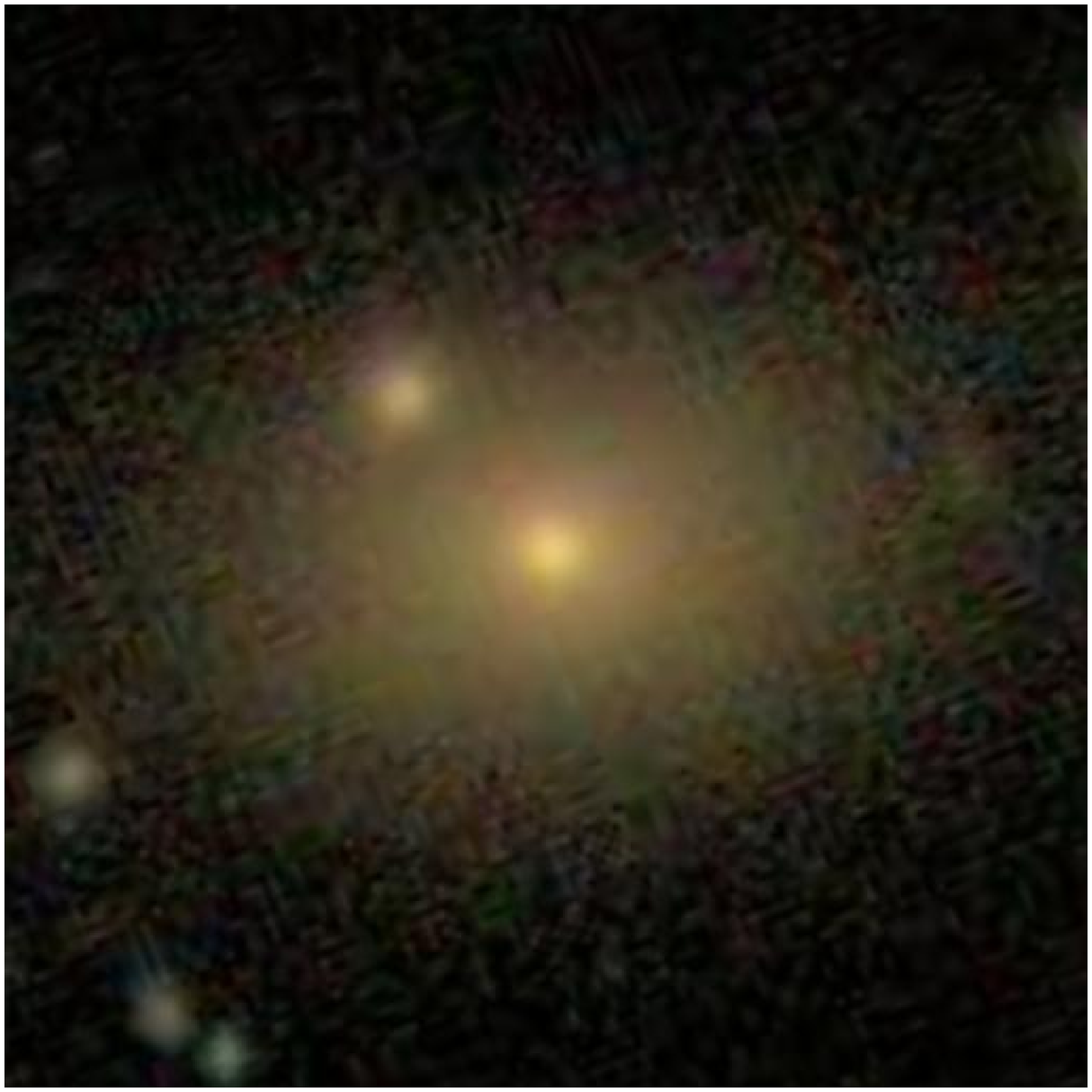}
\includegraphics[scale=0.175]{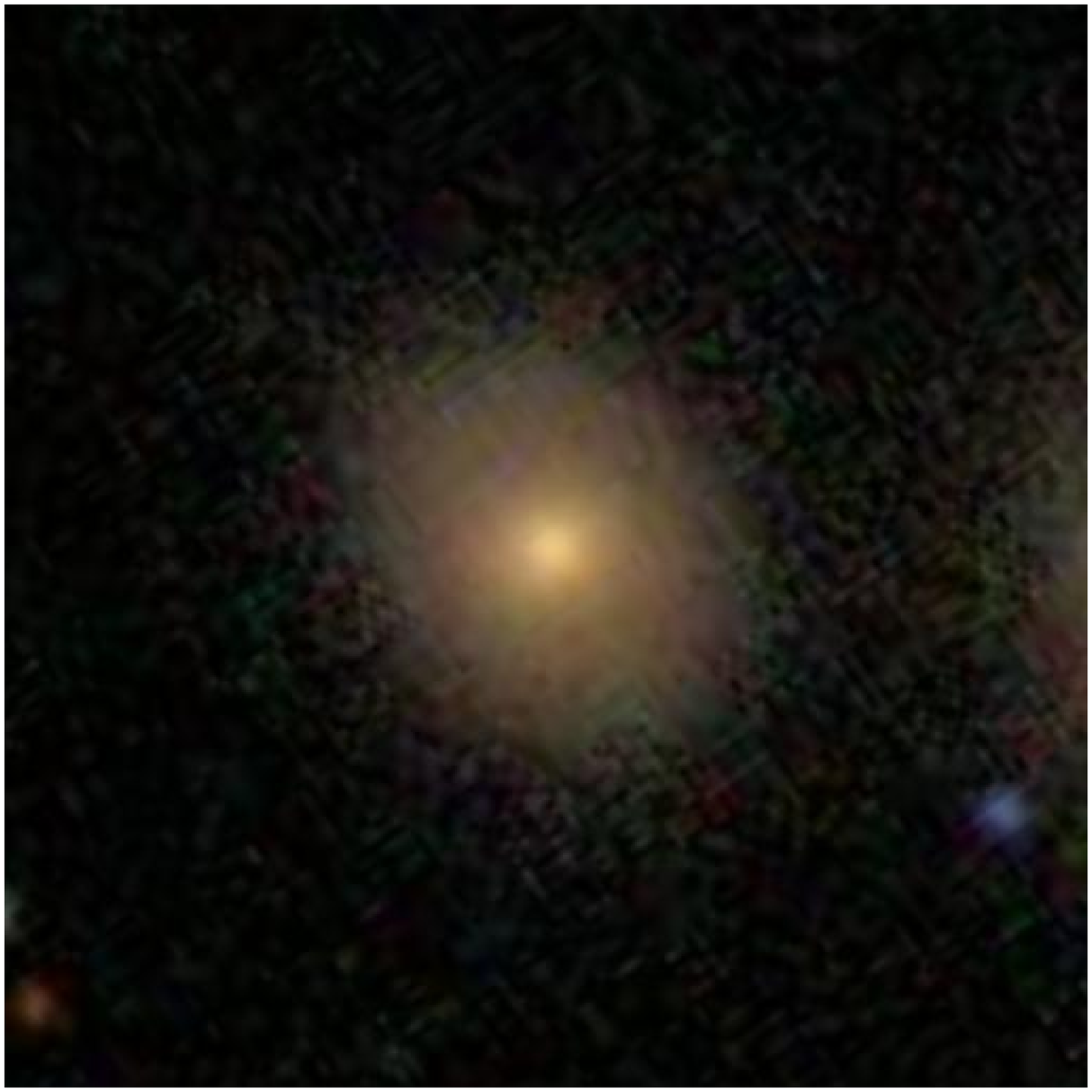}
\includegraphics[scale=0.175]{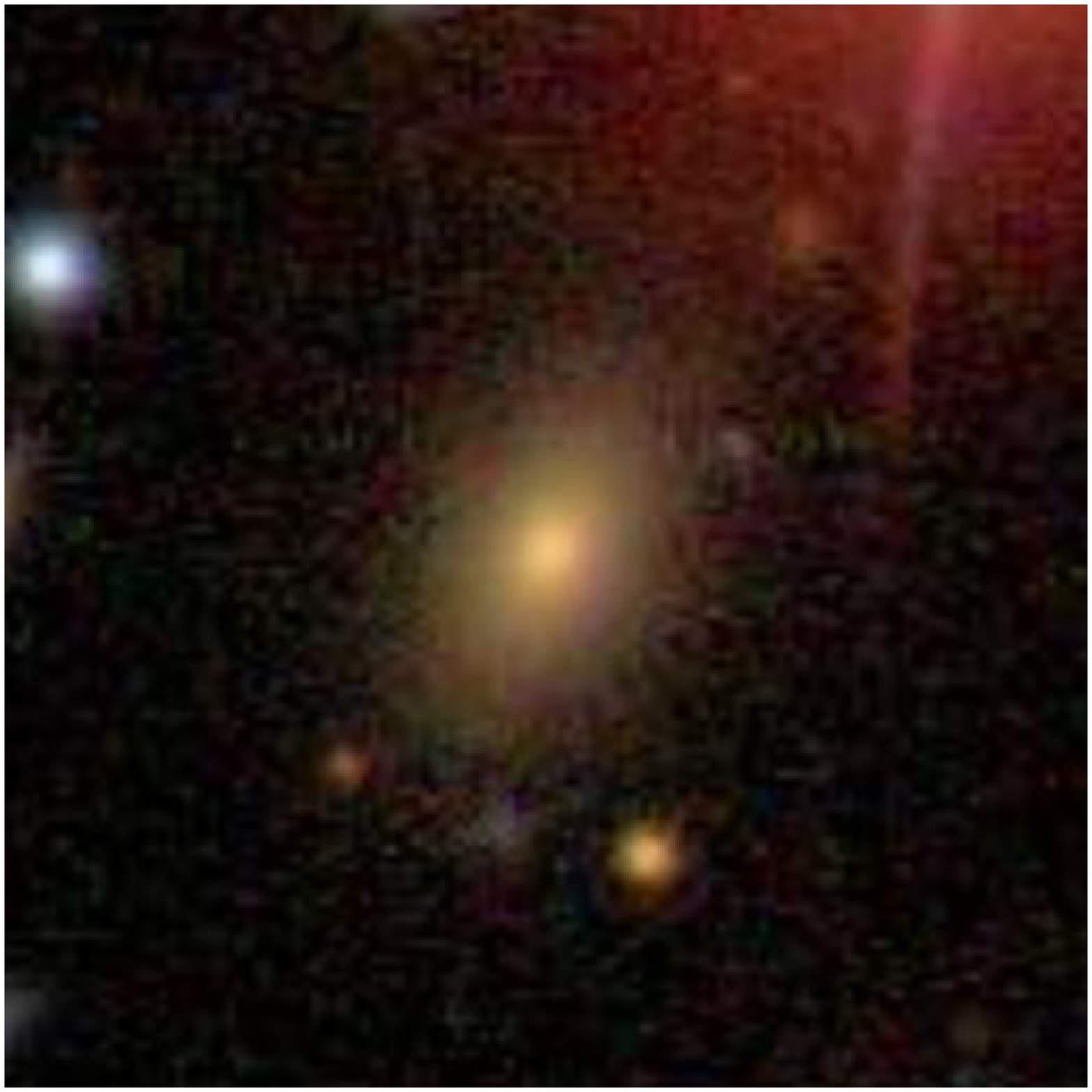}
\includegraphics[scale=0.175]{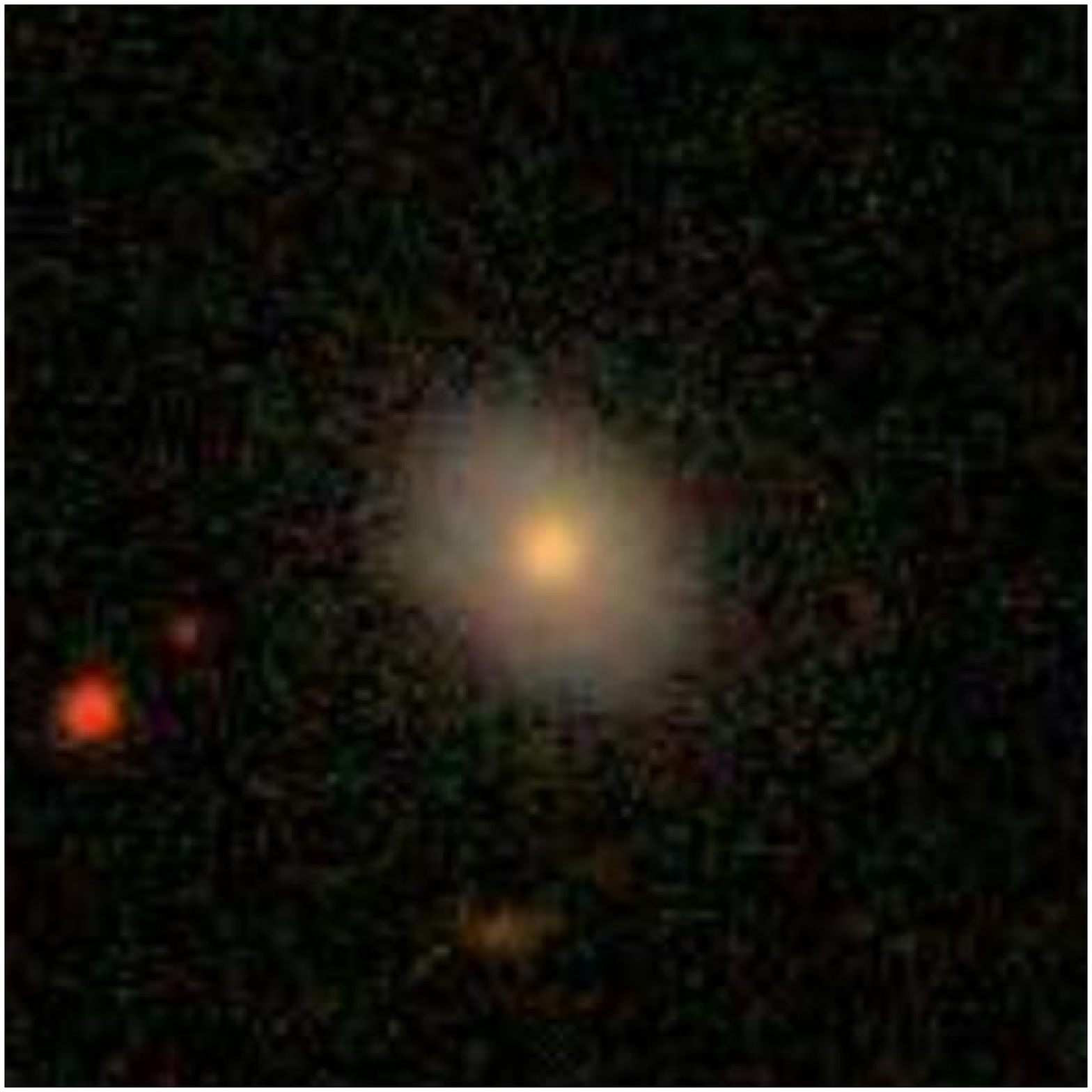}}

\caption{$50\arcsec \times 50\arcsec$ SDSS images of the 5 putative late type
  galaxies, with $C_{r}<2.6$, all but the one on the left clearly ETG.}
\label{spirals}
\end{figure*}

\section{Summary and conclusions}
\label{summary}

In a previous paper \citep{baldi09} we showed that the sources selected by
\citet{best05a} by cross-correlating the NVSS, FIRST, and SDSS surveys (the
SDSS/NVSS sample) do not follow the correlations between line and radio
luminosity defined by radio-galaxies part of samples with a higher flux
threshold, such as the 3CR, the B2, and the 2 Jy. They are characterized by a
large deficit of radio emission at a given level of emission line luminosity
with respect to these more powerful radio-sources. We argued that the
prevalence of sources with luminous extended radio structures in high flux
limited samples is due to a selection bias. The SDSS/NVSS sources form the
bulk of the local radio-loud AGN population but they are still poorly
explored.

We here analyze their spectro-photometric properties looking for differences
in their hosts and in their optical spectra that might explain their different
radio properties. The majority of the SDSS/NVSS sources have emission line
ratios characteristic of Low Excitation Galaxies. A comparison with the LEG
part of the 3CR sample show that the SDSS/NVSS hosts are indistinguishable
from their counterparts of higher radio power from the point of view of their
morphology, distribution of black hole and stellar masses, and broad band
colors. In fact the hosts of both samples are massive, red, early-type
galaxies, with black hole masses in the range $7.7 \lesssim {\rm log}
(M_{\rm{BH}}/M_{\odot}) \lesssim 9.5$. A small percentage ($\sim 5\%$) of
objects with signs of recent star formation is also present in both samples,
in a similar fraction with respect to the population of quiescent (from the
point of view of nuclear activity) galaxies. Thus the differences in radio
power between SDSS/NVSS and 3CR sources (matched at a given level of emission
line luminosity) cannot be ascribed to the fact that their hosts are
different.

Environment can also play a role. In fact, in a dense environment, the
confinement of the radio emitting plasma produced by the hot intra-group or
intra-cluster medium is more effective than in regions of lower galaxies
densities. In this latter case, the rapid expansion of the relativistic plasma
causes dramatic adiabatic losses, lowering the observed radio power. However,
\citet{best05a} show that SDSS/NVSS are already in general found in richer
environments than normal galaxies. A detailed comparison of the environment of
SDSS/NVSS and 3CR sources is needed to address this issue and it will be
performed in a forthcoming paper.

In \citet{baldi09} we argued that SDSS/NVSS sources could be relatively young
and their extended radio emission has yet to fully form. This requires that
the radio luminosity increases with time as the source expands, slowly
burrowing their way into the interstellar medium of the host galaxy. The
results presented here provide us with a further piece of evidence in favor of
a evolutionary sequence in radio-loud AGN. In fact we found that Unclassified
Galaxies are more often associated with extended radio sources with respect to
Low Excitation Galaxies (the percentages being 27 \% and 5 \% for UG and LEG,
respectively). This is counter-intuitive, since LEG are brighter in line
emission than UG; one might have expected a higher line luminosity, generally
indicative of a higher AGN power, to correspond more often to a well developed
radio structure. However, also the line luminosity evolves with time and with
the size of the radio source. The compression of the ambient medium due to the
passage of the jet head (and to the lateral expansion shocks) increases the
density of the line emitting clouds and consequently (given the dependence of
line luminosity on the density squared) their line luminosity. The line
luminosity increases with time, as the jets compress more clouds (see
e.g. \citealt{labiano08}) until the radio source is confined to within the
host galaxy. Even larger sources propagate in regions of lower gas
densities. At a given level of the ionizing continuum, they are expected to be
less luminous in line that smaller sources where the jet-cloud interactions
are more efficient. Thus a radio-galaxy moves with time in the $L_{1.4~{\rm
GHz}}-L_{\rm [O~III]}$ plane along a complex evolutionary track. The higher
fraction of UG with extended radio structures can be due to the evolution of
part of the LEG population, from small radio sources of high
$L_{\rm{[O~III]}}$, to larger size and smaller $L_{\rm{[O~III]}}$.

Additional indicators of the jet and AGN power are needed to further compare
high and low radio power sources. These can be obtained, for example, from
measurements of the radio cores (from images of higher resolution and at
higher frequency than the FIRST maps currently available) or of the higher
frequency nuclear emission by using, e.g., HST or Chandra imaging. In
particular it is crucial to test whether also the sources belonging to the
SDSS/NVSS sample obey to the relation between radio core and line emission
spanning from miniature to high power radio-galaxies \citep{baldi09}. This
would represent a clear indication that there is indeed a close connection
between the radiative and kinetic output in radio-loud AGN, but that the best
indicator of jet power is the radio core luminosity and not the total radio
emission. This interpretation is in line with the proportionality of core and
jet power derived from the estimates of the energy required to form the
cavities observed in X-ray images \citep{heinz07}. Better radio images will
also provide us with a measurement of the core dominance, to test the
prediction that the deficit of radio power of SDSS/NVSS sources is indeed
related to a higher core dominance with respect to, e.g., 3CR objects, in
analogy with the results we obtained for miniature radio-galaxies.

After a detailed analysis on a object by object basis, we concluded that no
bona-fide radio-loud AGN with low excitation spectra could be found in a
  late type host or with black holes with a mass substantially lower than
  $10^8 M_{\odot}$. The SMBH mass distributions of 3CR/LEG and in the
  SDSS/NVSS samples suggests that a RLAGN LEG can only be associated with
relatively large mass of the central object. This might represent a very
valuable information in our quest for the origin and evolution of RLAGN (see
e.g. \citealt{chiaberge08}) and of their host galaxies. Nonetheless,
  since a link between $M_{\rm{BH}}$ and the normalization of the radio
  luminosity function exists, RLAGN might be associated with low SMBH masses
  and not be represented in the sample because they are too rare. However, we
  estimated that 7 low $M_{\rm{BH}}$ objects would be expected, while none was
  found. This idea can be tested in greater depth by exploring the radio
  properties of lower mass ETG, harboring black holes around $10^7
  M_{\odot}$.

A minority ($\sim 10\%$) of the SDSS/NVSS sample shows rather different
properties, showing a spectrum typical of high excitation or star forming
galaxies. They strikingly differ from LEG: they have a wider distribution of
SMBH masses, ranging from $\sim 10^6 M_{\odot}$ to $\sim 10^9 M_{\odot}$ and
they are often found in spiral galaxies. Only a few of the HEG in the
SDSS/NVSS sample (less than $\sim$ 10 objects) have spectro-photometric
properties similar to the HEG found in the 3CR. Furthermore, there might be a
contamination from radio-quiet galaxies with high excitation spectra whose
radio emission is dominated by star formation.

Interestingly, however, in the SDSS/NVSS there is no clear discontinuity in
either the host or the AGN properties moving from galaxies lying in the region
characteristic of Seyfert to that covered by 3CR/HEG, objects that would be
classically defined as radio-quiet and radio-loud AGN. This issue certainly
deserves further studies, requiring to collect additional multiband data on
these galaxies.

It is possible that the same situation occurs also for LEG. The density of
SDSS/NVSS sources, at given line luminosity, increases toward lower radio
luminosity (with a dependence well described by a power law $N \sim
L_{1.4~{\rm GHz}}^{-1}$) down to the completeness limit imposed by the radio
flux limit. We cannot establish whether a discontinuity exists between the
RLAGN discussed here and the population of radio-quiet LEG, i.e. the LINER.
An extension of this study toward even lower radio flux limits, probably at
sub-mJy level, is required to evidence the presence of a possible
radio-quiet/radio-loud dichotomy in low excitation AGN.

Furthermore, if indeed galaxies similar from the point of view of their
nuclear properties give rise to radio-sources of widely different power and
morphology, the use of the total radio power as calorimeter of the jet is
questioned. This opens also the issue of how to define the radio-loudness of
AGN. Apparently, RL and RQ AGN are best separated considering their nuclear
properties, in particular by comparing the luminosity in the radio with X-ray
and optical data (e.g. \citealt{early1}), stressing again the importance of
high resolution multi-band measurements of their nuclear emission.

Finally, we note that an important issue in the study of these radio sources
of, generally, low power, is related to the identification of the dominant
process of the radio emission, i.e. AGN or star formation. \citet{best05a}
adopted a separation based on the radio luminosity expected for a stellar
population of a given age (derived from the value of the $D_n(4000)$ index)
and mass. While the proposed method is sufficiently accurate to build, e.g.,
radio luminosity functions, it cannot be safely applied to individual
galaxies. In fact, we found that many objects located just above the threshold
in the $D_{n}(4000)$ vs. $L_{\rm 1.4GHz}/M_{*}$ plane have spectro-photometric
characteristics very different from the bulk of the sample. This suggests that
the ``grey region'' where the two populations (with radio powered by an AGN or
by star formation) overlap covers a rather large range of these
parameters. This is likely due to the different scales probed by NVSS images
and SDSS spectra. For example, we found a spiral galaxy whose SDSS spectrum is
indicative of an old stellar population. However, the SDSS fiber covers only
its bulge and, consequently, the predicted radio emission from star formation
is relatively low, suggesting an AGN origin. However, no radio core is seen in
the FIRST images, suggesting that the radio emission might originate from
young stars in the spiral arms, not covered by the nuclear spectrum. Deeper
radio images and with better resolution than the FIRST and NVSS maps could be
used to isolate these objects based on their radio morphology and on the
spatial association of radio and optical structures.

\begin{acknowledgements}
  The authors acknowledge partial financial support from ASI grant I/023/050.
  This research has made use of the NASA/IPAC Extragalactic Database (NED)
  (which is operated by the Jet Propulsion Laboratory, California Institute of
  Technology, under contract with the National Aeronautics and Space
  Administration), of the LEDA database, and of the NASA's Astrophysics Data
  System (ADS).  The research makes use of the SDSS Archive, funding for the
  creation and distribution of which was provided by the Alfred P. Sloan
  Foundation, the Participating Institutions, the National Aeronautics and
  Space Administration, the National Science Foundation, the U.S. Department
  of Energy, the Japanese Monbukagakusho, and the Max Planck Society, and The
  Higher Education Funding Council for England. The research uses the NVSS and
  FIRST radio surveys, carried out using the NRAO Very Large Array: NRAO is
  operated by Associated Universities Inc., under co-operative agreement with
  the National Science Foundation.
\end{acknowledgements}

\appendix
\section{Appendix}

We report in this Appendix the analogous of Fig. \ref{diagn} (the
spectroscopic diagnostic diagrams), Fig. \ref{fcfha} (left panel) (\Ha\ flux
vs. the continuum at the \Ha\ line), Fig. \ref{dn4000} (concentration index
vs. 4000 \AA\ break), Fig. \ref{SFGselection} (D$_{n}$(4000) vs. L$_{\rm {1.4
GHz}}$/M$_{*}$), Fig. \ref{histombh} (distributions of the SMBH masses for
each spectroscopic class), and Fig. \ref{lextlo3} (total radio power
vs. [O~III] luminosities) for the objects with redshift $0.1<z<0.3$.

\begin{figure*}
\centerline{
\includegraphics[scale=0.75,angle=90]{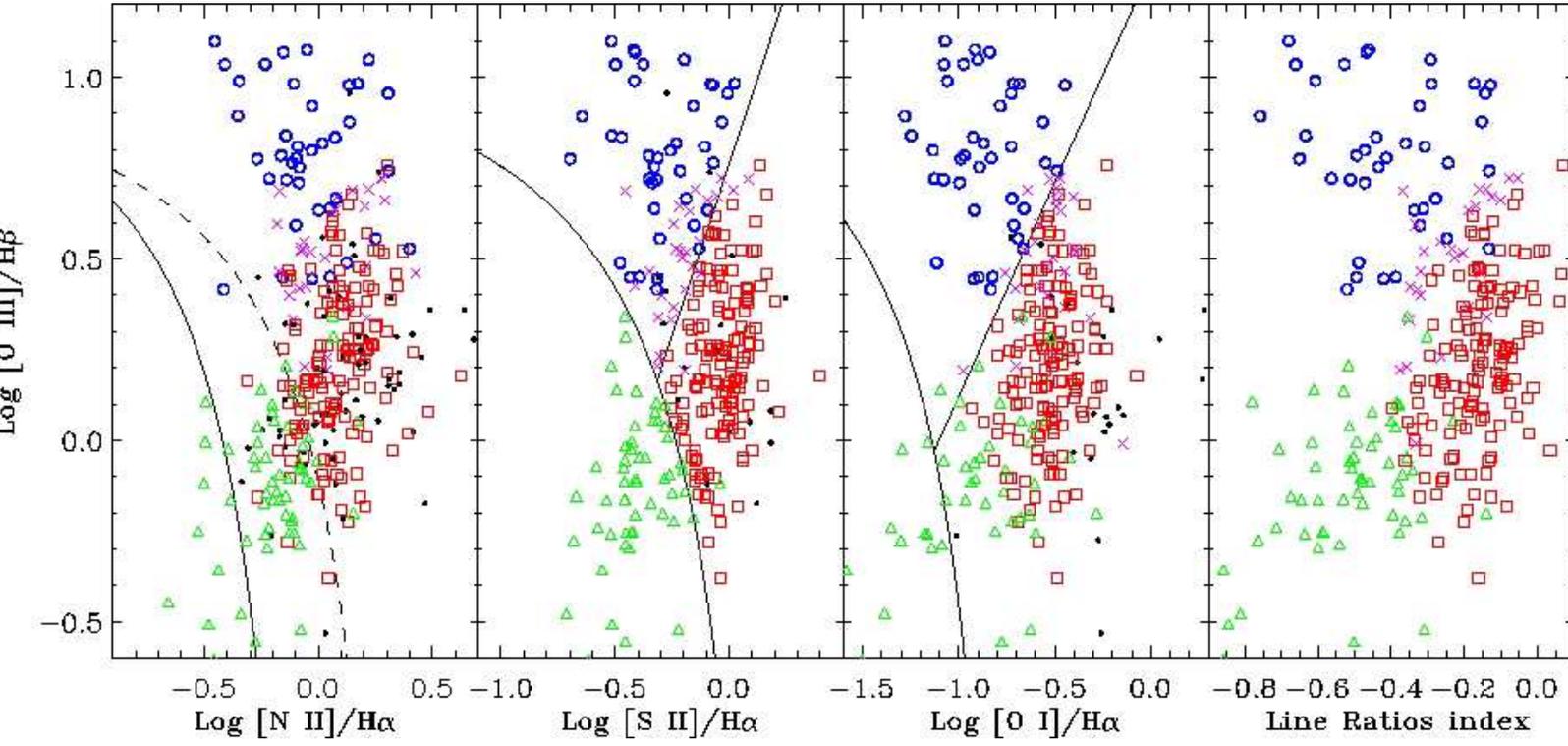}
}
\caption{Spectroscopic diagnostic diagrams for the objects with
  $0.1<z<0.3$, with the same structure and symbols of Fig. \ref{diagn}.}
\label{diagnhz}
\end{figure*}

\begin{figure*}
\centerline{
\includegraphics[scale=0.45,angle=90]{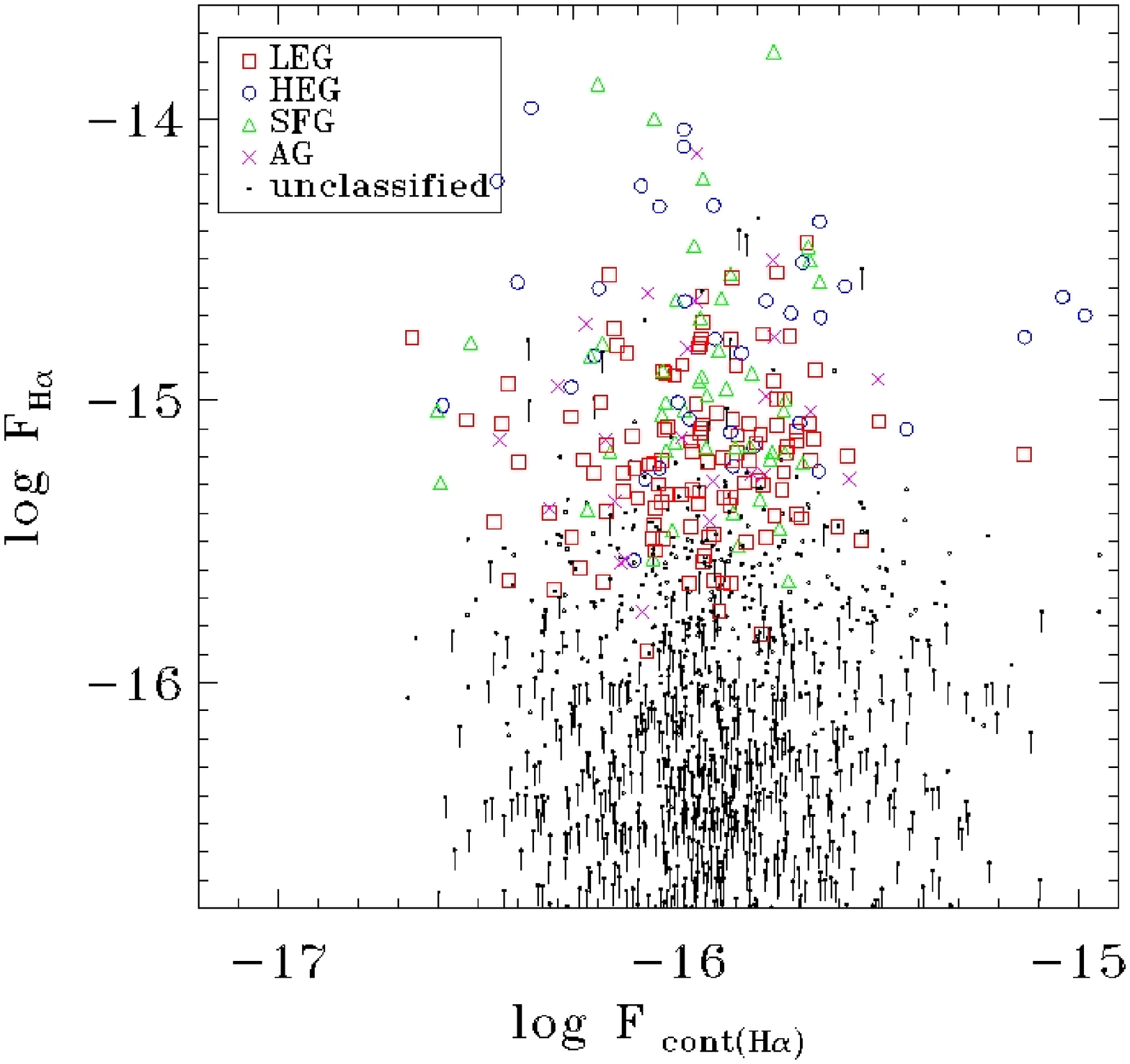}
\includegraphics[scale=0.45,angle=90]{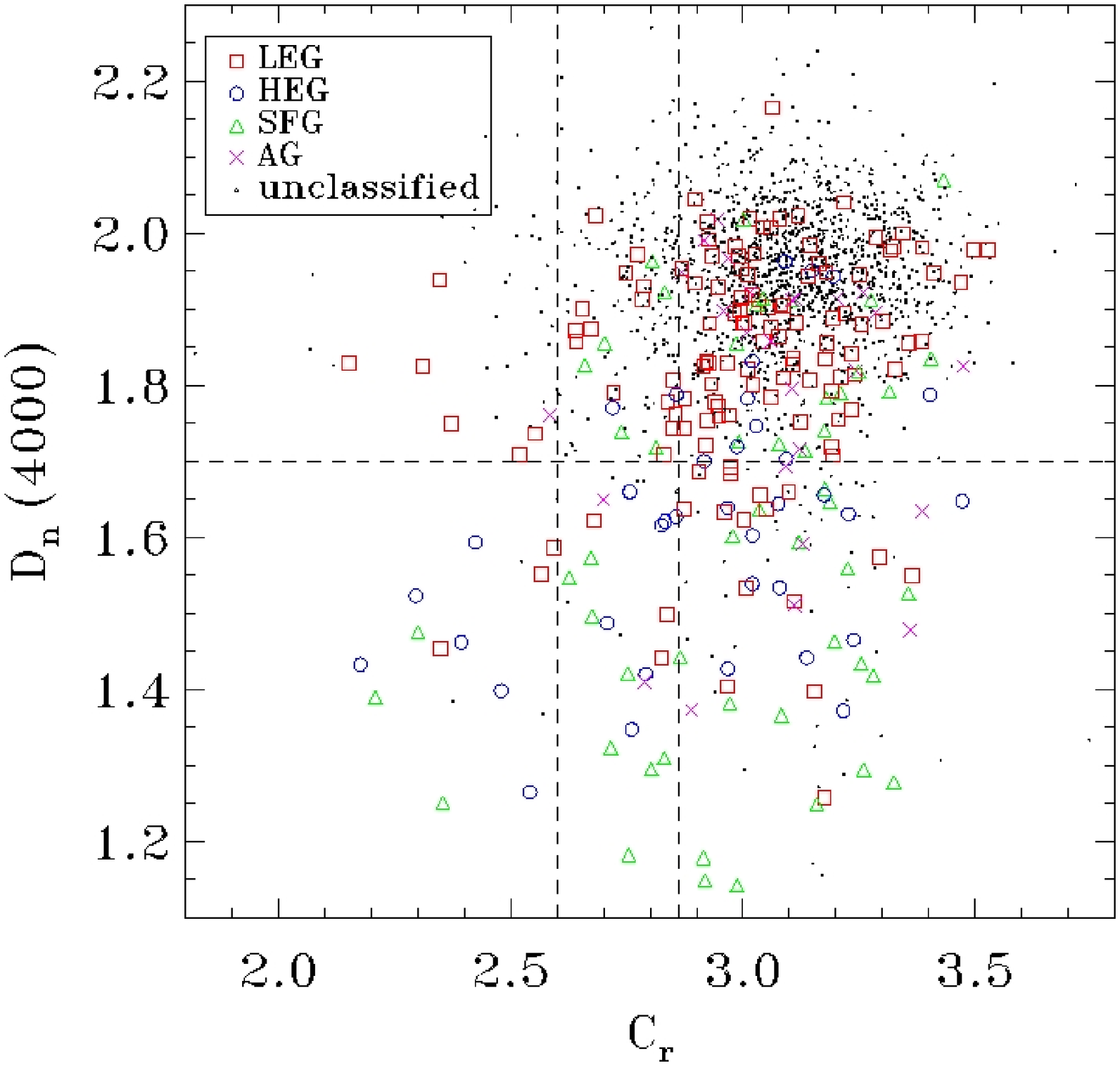}}
\centerline{
\includegraphics[scale=0.45,angle=90]{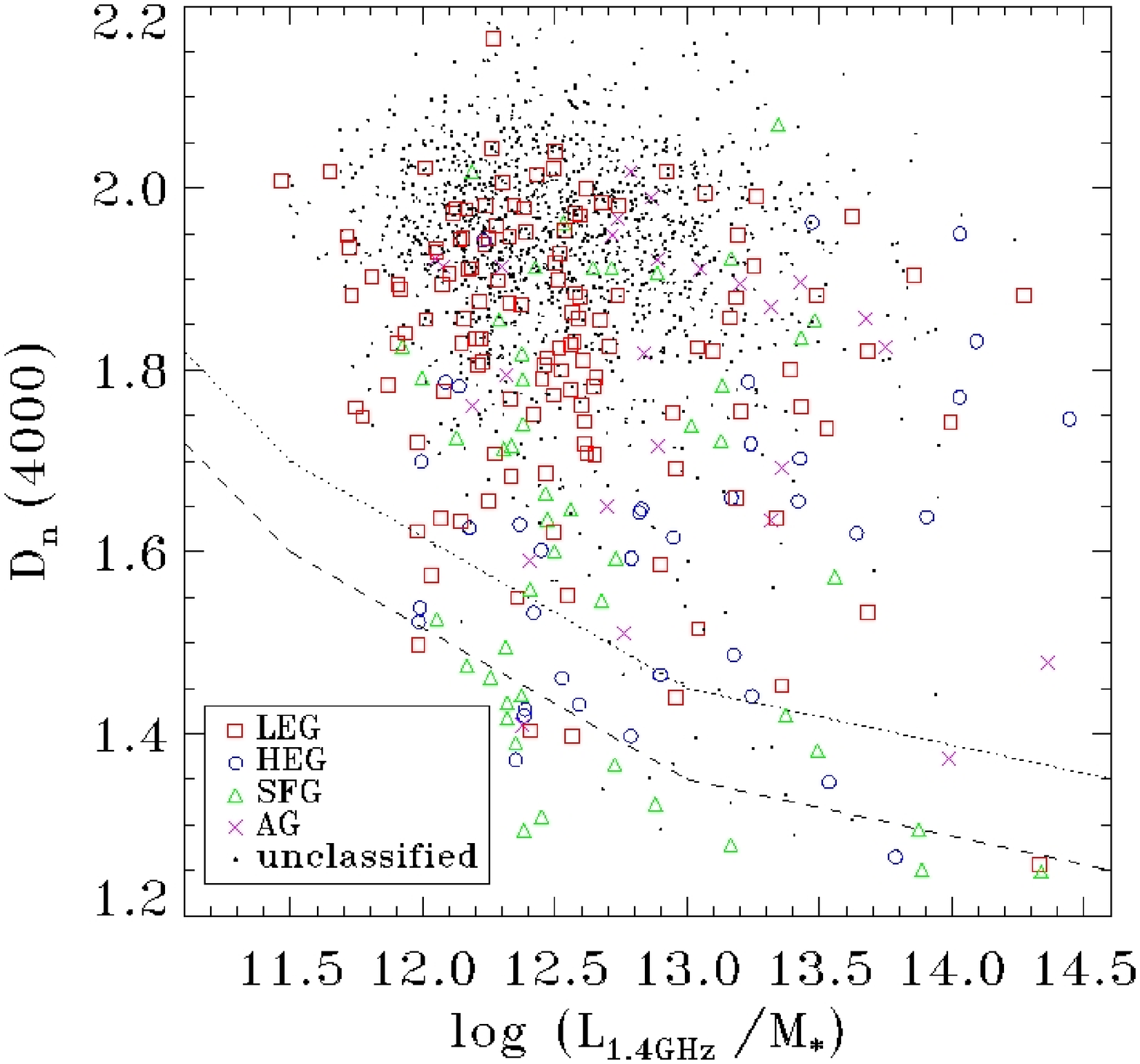}
\includegraphics[scale=0.45,angle=90]{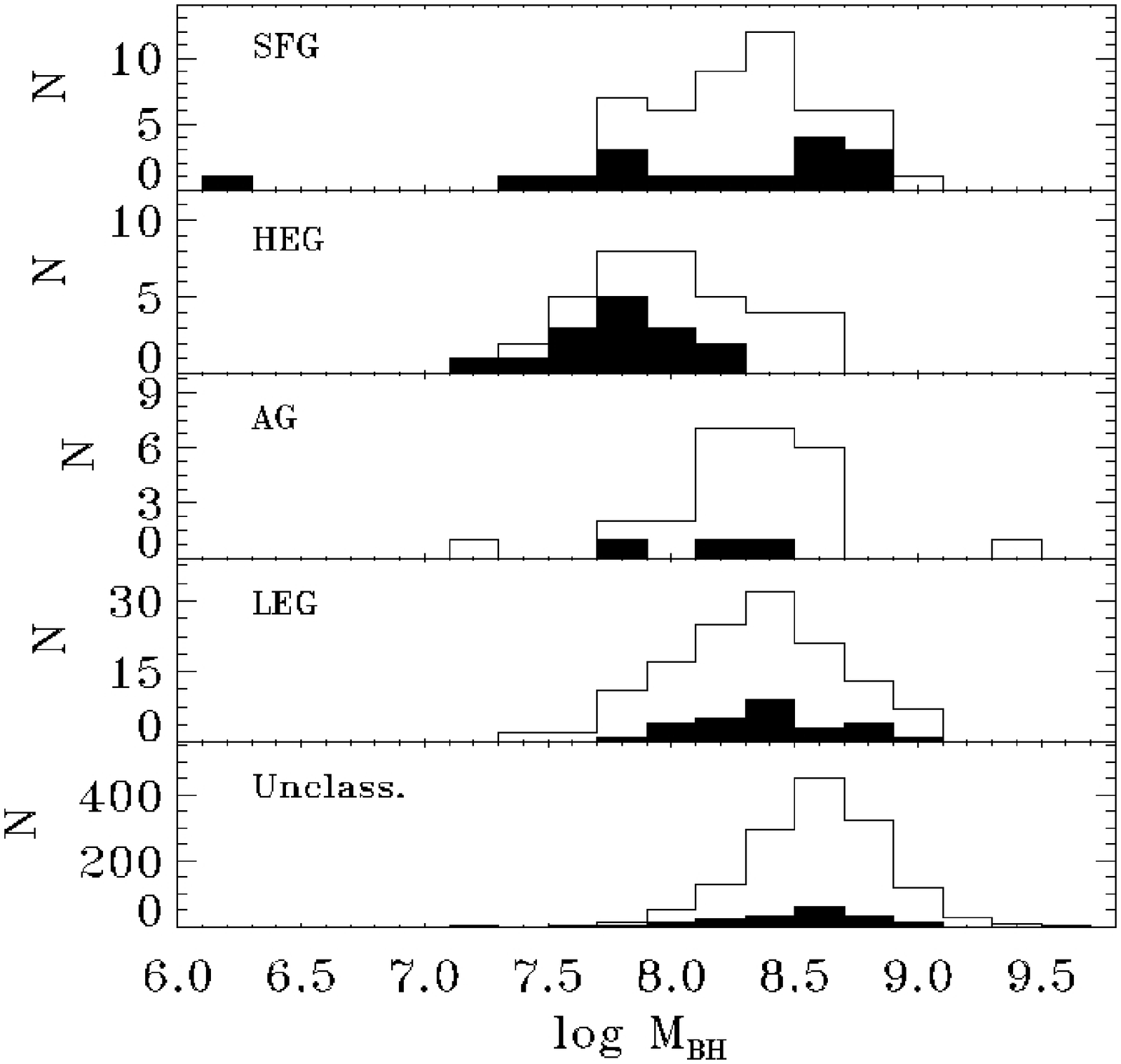}}
\caption{Analogous of Fig.\ref{fcfha} (left panel), Fig. \ref{dn4000},
  Fig. \ref{SFGselection}, and Fig. \ref{histombh} for the higher redshift
  objects ($0.1<z<0.3$) i.e. : (upper left panel) logarithm of the \Ha\ flux
  versus logarithm of the continuum at the \Ha\ line (in erg s$^{-1}$ and erg
  s$^{-1}$\AA$^{-1}$ units respectively); (upper right panel) concentration
  index $C_{r}$ vs. 4000 \AA\ break; (lower left panel) $D_{n}(4000)$
  vs. $L_{1.4 {\rm GHz}}/M_{*}$; (lower right panel) distributions of the
  $M_{\rm BH}$ (in solar units) for the various spectroscopic classes. The
  filled portion of the histograms represent the contribution of galaxies with
  $C_r<2.86$.}
\label{allhz1}
\end{figure*}

\begin{figure*}
\centerline{
\includegraphics[scale=0.70,angle=90]{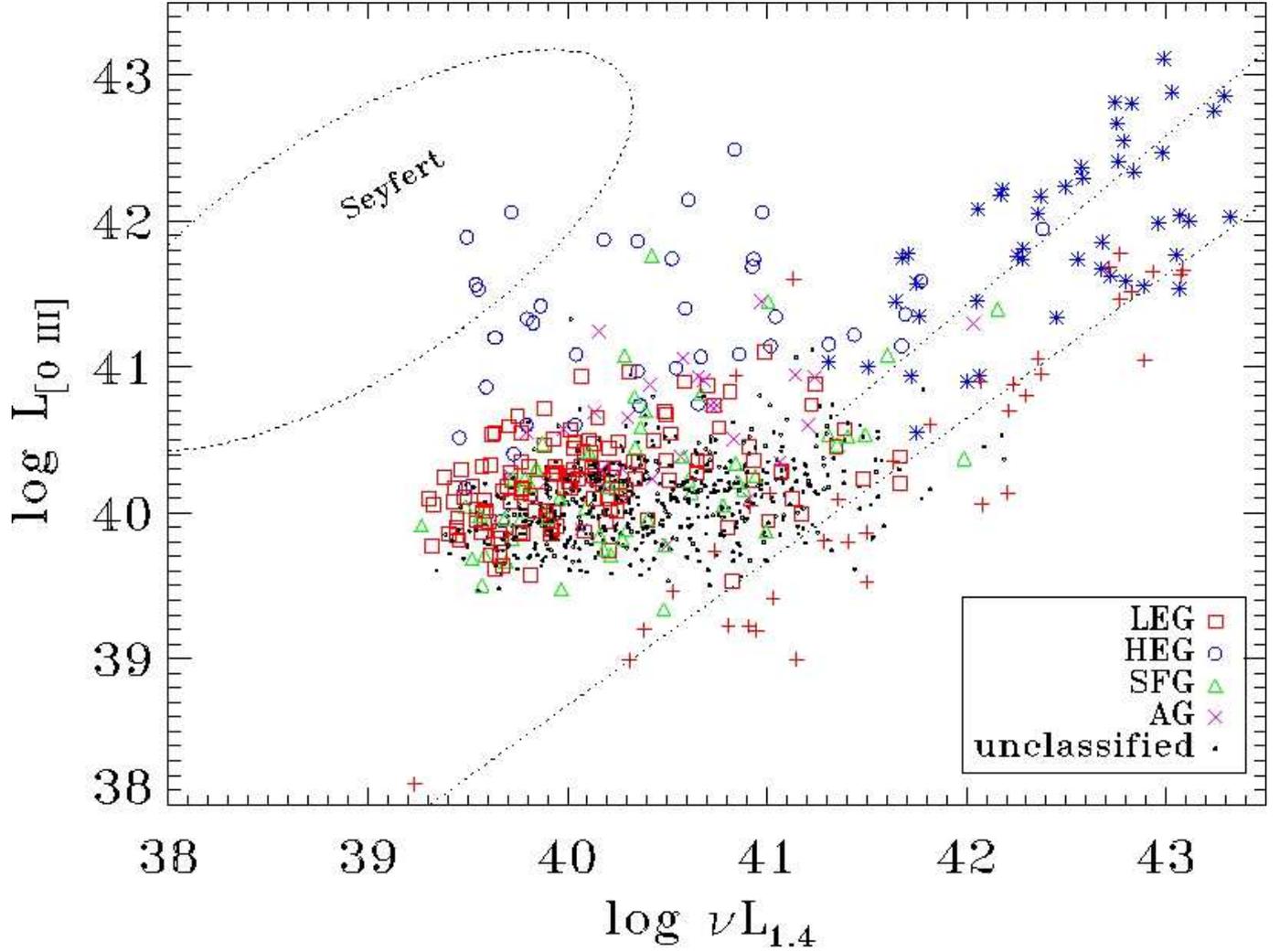}}
\caption{Analogous of Fig. \ref{lextlo3} for the higher redshift objects
  ($0.1<z<0.3$) i.e. logarithm of radio power $L_{1.4 {\rm GHz}}$ vs. [O~III]
  luminosities. For more clarity the large number of upper limits in $L_{\rm
    [O~III]}$ are not shown. The two dashed lines reproduce the line-radio
  correlation followed by the LEG and HEG of the 3CR sample (red pluses and
  blue asterisks respectively). The ellipse marks the boundaries of the
  location of Seyfert galaxies (e.g. \citealt{whittle85}).}
\label{allhz3}
\end{figure*}


\begin{thebibliography}{50}
\expandafter\ifx\csname natexlab\endcsname\relax\def\natexlab#1{#1}\fi

\bibitem[{{Baldi} {et~al.}(2010){Baldi}, {Chiaberge}, {Capetti}, {Macchetto},
  \& {Sparks}}]{baldi10}
{Baldi}, R., {Chiaberge}, M., {Capetti}, A., {Macchetto}, F., \& {Sparks}, W.
  2010, ApJ, submitted

\bibitem[{{Baldi} \& {Capetti}(2008)}]{baldi08}
{Baldi}, R.~D. \& {Capetti}, A. 2008, \aap, 489, 989

\bibitem[{{Baldi} \& {Capetti}(2009)}]{baldi09}
{Baldi}, R.~D. \& {Capetti}, A. 2009, ArXiv e-prints

\bibitem[{{Baldwin} {et~al.}(1981){Baldwin}, {Phillips}, \&
  {Terlevich}}]{baldwin81}
{Baldwin}, J.~A., {Phillips}, M.~M., \& {Terlevich}, R. 1981, \pasp, 93, 5

\bibitem[{{Balmaverde} \& {Capetti}(2006)}]{balmaverde06b}
{Balmaverde}, B. \& {Capetti}, A. 2006, \aap, 447, 97

\bibitem[{{Balogh} {et~al.}(1999){Balogh}, {Morris}, {Yee}, {Carlberg}, \&
  {Ellingson}}]{balogh99}
{Balogh}, M.~L., {Morris}, S.~L., {Yee}, H.~K.~C., {Carlberg}, R.~G., \&
  {Ellingson}, E. 1999, \apj, 527, 54

\bibitem[{{Baum} \& {Heckman}(1989)}]{baum89b}
{Baum}, S.~A. \& {Heckman}, T. 1989, \apj, 336, 702

\bibitem[{{Becker} {et~al.}(1995){Becker}, {White}, \& {Helfand}}]{becker95}
{Becker}, R.~H., {White}, R.~L., \& {Helfand}, D.~J. 1995, \apj, 450, 559

\bibitem[{{Bell} {et~al.}(2003){Bell}, {McIntosh}, {Katz}, \&
  {Weinberg}}]{bell03}
{Bell}, E.~F., {McIntosh}, D.~H., {Katz}, N., \& {Weinberg}, M.~D. 2003, \apjs,
  149, 289

\bibitem[{{Bell} {et~al.}(2007){Bell}, {Zheng}, {Papovich}, {Borch}, {Wolf}, \&
  {Meisenheimer}}]{bell07}
{Bell}, E.~F., {Zheng}, X.~Z., {Papovich}, C., {et~al.} 2007, \apj, 663, 834

\bibitem[{{Bernardi} {et~al.}(2009){Bernardi}, {Shankar}, {Hyde}, {Mei},
  {Marulli}, \& {Sheth}}]{bernardi09}
{Bernardi}, M., {Shankar}, F., {Hyde}, J.~B., {et~al.} 2009, ArXiv e-prints

\bibitem[{{Bernardi} {et~al.}(2003){Bernardi}, {Sheth}, {Annis}, {Burles},
  {Eisenstein}, {Finkbeiner}, {Hogg}, {Lupton}, {Schlegel}, {SubbaRao},
  {Bahcall}, {Blakeslee}, {Brinkmann}, {Castander}, {Connolly}, {Csabai},
  {Doi}, {Fukugita}, {Frieman}, {Heckman}, {Hennessy}, {Ivezi{\'c}}, {Knapp},
  {Lamb}, {McKay}, {Munn}, {Nichol}, {Okamura}, {Schneider}, {Thakar}, \&
  {York}}]{bernardi03}
{Bernardi}, M., {Sheth}, R.~K., {Annis}, J., {et~al.} 2003, \aj, 125, 1849

\bibitem[{{Best} {et~al.}(2005{\natexlab{a}}){Best}, {Kauffmann}, {Heckman},
  {Brinchmann}, {Charlot}, {Ivezi{\'c}}, \& {White}}]{best05b}
{Best}, P.~N., {Kauffmann}, G., {Heckman}, T.~M., {et~al.} 2005{\natexlab{a}},
  \mnras, 362, 25

\bibitem[{{Best} {et~al.}(2005{\natexlab{b}}){Best}, {Kauffmann}, {Heckman}, \&
  {Ivezi{\'c}}}]{best05a}
{Best}, P.~N., {Kauffmann}, G., {Heckman}, T.~M., \& {Ivezi{\'c}}, {\v Z}.
  2005{\natexlab{b}}, \mnras, 362, 9

\bibitem[{{Buttiglione} {et~al.}(2009){Buttiglione}, {Capetti}, {Celotti},
  {Axon}, {Chiaberge}, {Macchetto}, \& {Sparks}}]{buttiglione09}
{Buttiglione}, S., {Capetti}, A., {Celotti}, A., {et~al.} 2009, \aap, 495, 1033

\bibitem[{{Buttiglione} {et~al.}(2010){Buttiglione}, {Capetti}, {Celotti},
  {Axon}, {Chiaberge}, {Macchetto}, \& {Sparks}}]{buttiglione10}
{Buttiglione}, S., {Capetti}, A., {Celotti}, A., {et~al.} 2010, \aap, 509, A6+

\bibitem[{{Capetti} \& {Balmaverde}(2005)}]{early1}
{Capetti}, A. \& {Balmaverde}, B. 2005, \aap, 440, 73

\bibitem[{{Chiaberge}(2008)}]{chiaberge08}
{Chiaberge}, M. 2008, in COSPAR, Plenary Meeting, Vol.~37, 37th COSPAR
  Scientific Assembly, 532--+

\bibitem[{{Cid Fernandes} {et~al.}(2009){Cid Fernandes}, {Stasinska}, {Vale
  Asari}, {Mateus}, {Schlickmann}, \& {Schoenell}}]{cid09}
{Cid Fernandes}, R., {Stasinska}, G., {Vale Asari}, N., {et~al.} 2009, ArXiv
  e-prints

\bibitem[{{Condon} {et~al.}(1998){Condon}, {Cotton}, {Greisen}, {Yin},
  {Perley}, {Taylor}, \& {Broderick}}]{condon98}
{Condon}, J.~J., {Cotton}, W.~D., {Greisen}, E.~W., {et~al.} 1998, \aj, 115,
  1693

\bibitem[{{Heckman}(1980)}]{heckman80}
{Heckman}, T.~M. 1980, \aap, 87, 152

\bibitem[{{Heinz} {et~al.}(2007){Heinz}, {Merloni}, \& {Schwab}}]{heinz07}
{Heinz}, S., {Merloni}, A., \& {Schwab}, J. 2007, \apjl, 658, L9

\bibitem[{{Hine} \& {Longair}(1979)}]{hine79}
{Hine}, R.~G. \& {Longair}, M.~S. 1979, \mnras, 188, 111

\bibitem[{{Hyde} \& {Bernardi}(2009)}]{hyde09}
{Hyde}, J.~B. \& {Bernardi}, M. 2009, \mnras, 394, 1978

\bibitem[{{Jackson} \& {Rawlings}(1997)}]{jackson97}
{Jackson}, N. \& {Rawlings}, S. 1997, \mnras, 286, 241

\bibitem[{{Kauffmann} {et~al.}(2003){Kauffmann}, {Heckman}, {White}, {Charlot},
  {Tremonti}, {Brinchmann}, {Bruzual}, {Peng}, {Seibert}, {Bernardi},
  {Blanton}, {Brinkmann}, {Castander}, {Cs{\'a}bai}, {Fukugita}, {Ivezic},
  {Munn}, {Nichol}, {Padmanabhan}, {Thakar}, {Weinberg}, \&
  {York}}]{kauffmann03b}
{Kauffmann}, G., {Heckman}, T.~M., {White}, S.~D.~M., {et~al.} 2003, \mnras,
  341, 33

\bibitem[{{Kewley} {et~al.}(2006){Kewley}, {Groves}, {Kauffmann}, \&
  {Heckman}}]{kewley06}
{Kewley}, L.~J., {Groves}, B., {Kauffmann}, G., \& {Heckman}, T. 2006, \mnras,
  372, 961

\bibitem[{{Labiano}(2008)}]{labiano08}
{Labiano}, A. 2008, \aap, 488, L59

\bibitem[{{Laing} {et~al.}(1994){Laing}, {Jenkins}, {Wall}, \&
  {Unger}}]{laing94}
{Laing}, R.~A., {Jenkins}, C.~R., {Wall}, J.~V., \& {Unger}, S.~W. 1994, in The
  First Stromlo Symposium: The Physics of Active Galaxies. ASP Conference
  Series, Vol. 54, 1994, G.V. Bicknell, M.A. Dopita, and P.J. Quinn, Eds.,
  p.201, 201--+

\bibitem[{{Lintott} {et~al.}(2008){Lintott}, {Schawinski}, {Slosar}, {Land},
  {Bamford}, {Thomas}, {Raddick}, {Nichol}, {Szalay}, {Andreescu}, {Murray}, \&
  {Vandenberg}}]{lintott08}
{Lintott}, C.~J., {Schawinski}, K., {Slosar}, A., {et~al.} 2008, \mnras, 389,
  1179

\bibitem[{{Madrid} {et~al.}(2006){Madrid}, {Chiaberge}, {Floyd}, {Sparks},
  {Macchetto}, {Miley}, {Axon}, {Capetti}, {O'Dea}, {Baum}, {Perlman}, \&
  {Quillen}}]{madrid06}
{Madrid}, J.~P., {Chiaberge}, M., {Floyd}, D., {et~al.} 2006, \apjs, 164, 307

\bibitem[{{Mannucci} {et~al.}(2001){Mannucci}, {Basile}, {Poggianti},
  {Cimatti}, {Daddi}, {Pozzetti}, \& {Vanzi}}]{mannucci01}
{Mannucci}, F., {Basile}, F., {Poggianti}, B.~M., {et~al.} 2001, \mnras, 326,
  745

\bibitem[{{Marconi} \& {Hunt}(2003)}]{marconi03}
{Marconi}, A. \& {Hunt}, L.~K. 2003, \apjl, 589, L21

\bibitem[{{Martel} {et~al.}(1999){Martel}, {Baum}, {Sparks}, {Wyckoff},
  {Biretta}, {Golombek}, {Macchetto}, {de Koff}, {McCarthy}, \&
  {Miley}}]{martel99}
{Martel}, A.~R., {Baum}, S.~A., {Sparks}, W.~B., {et~al.} 1999, \apjs, 122, 81

\bibitem[{{Martin} {et~al.}(1976){Martin}, {Angel}, \& {Maza}}]{martin76}
{Martin}, P.~G., {Angel}, J.~R.~P., \& {Maza}, J. 1976, \apjl, 209, L21+

\bibitem[{{Morganti} {et~al.}(1997){Morganti}, {Tadhunter}, {Dickson}, \&
  {Shaw}}]{morganti97}
{Morganti}, R., {Tadhunter}, C.~N., {Dickson}, R., \& {Shaw}, M. 1997, \aap,
  326, 130

\bibitem[{{Nakamura} {et~al.}(2003){Nakamura}, {Fukugita}, {Yasuda}, {Loveday},
  {Brinkmann}, {Schneider}, {Shimasaku}, \& {SubbaRao}}]{nakamura03}
{Nakamura}, O., {Fukugita}, M., {Yasuda}, N., {et~al.} 2003, \aj, 125, 1682

\bibitem[{{O'Dea} {et~al.}(2001){O'Dea}, {Koekemoer}, {Baum}, {Sparks},
  {Martel}, {Allen}, {Macchetto}, \& {Miley}}]{odea:3c236}
{O'Dea}, C.~P., {Koekemoer}, A.~M., {Baum}, S.~A., {et~al.} 2001, \aj, 121,
  1915

\bibitem[{{Prandoni} {et~al.}(2009){Prandoni}, {de Ruiter}, {Ricci}, {Parma},
  {Gregorini}, \& {Ekers}}]{prandoni09}
{Prandoni}, I., {de Ruiter}, H.~R., {Ricci}, R., {et~al.} 2009, ArXiv e-prints

\bibitem[{{Rawlings} {et~al.}(1989){Rawlings}, {Saunders}, {Eales}, \&
  {Mackay}}]{rawlings89}
{Rawlings}, S., {Saunders}, R., {Eales}, S.~A., \& {Mackay}, C.~D. 1989,
  \mnras, 240, 701

\bibitem[{{Schawinski} {et~al.}(2009){Schawinski}, {Lintott}, {Thomas},
  {Sarzi}, {Andreescu}, {Bamford}, {Kaviraj}, {Khochfar}, {Land}, {Murray},
  {Nichol}, {Raddick}, {Slosar}, {Szalay}, {Vandenberg}, \&
  {Yi}}]{schawinski09}
{Schawinski}, K., {Lintott}, C., {Thomas}, D., {et~al.} 2009, \mnras, 396, 818

\bibitem[{{Shen} {et~al.}(2003){Shen}, {Mo}, {White}, {Blanton}, {Kauffmann},
  {Voges}, {Brinkmann}, \& {Csabai}}]{shen03}
{Shen}, S., {Mo}, H.~J., {White}, S.~D.~M., {et~al.} 2003, \mnras, 343, 978

\bibitem[{{Smol{\v c}i{\'c}}(2009)}]{smolcic09}
{Smol{\v c}i{\'c}}, V. 2009, \apjl, 699, L43

\bibitem[{{Stoughton} {et~al.}(2002){Stoughton}, {Lupton}, {Bernardi},
  {Blanton}, {Burles}, {Castander}, {Connolly}, {Eisenstein}, {Frieman},
  {Hennessy}, {Hindsley}, {Ivezi{\'c}}, {Kent}, {Kunszt}, {Lee}, {Meiksin},
  {Munn}, {Newberg}, {Nichol}, {Nicinski}, {Pier}, {Richards}, {Richmond},
  {Schlegel}, {Smith}, {Strauss}, {SubbaRao}, {Szalay}, {Thakar}, {Tucker},
  {Vanden Berk}, {Yanny}, {Adelman}, {Anderson}, {Anderson}, {Annis},
  {Bahcall}, {Bakken}, {Bartelmann}, {Bastian}, {Bauer}, {Berman},
  {B{\"o}hringer}, {Boroski}, {Bracker}, {Briegel}, {Briggs}, {Brinkmann},
  {Brunner}, {Carey}, {Carr}, {Chen}, {Christian}, {Colestock}, {Crocker},
  {Csabai}, {Czarapata}, {Dalcanton}, {Davidsen}, {Davis}, {Dehnen},
  {Dodelson}, {Doi}, {Dombeck}, {Donahue}, {Ellman}, {Elms}, {Evans}, {Eyer},
  {Fan}, {Federwitz}, {Friedman}, {Fukugita}, {Gal}, {Gillespie}, {Glazebrook},
  {Gray}, {Grebel}, {Greenawalt}, {Greene}, {Gunn}, {de Haas}, {Haiman},
  {Haldeman}, {Hall}, {Hamabe}, {Hansen}, {Harris}, {Harris}, {Harvanek},
  {Hawley}, {Hayes}, {Heckman}, {Helmi}, {Henden}, {Hogan}, {Hogg}, {Holmgren},
  {Holtzman}, {Huang}, {Hull}, {Ichikawa}, {Ichikawa}, {Johnston}, {Kauffmann},
  {Kim}, {Kimball}, {Kinney}, {Klaene}, {Kleinman}, {Klypin}, {Knapp},
  {Korienek}, {Krolik}, {Kron}, {Krzesi{\'n}ski}, {Lamb}, {Leger},
  {Limmongkol}, {Lindenmeyer}, {Long}, {Loomis}, {Loveday}, {MacKinnon},
  {Mannery}, {Mantsch}, {Margon}, {McGehee}, {McKay}, {McLean}, {Menou},
  {Merelli}, {Mo}, {Monet}, {Nakamura}, {Narayanan}, {Nash}, {Neilsen},
  {Newman}, {Nitta}, {Odenkirchen}, {Okada}, {Okamura}, {Ostriker}, {Owen},
  {Pauls}, {Peoples}, {Peterson}, {Petravick}, {Pope}, {Pordes}, {Postman},
  {Prosapio}, {Quinn}, {Rechenmacher}, {Rivetta}, {Rix}, {Rockosi}, {Rosner},
  {Ruthmansdorfer}, {Sandford}, {Schneider}, {Scranton}, {Sekiguchi}, {Sergey},
  {Sheth}, {Shimasaku}, {Smee}, {Snedden}, {Stebbins}, {Stubbs}, {Szapudi},
  {Szkody}, {Szokoly}, {Tabachnik}, {Tsvetanov}, {Uomoto}, {Vogeley}, {Voges},
  {Waddell}, {Walterbos}, {Wang}, {Watanabe}, {Weinberg}, {White}, {White},
  {Wilhite}, {Wolfe}, {Yasuda}, {York}, {Zehavi}, \& {Zheng}}]{stoughton02}
{Stoughton}, C., {Lupton}, R.~H., {Bernardi}, M., {et~al.} 2002, \aj, 123, 485

\bibitem[{{Strateva} {et~al.}(2001){Strateva}, {Ivezi{\'c}}, {Knapp},
  {Narayanan}, {Strauss}, {Gunn}, {Lupton}, {Schlegel}, {Bahcall}, {Brinkmann},
  {Brunner}, {Budav{\'a}ri}, {Csabai}, {Castander}, {Doi}, {Fukugita}, {Gy{\H
  o}ry}, {Hamabe}, {Hennessy}, {Ichikawa}, {Kunszt}, {Lamb}, {McKay},
  {Okamura}, {Racusin}, {Sekiguchi}, {Schneider}, {Shimasaku}, \&
  {York}}]{strateva01}
{Strateva}, I., {Ivezi{\'c}}, {\v Z}., {Knapp}, G.~R., {et~al.} 2001, \aj, 122,
  1861

\bibitem[{{Tadhunter} {et~al.}(1998){Tadhunter}, {Morganti}, {Robinson},
  {Dickson}, {Villar-Martin}, \& {Fosbury}}]{tadhunter98}
{Tadhunter}, C.~N., {Morganti}, R., {Robinson}, A., {et~al.} 1998, \mnras, 298,
  1035

\bibitem[{{Tremaine} {et~al.}(2002){Tremaine}, {Gebhardt}, {Bender}, {Bower},
  {Dressler}, {Faber}, {Filippenko}, {Green}, {Grillmair}, {Ho}, {Kormendy},
  {Lauer}, {Magorrian}, {Pinkney}, \& {Richstone}}]{tremaine02}
{Tremaine}, S., {Gebhardt}, K., {Bender}, R., {et~al.} 2002, \apj, 574, 740

\bibitem[{{Whittle}(1985)}]{whittle85}
{Whittle}, M. 1985, \mnras, 213, 33

\bibitem[{{Willott} {et~al.}(1999){Willott}, {Rawlings}, {Blundell}, \&
  {Lacy}}]{willott99}
{Willott}, C.~J., {Rawlings}, S., {Blundell}, K.~M., \& {Lacy}, M. 1999,
  \mnras, 309, 1017

\bibitem[{{York} {et~al.}(2000){York}, {Adelman}, {Anderson}, {Anderson},
  {Annis}, {Bahcall}, {Bakken}, {Barkhouser}, {Bastian}, {Berman}, {Boroski},
  {Bracker}, {Briegel}, {Briggs}, {Brinkmann}, {Brunner}, {Burles}, {Carey},
  {Carr}, {Castander}, {Chen}, {Colestock}, {Connolly}, {Crocker}, {Csabai},
  {Czarapata}, {Davis}, {Doi}, {Dombeck}, {Eisenstein}, {Ellman}, {Elms},
  {Evans}, {Fan}, {Federwitz}, {Fiscelli}, {Friedman}, {Frieman}, {Fukugita},
  {Gillespie}, {Gunn}, {Gurbani}, {de Haas}, {Haldeman}, {Harris}, {Hayes},
  {Heckman}, {Hennessy}, {Hindsley}, {Holm}, {Holmgren}, {Huang}, {Hull},
  {Husby}, {Ichikawa}, {Ichikawa}, {Ivezi{\'c}}, {Kent}, {Kim}, {Kinney},
  {Klaene}, {Kleinman}, {Kleinman}, {Knapp}, {Korienek}, {Kron}, {Kunszt},
  {Lamb}, {Lee}, {Leger}, {Limmongkol}, {Lindenmeyer}, {Long}, {Loomis},
  {Loveday}, {Lucinio}, {Lupton}, {MacKinnon}, {Mannery}, {Mantsch}, {Margon},
  {McGehee}, {McKay}, {Meiksin}, {Merelli}, {Monet}, {Munn}, {Narayanan},
  {Nash}, {Neilsen}, {Neswold}, {Newberg}, {Nichol}, {Nicinski}, {Nonino},
  {Okada}, {Okamura}, {Ostriker}, {Owen}, {Pauls}, {Peoples}, {Peterson},
  {Petravick}, {Pier}, {Pope}, {Pordes}, {Prosapio}, {Rechenmacher}, {Quinn},
  {Richards}, {Richmond}, {Rivetta}, {Rockosi}, {Ruthmansdorfer}, {Sandford},
  {Schlegel}, {Schneider}, {Sekiguchi}, {Sergey}, {Shimasaku}, {Siegmund},
  {Smee}, {Smith}, {Snedden}, {Stone}, {Stoughton}, {Strauss}, {Stubbs},
  {SubbaRao}, {Szalay}, {Szapudi}, {Szokoly}, {Thakar}, {Tremonti}, {Tucker},
  {Uomoto}, {Vanden Berk}, {Vogeley}, {Waddell}, {Wang}, {Watanabe},
  {Weinberg}, {Yanny}, \& {Yasuda}}]{york00}
{York}, D.~G., {Adelman}, J., {Anderson}, Jr., J.~E., {et~al.} 2000, \aj, 120,
  1579

\end{thebibliography}
\end{document}